\documentclass[a4paper,11pt]{article}
\usepackage{jcappub} % for details on the use of the package, please see the JINST-author-manual
\usepackage{xcolor}
\usepackage{systeme}
\usepackage{booktabs}
\usepackage{multirow}
\usepackage{adjustbox}
\usepackage{hyperref}

\title{\boldmath On the impact of the supernova subsamples in reducing the Hubble tension}

\author[a,b]{Gonçalo Martins,}
\author[b]{Santiago González-Gaitán,}
\author[a]{João Duarte,}
\author[a]{and Ana M. Mourão}
\affiliation[a]{CENTRA, Instituto Superior Técnico, Universidade de Lisboa,\\ Av. Rovisco Pais 1, 1049-001 Lisboa, Portugal}
\affiliation[b]{Instituto de Astrofísica e Ciências do Espaço, Faculdade de Ciências, Universidade de Lisboa, Ed. C8, Campo Grande, 1749-016
 Lisbon, Portugal}

\emailAdd{goncalo.jose.lourenco.martins@tecnico.ulisboa.pt}

\abstract{The persistent 4–6$\sigma$ discrepancy between early- and late-time measurements of the Hubble constant ($H_{0}$) is known as the "Hubble tension" and represents one of the major open problems in modern cosmology. In this work, we investigate how differences in light-curve parameters such as colour ($c$) and stretch ($x_{1}$),  and host galaxy properties such as the host stellar mass ($M$) and specific star formation rate (sSFR) between the calibration and Hubble Flow (HF) Type Ia supernovae (SNe Ia) samples used by SH0ES affect the SN luminosity standardization and the $H_{0}$ estimation. To do that,  we generate subsamples from both samples and use them to estimate $H_{0}$, $M_{B}$, $\alpha$, $\beta$, $\Delta_{host}$ and $\sigma_{int}$. Both one- and multi-dimensional Kolmogorov–Smirnov tests are used to evaluate the consistency between the subsamples property distributions and analyze how the estimated parameters vary with the better matching of the subsamples. We find that the calibration sample is not fully representative of the HF sample, particularly in the host's $M$ and sSFR. Improving consistency between subsamples leads to changes in $H_{0}$
  , $M_{B}$ and $\sigma_{int}$ %
  as the samples become more consistent, although overall values remain broadly stable. More consistent subsamples also tend to produce a mass step consistent with zero within 1$\sigma$. When disentangling SN subpopulations using different approaches, we identify persistent differences in $H_{0}$ ($\sim$ 2–3$\sigma$) and $M_{B}$ ($\sim$ 2$\sigma$) between low- and high-stretch SN~Ia subpopulations: we find $H_{0} = 75.27 \pm 1.18$ km s$^{-1}$ Mpc$^{-1}$ for low-stretch and $H_{0}$ = 71.25 $\pm$ 1.59 km s$^{-1}$ Mpc$^{-1}$ for high-stretch, resulting in a Hubble tension of 6.07 $\sigma$ and 2.52 $\sigma$, respectively. These differences highlight how multiple SNIa sub-populations likely present variations in dust properties and intrinsic colour that are not captured by the standard single $\beta$ parameterization and that affect cosmology. By considering these two SN subpopulations when estimating a single Hubble parameter for all SNe, we obtain $H_{0} = 73.78 \pm 2.17\ \mathrm{km\ s^{-1}\ Mpc^{-1}}$, with a much larger uncertainty than normally quoted and resulting in decreasing the tension with the Planck measurement from $5.87\sigma$ to about $2.86\sigma$.
  Moreover, our results also suggest that the mass step may arise from an over-correction of more than one SN subpopulation associated to different environments.}

\begin{document}
\maketitle
\flushbottom

\section{Introduction}
\label{sec:intro}
According to the Friedmann-Lemaître equations, the expansion of the Universe is redshift-dependent and can be characterized by the Hubble parameter $H(z)$. The Hubble constant ($H_{0}$) corresponds to the present time ($z$=0) value of this parameter, giving us the Universe's local expansion rate and providing crucial information about its age and history \citep{Freedman_2010}.

Given the importance of this parameter to understand our Universe, two main pathways of calculating the value of $H_{0}$ were developed and improved over the years. The first approach relies on early universe measurements such as the Cosmic Microwave Background (CMB), which assumes a flat
 $\Lambda$CDM (Lambda Cold Dark Matter) cosmological model  \citep{cdm-model}, to infer local values of expansion, i.e. $H_{0}$ = 67.4 $\pm$ 0.5 km s$^{-1}$ Mpc$^{-1}$ \citep{Planck_2018}. The second approach relies on later measurements from the local universe based on the cosmic distance ladder involving typically observations of SNe Ia and closer luminous objects such as Cepheids to calibrate their distances, favoring higher values  of $H_{0}$ = 73.04 $\pm$ 1.04 km s$^{-1}$ Mpc$^{-1}$ \citep{Riess_2022}. 

Analyzing the different estimations of $H_{0}$ from the early and late-time measurements it is possible to identify a consistent statistically significant tension that varies from 4 to 6$\sigma$ disagreement between both predictions, with higher values of $H_{0}$ obtained locally. This discrepancy is known as the "Hubble tension" and is considered one of the major open problems in modern cosmology. This  difference can indicate the need for new physics \citep[e.g.,][]{hu2023hubbletensionevidencenew} or, at the very least, unexpectedly large systematic errors in either or both of the two principal measurements.

The cosmic distance ladder is a technique used to determine cosmological
distances in the nearby universe, starting by using geometric distance measurements, such as parallax, to calibrate nearby standard candles located in our galaxy such as Cepheid variables. These are luminous variable stars of different brightnesses but with a well-defined Period-Luminosity relation \citep{Leavitt_1908}. Modern calibrations have refined this relation to include metallicity effects, leading to the Period-Luminosity-Metallicity \citep[PLZ;][]{Groenewegen_2018}. Once calibrated, this relation can be used to obtain the distances to Cepheids located in nearby galaxies, which can not be precisely  computed through parallax. These nearby galaxies can also contain type Ia Supernovae (SNe Ia), which are very bright explosions of carbon-oxygen white dwarfs in binary systems \citep{1960ApJ...132..565H} allowing us to detect them in even more distant galaxies.
Even though SNe Ia do not have a known universal peak magnitude, they serve as standardizable candles whose observed light curves can be used to obtain their distances. The observation of nearby events of this type showed that all explosions had quite similar peak luminosities and light curves with relatively small variations \citep{Branch_1992} that can be standardized by applying some empirical corrections based on the light curve shape-luminosity \citep{1993ApJ...413L.105P} and colour-luminosity relations expressed in the two-parameter correction model proposed by \cite{Tripp1998}. However, as observed by \cite{Sullivan2010} and \cite{Kelly2010}, after the application of these two corrections,
SNe Ia detected in higher-mass galaxies appear consistently
over-luminous, such that a significant mass step is evident in the difference between their corrected and predicted luminosities. The nature of this third correction is still not fully understood. 

It is important to make a clear distinction between the two samples of SNe Ia involved in the distance ladder.  On the one hand, we have the calibration sample, which includes SNe Ia located in nearby galaxies that also host Cepheid variables. This sample allows us to break the well-known degeneracy between the estimated absolute peak magnitude of SNe in the B-band ($M_{B}$) and $H_{0}$ \citep[e.g.,][]{Camarena_2021}. On the other hand, we have the Hubble Flow (HF) sample, which consists of SNe Ia in more distant galaxies. Together with the calibration sample, this sample enables us to estimate all the correction parameters used in the SNe Ia luminosities standardization, as well as the $M_{B}$ and $H_{0}$ parameters \citep{Riess_2016}. 

Although the current study will only focus on the traditional Cepheid-based distance ladder, several alternative methods have been adopting Brightness Fluctuations \citep[SBF;][]{Khetan_2021}, J-region Asymptotic Giant Branch \citep[JAGB;][]{Huang_2020}, and the Tip of the Red Giant Branch \citep[TRGB;][]{Freedman_2019} stars as complementary or substitute nearby distance indicators to Cepheids. Among all $H_0$ estimations relying on a local calibration, the TRGB-based approach, which  results in a value of $H_0$ = 69.8 $\pm$ 1.7 km s$^{-1}$ Mpc$^{-1}$ \citep{Freedman_2019}, is the one that exhibits the largest discrepancy (at the level of $\sim 1.7 $ $\sigma$) compared to the Cepheid-based determination by \cite{Riess_2022}. Additional efforts have explored the use of Type II supernovae as alternative cosmological standard candles \citep{de2020measurement,vogl2024rungsattacheddistanceladderfree}, as well as entirely independent ladder methods such as standard sirens \citep{sirens} and gravitational lensing time-delay cosmography \citep{birrer2024timedelaycosmographymeasuringhubble}.

A variety of potential systematic effects associated with the general framework of supernova analysis such as photometric calibrations, light curve fitting, Milky Way extinction, and the accuracy of redshifts, including peculiar velocity corrections have been demonstrated to have negligible influence on the determination of the Hubble constant, as shown by \cite{Brout_2022} and \cite{Steinhardt_2020}.  A detailed review of proposed solutions for the Hubble tension can be found in \cite{Di_Valentino_2021}. 

Cepheid variables are predominantly found in late-type, star-forming galaxies such as the Milky Way, and are normally observed in face-on galaxies to reduce the effects from enhanced crowding and reddening issues \citep[e.g.,][]{anderson2024cepheid}. To mitigate potential biases in dust extinction and reddening between the calibration sample SNe and those from the Hubble Flow hosted in passive environments
\citep[e.g.,][]{Rigault_2015},  \cite{Riess_2022} selected similar late-type galaxies for the calibration and HF samples, assuming that the 
dust extinction laws are correlated with the
star formation rate (SFR) and the stellar mass formed in each galaxy \citep[e.g.,][]{Salim_2018} and so that the extinction law can be customized based on the galaxy type. Following this assumption, large, star-forming spiral hosts of SNe Ia  have,  presumably, a similar dust extinction law to the Milky Way, which helps maintain consistency across all rungs of the local distance ladder.  However, this selection alone may not be sufficient to ensure that the dust properties are consistent between the samples. Moreover,  the extinction model adopted in the simulations that are used for bias
corrections in the SNe Ia luminosity standardization in studies such as the one by \cite{Brout_2022} is trained on the unrestricted Hubble flow sample, which includes galaxies of all morphological types, and then extrapolated to the calibration galaxies \citep[e.g.,][]{Popovic_2023}, potentially introducing a significant bias in the correction.

As shown in several studies such as \cite{Sullivan2010} and \cite{pruzhinskaya2020dependence}, there is a dependency between the SNe light curve properties and their host environment and morphology. Studies by \cite{Nicolas_2021} and more recently by \cite{Ginolin_2024} suggest that the stretch-magnitude relation might be non-linear with a strong correlation with the SN environment and redshift, contrarily to the relation commonly adopted in cosmological analyses. This non-linearity may reflect the presence of distinct SN Ia populations, as revealed by the bimodal behavior identified in the stretch distribution of both studies. Consistently, \cite{wojtak2025stretchstretchdustdust} show that this non-linearity in stretch is an emergent property resulting from mixing two populations with effective magnitude offset (see their Fig 3). Specifically, a low-stretch mode is associated with SNe Ia in old stellar populations, typically hosted by redder and/or more massive galaxies, unlike the high-stretch mode that is shared by both young and old environments. These results  are also in agreement with the findings by \cite{Rigault_2020},
which confirms that the SN Ia light curve stretch distribution reflects an intrinsic SN property that depends on the progenitor age, with younger SNe Ia predominantly populating the higher stretch mode and constituting an intrinsically more homogeneous population. 
The work done by \cite{Wojtak_2022} also claims for a strong evidence that SN standardization within the calibration sample requires a significantly steeper colour correction slope compared to the Hubble Flow sample. This discrepancy points to an intrinsic tension between the calibration and HF samples and assumes the existence of overlapping SN Ia subpopulations, primarily distinguished by differences in their mean stretch parameter. According to \cite{10.1093/mnras/stad2590}, SNe belonging to the high-stretch population tend to be intrinsically bluer and exhibit approximately twice the level of dust reddening along their lines of sight relative to those from the low-stretch population.

Other studies such as the one by  \cite{Duarte_2023} reinforce the idea that the mass step correction is partially driven by differences in supernova intrinsic and extrinsic properties, finding that this effect cannot be entirely explained as a dust systematic, as suggested by \cite{Brout_2021}. There are also some works that report a relation between the galaxies ages and the mass step, as the one by \cite{Chung_2023} and
\cite{10.1093/mnras/stad488}, and that the mass step can vary based on whether a supernova is located in the inner or outer region of the galaxy \citep{Toy_2024}.

Considering all the results above, it is possible to conclude that the accuracy of the distance ladder method also relies on the precise matching of the supernovae population properties and their host environments both in the calibration and in the Hubble Flow galaxies, which is taken for granted in almost all studies that use this technique to estimate the $H_{0}$ values. However, we can have a calibration sample whose SNe Ia are not representative of the ones that we can find in the HF sample, both in terms of their intrinsic properties as in terms of their host galaxies properties, which can introduce a possible bias in the calibration. Furthermore, supernovae that appear very similar in terms of their light-curve properties may still differ in their intrinsic and environmental properties. A clear example of this effect is the apparent color of SNe Ia that results from a combination of the intrinsic SN colour and intervening dust along
the line-of-sight (e.g. \cite{Santiago_2021,10.1093/mnras/stac2714}). However, intrinsic SN colours are expected to vary (e.g.,~\cite{Burns_2014}), while dust reddening effects also appear to vary with the environment (e.g.,~\cite{duarte2025assessing}).

In this study, we propose to investigate and compare the supernovae light curve parameters as well as their host galaxy properties of both the calibration and Hubble Flow samples, see if there are subsamples from the calibration and HF samples that better match each other,  understand the impact that this sample matching has in the Hubble parameter estimation and derive a more accurate calibration for SNe Ia luminosities. The dataset used in this study and its treatment are detailed in Section \ref{sec:data}. The methodology used to obtain the correction parameters, as well as the values of the SNe absolute magnitude in the B-band (M$_{B}$) and $H_{0}$ is presented in Section \ref{sec:methodology}. The analysis and results can be found in Section \ref{sec:analysis}. In Subsection \ref{sub:full_sample}, we compare both calibration and HF SNe samples property distributions, while in Subsection \ref{sub:subsample_cal}  we generate pairs of random subsamples from the calibration and Hubble flow samples. In Subsection \ref{sub:matching} we see if there is any relation between the better matching of the generated subsamples property distributions and the estimated $H_{0}$, $M_{B}$, $\alpha$, $\beta$, $\Delta_{host}$ and $\sigma_{int}$. We further investigate the impact of accounting for the properties of the SNe within the generated subsamples in our analysis in Subsection \ref{sub:subsamples_props}, as well as the effect of using SNe from different property bins on the estimated $H_{0}$, $M_{B}$, and the correction parameters in Subsection \ref{sub:bins}, as a way to assess the different SN subpopulations within the sample. In particular, we focus on stretch-based subpopulations and construct subsamples from distinct stretch bins in Subsection \ref{sub:bins}.  Finally, we present a final discussion and the main conclusions of this study in Sections \ref{sec:discussion} and \ref{sec:conclusion}, respectively. 

\section{Data}
\label{sec:data}
We use the Pantheon+SH0ES compilation of SNe Ia data\footnote{ \url{https://github.com/PantheonPlusSH0ES}} from the analysis by  \cite{Riess_2022} and \cite{Brout_2022}. The SNe light curves covered by this compilation are fitted using the SALT2 model originally developed by  \cite{Guy_2010} returning parameters such as the peak magnitude in the B-band ($m_{B}$), the colour parameter defined as the difference between the magnitude in the B and V bands at the fitted epoch of peak brightness ($c$) and the  stretch ($x_{1}$), which corresponds to a dimensionless parameter that quantifies the width of the light curve. The compilation also includes the log($M/M_{\odot}$) and log(sSFRyr) values for each SN host galaxy, estimated from multi-band global photometry and galaxy SED fitting, as described in \cite{Scolnic_2022}. In addition, a covariance matrix ($C$) resulting from the sum of the statistical and systematic covariance matrixes including the uncertainties of supernova corrected magnitudes and distances moduli from Cepheids is also provided in the released data.

This compilation includes repeated SNe corresponding to the same event observed in different surveys which introduces non-diagonal terms for the same SN in the statistical covariance matrix representing measurement noise from components other than the light curve itself. To simplify the analysis and remove these non-vanishing off-diagonal terms, we combined the repeated observations of each SN into a single representative measurement per SN, obtaining one set of light-curve parameters $c$, $m_{B} $ and $x_{1}$, their corresponding uncertainties, and the covariance between them as follows. Assuming that the light curve parameters follow Gaussian distributions for each repeated observation $i$ of a given SN $j$, the mean values of these parameters can be determined from the Gaussian distribution resulting from the product of all individual distributions of the repeated observations \citep{PP}. The mean parameter values vector ($\nu_{j}$) and the corresponding covariance matrix ($\Sigma_{j}$) of the resulting distribution for a given SN $j$ with $N$ observations can be computed as
\begin{equation}
\centering
    \nu_{j} = \Sigma_{j} \cdot \left(\sum^{N}_{i=1}\Sigma^{-1}_{i} \cdot \nu_{i} \right) \hspace{7mm}, \hspace{3mm}\quad
   \Sigma_{j} = \left(\sum^{N}_{i=1}\Sigma^{-1}_{i}\right)^{-1}
\end{equation}

\noindent where $\nu_{i}$ represents the mean parameter values vector for each observation $i$ of the SN $j$ and $\Sigma_{i}$ the corresponding covariance matrix for that observation, both defined as:

\begin{equation}
\centering
    \nu_{i} =
    \begin{bmatrix}
    m_{B}\\
     x_{1}\\ c
    \end{bmatrix} \hspace{8mm} ,\quad 
    \Sigma_{i} = \begin{bmatrix}
    \sigma^{2}_{m_{B}} & \sigma_{m_{B},x_{1}} & \sigma_{m_{B},c}\\
    \sigma_{x_{1},m_{B}} & \sigma^{2}_{x_{1}} & \sigma_{x_{1},c}\\ \sigma_{c,m_{B}}& \sigma_{c,x_{1}}& \sigma^{2}_{c}
    \end{bmatrix} 
\end{equation}

\noindent where the diagonal terms of the covariance matrix $\sigma_{m_{B}}$, $\sigma_{x_{1}}$ and $\sigma_{c}$ correspond to the uncertainties associated with each of the light curve fit parameters while the non-diagonal elements correspond to the covariance between them. All parameters defining $\nu_{i}$ and $\Sigma_{i}$ for the repeated SNe are provided in the Pantheon+SH0ES compilation. 

There are 7 SNe from the analysis of \cite{Riess_2022} that present values of log(sSFRyr) $<$ - 15, which seem to be outliers lower limits on the true values of this parameter, which is a common interpretation in the literature \citep[e.g.,][]{Feldmann_2017,Hahn_2019}. To ensure a more accurate comparison and analysis of the sSFR distributions between the two samples, we chose to retain only those with sSFR values that lie within one standard deviation ($1\sigma$) of the mean for both the calibration and Hubble Flow distributions  (a brief discussion about the effect of including these SNe is provided in the appendix \ref{ap:ssfr_effect}). Although we adopt the stellar mass values provided by \cite{Brout_2022} and used in the analysis of \cite{Riess_2022}, it is worth noting that some of these masses also appear to be in conflict with independent estimates from the literature, as highlighted by \cite{wojtak2024consistentextinctionmodeltype} and \cite{2025arXiv251220834P}.

\begin{figure}[h]
\centering  
\includegraphics[width=0.7\linewidth]{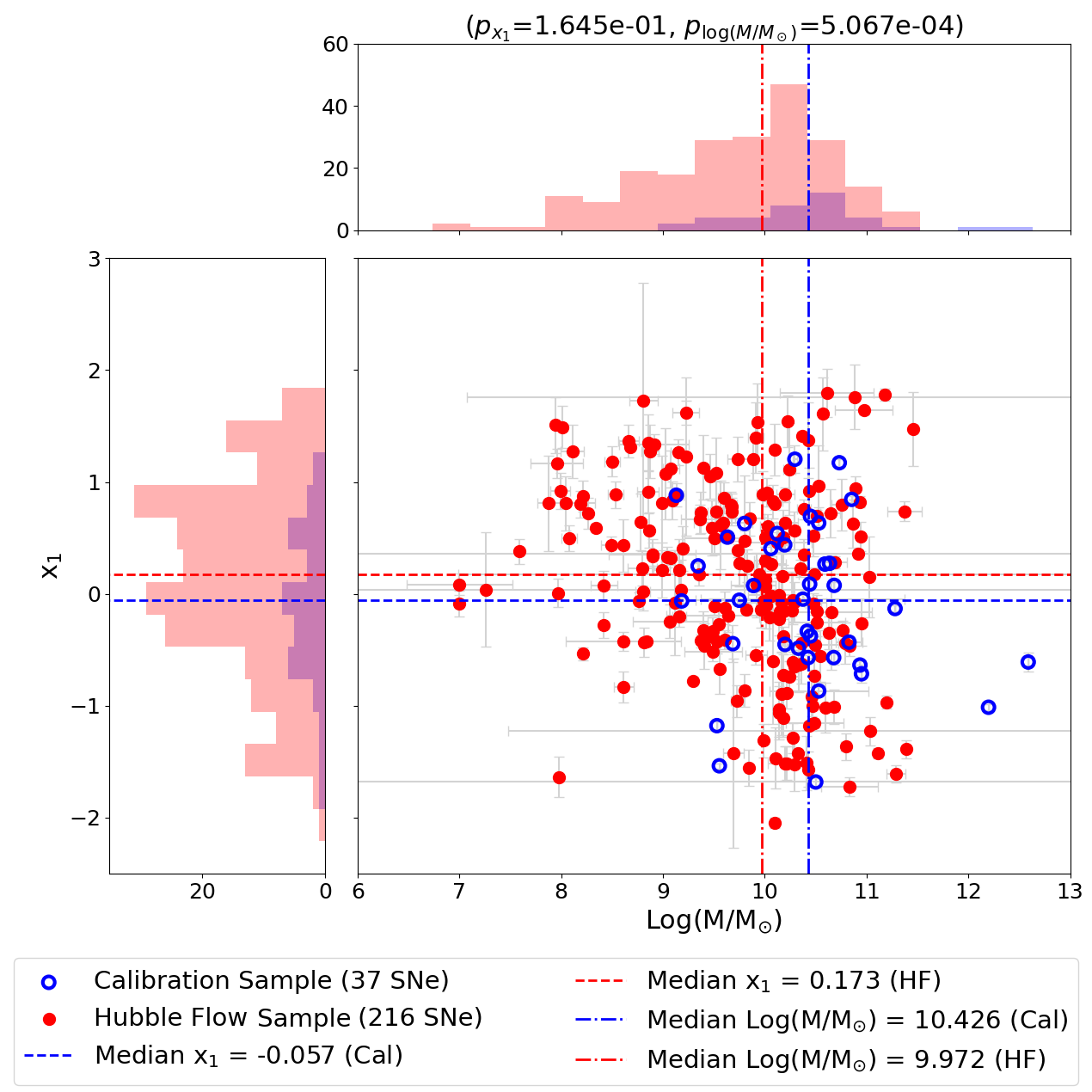}
    \caption{Strech ($x_{1}$) as a function of the logarithm of the stellar mass ($M$) in solar units ($M_\odot$) for supernovae from both calibration (blue, open circles) and Hubble flow (red, filled circles) samples. The dashed horizontal lines correspond to the median value of the light curves parameters distributions and the dashed-dotted vertical lines to the median value of the host properties distributions for each sample depending on the colour. The K-S test $p$-values obtained by comparing the parameters distributions of the calibration and Hubble flow samples are also displayed at the top of the Figure.}
    \label{fig:x1_vs_mass}
    
\end{figure}

\begin{figure}[]
\centering  
\includegraphics[width=0.7\linewidth]{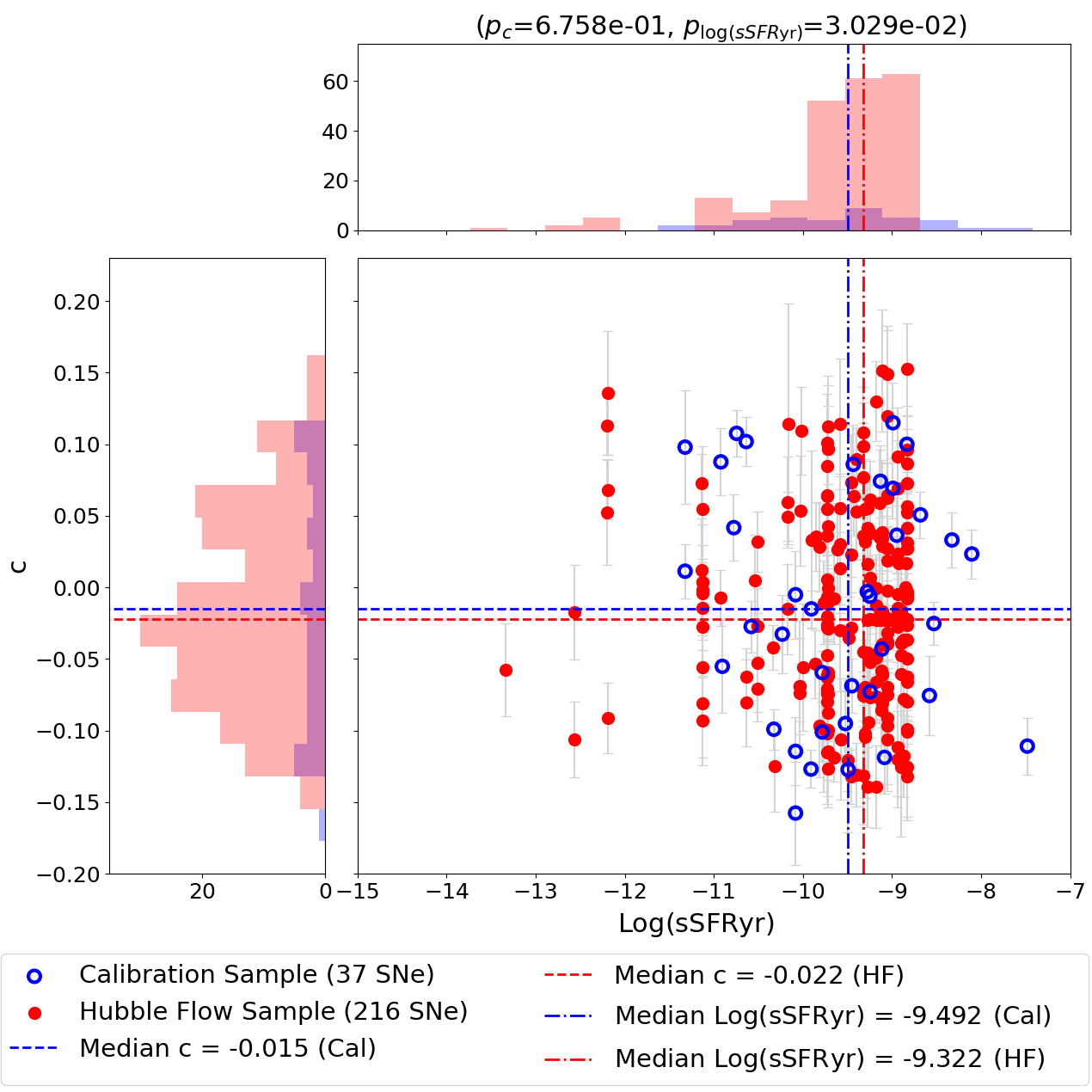}
    \caption{Similar to Fig. \ref{fig:x1_vs_mass} for colour ($c$) and the logarithm of the specific star formation rate (sSFR).}
\label{fig:c_vs_ssfr}
\end{figure}

\begin{figure}[]
    \centering
    \includegraphics[width=0.7\linewidth]{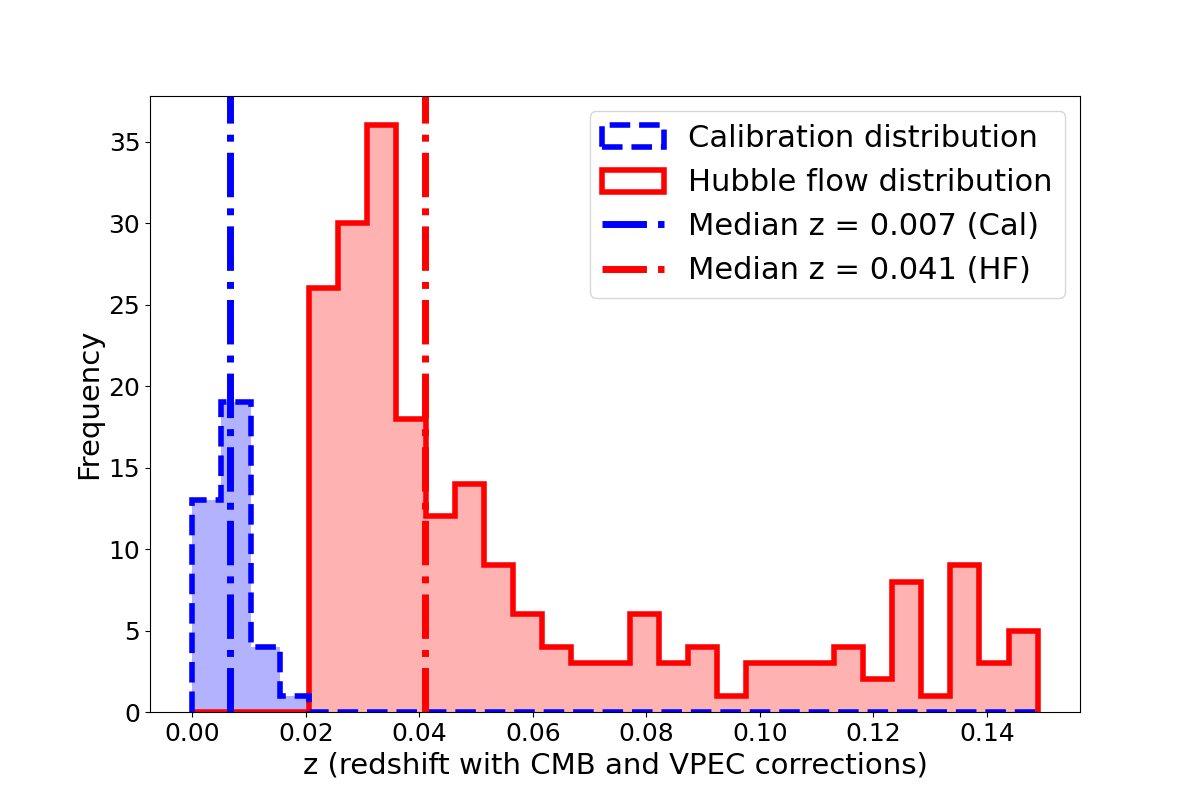}
    \caption{Redshift ($z$) distribution for both calibration (blue, delineated by dashed lines) and Hubble Flow (red, delineated by solid lines) SNe samples. The redshifts are corrected for both the CMB and peculiar velocities (VPEC) and are provided in the Pantheon+SH0ES compilation \citep{Riess_2022,Brout_2022}. The vertical dash-dotted lines indicate the redshift medians of each sample.}
    \label{fig:redshift}
\end{figure}

This treatment left us with 37 different SNe in the calibration sample and 216 SNe in the Hubble Flow sample.
 The light curve parameters $c$ and $x_{1}$ and host galaxies' stellar mass ($M$) in solar mass units ($M_{\odot}$) and specific star formation rate (sSFR) in $yr^{-1}$ distributions are shown in Figures \ref{fig:x1_vs_mass} and \ref{fig:c_vs_ssfr}. The redshift ($z$) distribution of both samples is also presented in Figure \ref{fig:redshift}.

% ===== RIGHT COLUMN ===== %
%\switchcolumn

% --- Top image: Figure 2 --- %

% --- Redshift image directly below --- %
\section{Methodology}
\label{sec:methodology}
The corrected SN magnitude using the \cite{Tripp1998} correction model accounting for the mass step \citep[e.g.,][]{Sullivan2010} is given by

\begin{equation}
m_{B}^{corr}=m_{B}+\alpha x_{1}-\beta c + \delta_{host}
\label{trippformula}
\end{equation}
where $\alpha$ and $\beta$ are the model coefficients that describe the shape-luminosity and colour-luminosity relations, respectively, while $\delta_{host}$ represents the mass step correction term depending on the SNe host stellar mass $M$ as
\begin{equation}
\label{eq:mass-step}
\centering
 \delta_{host}=
\systeme{
 + \Delta_{host} \text{ if $\log(M/M_{\odot}$) $<$ } \text{$\log({M_{step}}/M_{\odot})$},
 - \Delta_{host} \text{ if $\log(M/M_{\odot}$) $\geq$ } \text{$\log({M_{step}}/M_{\odot})$}}
\end{equation} 
with $\Delta_{host}$ corresponding to the step magnitude, $M_{step}$ the stellar mass threshold at which the step correction is applied and $M_{\odot}$ the solar mass. While most studies adopt a fixed value of $M_{step}$ = 10 $M_{\odot}$, in this analysis we will set $M_{step}$ = $\tilde{M}$, with the latter representing the median value of $M$ for the subsamples used to obtain the parameters.

To properly account for both intrinsic and dust reddening effects on luminosity, some models assume that the observed SN colour $c$ is the sum of two components: the intrinsic colour $c 
_{int}$, which is related to luminosity through an intrinsic colour–luminosity relation with slope $\beta_{int}$, and the colour excess due to dust, commonly referred to as dust reddening, $E(B - V)$ which is related to luminosity through the total-to-selective extinction coefficient 
$R_{B}$ = $R_{V}$ + 1, representing the external host-galaxy dust extinction component \citep{Brout_2021}. 
Some studies also include an additional bias correction term $\delta_{\text{bias}}$ in equation \ref{trippformula} to account for observational selection biases determined from simulations, as well as second order corrections arising from full forward modelling of SN colour parameters and observed magnitudes based
on physically motivated models of dust and supernova intrinsic colour \citep[e.g.,][]{Popovic_2023}.

By adopting a simplified $\beta$-colour correction, excluding the $\delta_{\text{bias}}$ correction term from our analysis and selecting subsamples of the calibration and HF samples that are closely matched in their properties (see Section \ref{sec:analysis}), we are implicitly assuming that extinction and intrinsic colour effects are indistinguishable and identical for all SNe within the subsample. This approach avoids relying on dust and colour correction models that might not capture the full complexity of dust extinction of SNe Ia and allows us to assess the raw impact of differences in the property distributions of the calibration and Hubble flow SNe samples, as well as the effects of distinct SN subpopulations, on luminosity standardization. These effects may otherwise be overlooked by such corrections, which can themselves be incorrectly assumed to apply equally to both samples \citep[e.g.,][]{Wojtak_2022,wojtak2024consistentextinctionmodeltype}. The motivation for not including bias correction, as well as the resulting effects, are discussed in more detail in Appendix \ref{ap:bias}.

Knowing that the distance modulus $\mu$ is defined as the difference between the standardized apparent and absolute peak magnitudes in the B-band, $m_{B}^{corr}$ and $M_{B}$, and recalling the expression (\ref{trippformula}) we can define: 
\begin{equation}
\centering
\mu=m_{B}^{corr}-M_{B}=m_{B}-M_{B}+\alpha x_{1}-\beta c + \delta_{host} .
\label{eq:mu}
\end{equation}

The distance modulus can also be expressed in terms of the luminosity distance $d_{L} (z)$, and the latter can be estimated using the Hubble parameter $H(z)$.  We can then relate $\mu$(z) for a given redshift with the local Hubble constant $H_{0}$ by expanding $H(z)$ to the first orders for low redshifts, leading to \citep[see][]{1993ppc..book.....P}

\begin{align}
%\begin{split}
\centering
\mu(z) &\approx 5 \log_{10}\Bigg(\frac{c}{H_{0}}[z+\frac{1}{2}(1-q_{0})z^{2}
-\frac{1}{6}(1-q_{0}-3q_{0}^2 + j_{0})z^{3}] \Bigg) + 25
%\end{split}
\label{mu_red}
\end{align}
with $q_{0}$ = - 0.51 being the deceleration parameter obtained by \cite{Riess_2022}, and $j_{0}$ = 1 representing the jerk \citep{Visser_2004}. 

To constrain the $H_{0}$ and $M_{B}$, as well as all the correction parameters $\alpha$, $\beta$ and $\Delta_{host}$, we apply a Bayesian fitting procedure to minimize the value of the likelihood function described by

\begin{equation}
\label{eq:new_likelihood}
\centering
ln(\mathcal{L})=-\frac{1}{2}\sum^{N}_{i=0}\left[ (\Delta\xi_{i})^{2} + ln(2\pi\sigma_{i}^2 )\right],
\end{equation} 
where $\Delta \xi_{i}$ corresponds to the $i$-th component of the solution vector obtained from the system of linear equations defined for each SN $i$ given by

\begin{equation}
\centering\Delta \mu = C \Delta\xi \Leftrightarrow \Delta\xi =C^{-1} \Delta \mu  
\end{equation} 
that takes into account all the covariance matrix terms and the vector $\Delta\mu$, which consists of the Hubble residuals for each SN $i$, defined as $\Delta\mu_{i}=\mu_{i} - \mu_{i,model}$,   where $\mu_{i}$ is given by (\ref{eq:mu}). If a SN $i$ belongs to the calibration sample, then its distance comes from Cepheids and $\mu_{i,model}$ = $\mu_{i,ceph}$, whereas if it is in the Hubble Flow sample we use $\mu_{i,model}$ = $\mu_{i}(z)$ given by equation (\ref{mu_red}).

We have also included the Gaussian normalization term, $\ln(2\pi\sigma_{i}^{2})$, in equation (\ref{eq:new_likelihood}), as some studies \citep[e.g.,][]{Kessler_2017} point out that leaving this term out from the likelihood function can introduce significant biases in the recovered fit parameters. The $\sigma_{i}$ denotes the $i$-th term of the diagonal of the statistical and systematic covariance matrix $C$ provided in the released data and is defined as

\begin{equation}
\centering
\sigma_{i}^{2}=\sigma^{2}_{m_{B}}+(\alpha \sigma_{x_{1}})^{2}+(\beta \sigma_{c})^{2}-2\beta\sigma_{m_{B},c} + 2\alpha\sigma_{m_{B},x_{1}}
-2\alpha\beta\sigma_{x_{1},c}+\sigma^{2}_{int}+\sigma^{2}_{lens}+\sigma^{2}_{z}+\sigma^{2}_{vpec} 
\label{eq:sigma}
\end{equation} 
with $\sigma_{m_{B}}$, $\sigma_{x_{1}}$ and $ \sigma_{c}$ corresponding to the uncertainties associated with the light curve fit parameters of each SN, $\sigma_{m_{B},c}$, $\sigma_{m_{B},x_{1}}$ and  $\sigma_{x_{1},c}$ representing their covariance terms and $\sigma_{int}$ being a free parameter that accounts for possible intrinsic variations in the SNe luminosities.  The parameters $\sigma_{lens}$, $\sigma_{z}$ and $\sigma_{vpec}$ represent the uncertainty contribution from gravitational lensing, as given by \cite{Jonsson_2010}, and the redshift and peculiar velocity measurements uncertainties, respectively.

To constrain and estimate our own correction parameters $\alpha$, $\beta$, $\Delta_{host}$ and $\sigma_{int}$ that appear both in the modified Tripp formula (\ref{trippformula}) and in the diagonal terms of the covariance matrix (\ref{eq:sigma}), we subtracted the diagonal terms that explicitly depend on them, adopting the values of $\alpha$ = 0.148 and $\beta$ = 3.112 from \cite{Brout_2022}, along with the $\delta_{host}$, $\sigma_{int}$, and covariances values between $m_{B}$, $c$ and $x_{1}$ provided in the data compilation. We then reintroduced these terms into the covariance matrix setting the correction parameters as free parameters in the Bayesian fitting procedure that relies on the Monte Carlo Markov
Chain (MCMC) technique, a sampling method used to generate distributions of values by drawing samples from a probability distribution based on independent observations implemented in the \textit{emcee}\footnote{\url{https://github.com/dfm/emcee}} python package \citep{emcee}. The parameter space was sampled using 128 walkers and 800 iterations for each fit, with an initial burn-in of 64 steps which ensures parameter convergence for all cases. For the covariance values, we used those derived during the unification of SNe parameter values, as described in Section \ref{sec:data}. 

\section{Analysis}  
\label{sec:analysis}

\subsection{Comparing the full calibration and Hubble flow samples}
\label{sub:full_sample}

To analyze and compare the light curve parameters and host properties distributions between the calibration and Hubble Flow samples  we used the python implementation\footnote{\url{https://docs.scipy.org/doc/scipy/reference/generated/scipy.stats.ks_2samp.html}} of the one-dimensional two-sample Kolmogorov–Smirnov (K–S) test provided by the SciPy library \cite{2020SciPy-NMeth} and introduced by \cite{an1933sulla} and \cite{smirnov1939estimation}. This test is used to compare two independent samples to assess whether they come from the same distribution, evaluating the maximum vertical difference between the empirical cumulative distribution functions of both. The decision is translated on the resulting $p$-value. If the $p$-value is less than the significance level of 0.05, the null hypothesis that the two samples come from the same distribution might be rejected. Otherwise, we can say that both distributions are in agreement and might come from the same distribution.

As we can see in the top of Figures \ref{fig:x1_vs_mass} and \ref{fig:c_vs_ssfr} , the  $x_{1}$ and $c$ distributions of the calibration and Hubble Flow samples seem to be in 
concordance with each other according to the obtained $p$-values of 1.645 $\times$ $10^{-1}$ and 6.758 $\times$ $10^{-1}$, respectively. However, for the host $M$ and sSFR distributions, we can see that the estimated $p$-values shown in the upper parts of both Figures are lower than 0.05, with values of 5.067 $\times$ 10$^{-4}$ and 3.029 $\times$ 10$^{-2}$, respectively. This means that these properties distributions are not in agreement and most likely come from different distributions, leading us to the conclusion that SNe in the calibration sample are not representative of all the SNe found in the HF sample in terms of these two host properties. The calibration sample has no SN located in lower mass galaxies  ($\log$($M/M_{\odot}$) $<$ 9), contrarily to the Hubble Flow sample. The calibration sample also seems to have SN hosts with $\log(\mathrm{sSFRyr}) \gtrsim -8.8$, which are not present in the HF sample. Although high stellar masses are generally expected for quiescent galaxies (e.g., \cite{Peng_2010, zahid2012}), we find that the star-forming galaxies in the calibration sample are predominantly massive. This likely reflects the known correlation between stellar mass and star-formation rate in star-forming galaxies, where higher-mass galaxies tend to have higher star-formation (e.g, \cite{Brinchmann_2004,Noeske_2008}). Consequently, the calibration sample is naturally biased toward very massive star-forming galaxies, which host large populations of young stars, including Cepheids.

To estimate the Hubble constant, as well as all the other standardization parameters, we will use the approach described in Section \ref{sec:methodology}. The parameters estimated using the full sample can be found in Table \ref{tab:parameters}. The estimated value of $H_0$ is consistent with the value of 73.04 $\pm$ 1.04 km s$^{-1}$ Mpc$^{-1}$ obtained by \cite{Riess_2022}, with a concordance of 0.53$\sigma$. The differences in the obtained values are likely driven by the change in the fitting methodology, the omission of bias corrections, the use of a single representative parameter value for repeatedly observed SNe, and the exclusion of certain SNe from the sample. 

\begin{table}[h]
\centering
\resizebox{\columnwidth}{!}{%
\begin{tabular}{ccccccc}

\hline
Sample &
$H_{0}$ & M$_{B}$ & $\alpha$ & $\beta$  & $\Delta_{host}$  & $\sigma_{int}$ \\ \hline

\\ Full sample &
73.781$^{+0.970}_{-0.944}$& -19.199$^{+0.027}_{-0.027}$	& 
0.136$^{+0.010}_{-0.010}$& 
2.777$^{+0.116}_{-0.111}$& 
-0.020$^{+0.009}_{-0.008}$& 
0.092$^{+0.009}_{-0.009}$ \\ 

\\ \hline

\end{tabular}
}
\caption{Median parameters values obtained using the full Calibration and Hubble flow sample and their respective difference to the 16th and 84th  percentiles.}
\label{tab:parameters}
\end{table}

\subsection{Comparing calibration and Hubble Flow subsamples}

\label{sub:subsample_cal}

In order to evaluate the matching effect on the estimation of the $H_{0}$ and the correction parameters $\alpha$, $\beta$, $\Delta_{host}$ and $\sigma_{int}$ we base our approach on two considerations. First, as shown in the previous Subsection \ref{sub:full_sample}, the full calibration sample is not fully representative of the HF sample in terms of stellar mass and sSFR distributions and we want to obtain more consistent properties in the calibration and HF samples. Second, there is growing evidence that SNe Ia have multiple subpopulations \citep[e.g.,][]{wang2013evidence,10.1093/mnras/stad2590, Ginolin_2024} that may exist even in the calibration sample. For these reasons, we generate a total of 4000 paired subsamples, each consisting of one calibration subsample and one Hubble Flow  subsample. For each pair, we obtain the $p$-value for the $c$, x$_{1}$, $\log$($M/M_{\odot}$) and $\log$(sSFRyr) distributions by applying the K-S test between their properties distributions, and estimated the standardization parameters for each one, as well as the $H_{0}$ to see if there is any correlation between the concordance of the distributions and the parameters estimated.

To ensure statistical significance when defining the minimum calibration subsample size, we adopted the number of SNe used in the calibration sample by \cite{Dhawan_2018} to measure the $H_{0}$ as a reference, setting a lower limit of 10 SNe for each calibration subsample and an upper limit corresponding to the full calibration sample of 37 SNe. For the HF subsamples, the lower limit was set equal to the number of SNe in the corresponding calibration subsample, while the upper limit was defined as the integer part of 5.6 times that number. This factor corresponds to the approximated ratio between the total number of SNe in the full Hubble Flow sample (216 SNe) and that in the full calibration sample (37 SNe). To ensure that no subsamples were repeated, we excluded those with identical $p$-values for two or more of the compared distributions, resulting in 3940 unique subsamples.

\begin{figure*}[h]
    \centering
        \includegraphics[width=\linewidth]{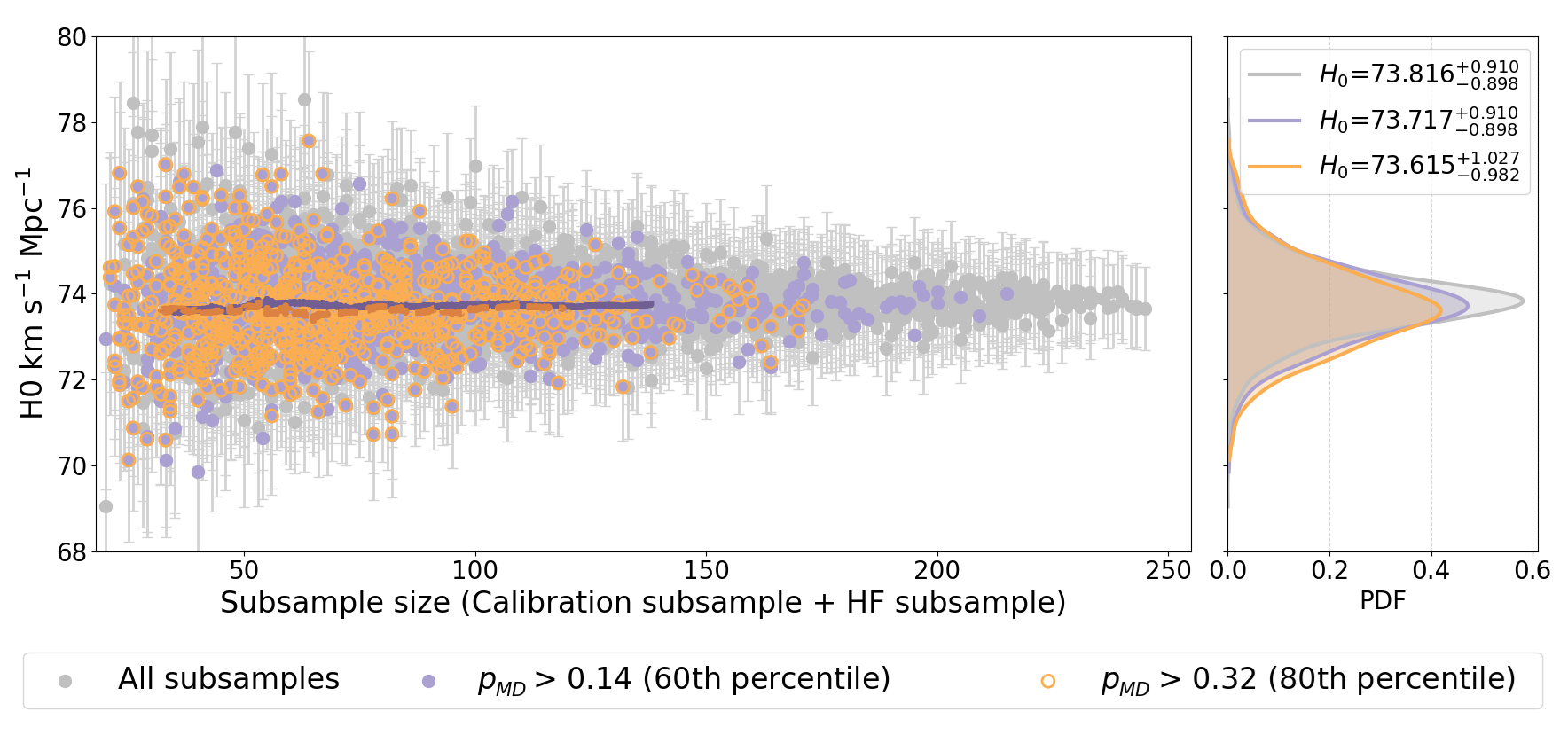}
    
    \caption{$H_{0}$ as a function of the generated subsample size, using paired subsamples drawn from the calibration and HF samples. Grey dots show the full subsample distribution, lilac dots highlight subsamples with $p_{MD}$ above the 60th percentile, and orange open dots indicate those above the 80th percentile.}
    \label{fig:size_pvalues_subsamples}
\end{figure*}

\begin{figure*}[h]
    \centering
        \includegraphics[width=\linewidth]{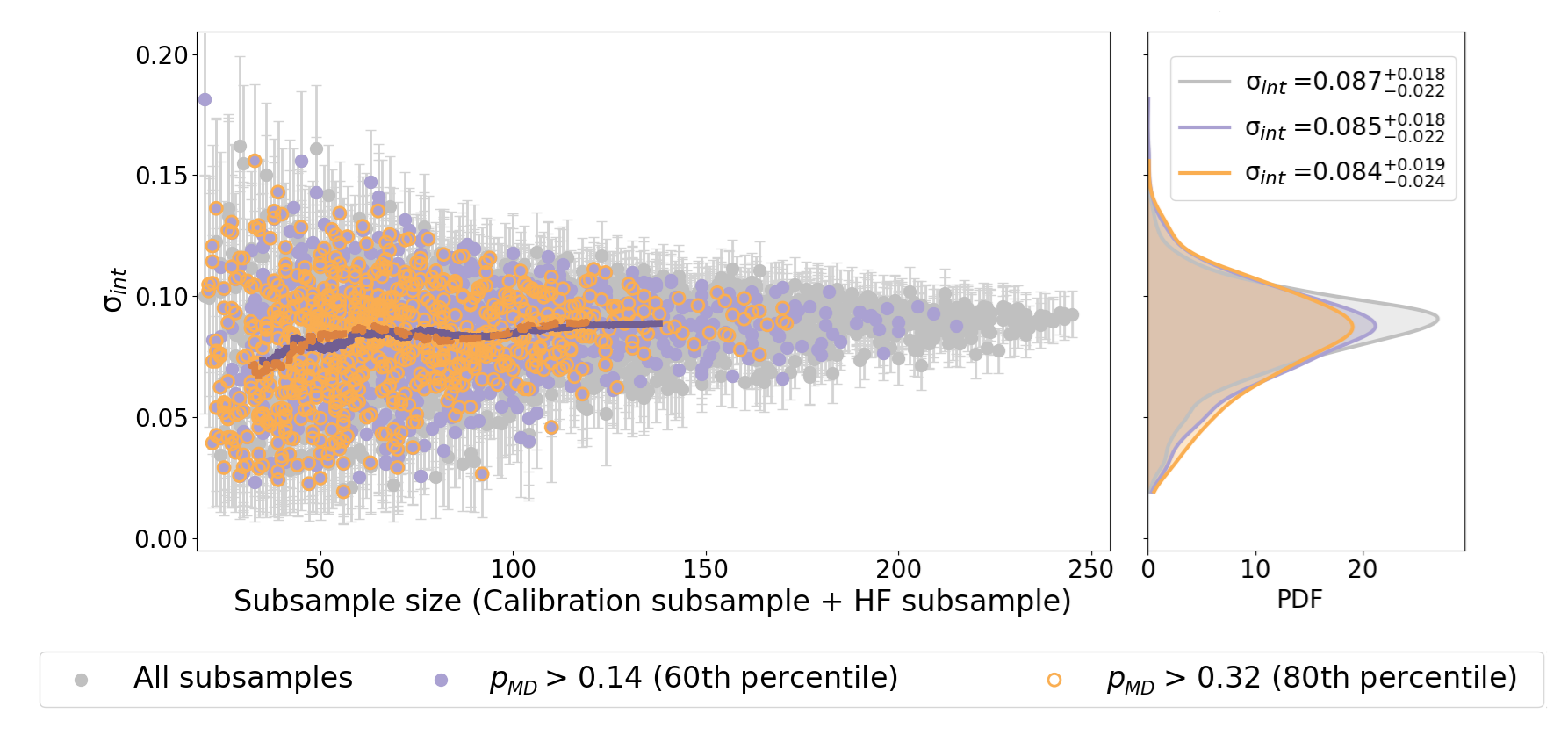}
    
    \caption{Similar to Fig. \ref{fig:size_pvalues_subsamples} for $\sigma_{int}$.}
    \label{fig:size_pvalues_subsamples_sigma}
\end{figure*}

In Figure \ref{fig:size_pvalues_subsamples} and \ref{fig:size_pvalues_subsamples_sigma}, we present the estimated values of $H_{0}$ and $\sigma_{int}$, respectively, derived from the 3940 randomly generated subsamples. In this analysis, we use the python implementation\footnote{\url{https://github.com/wmpg/fasano-franceschini-test}} \citep{CHOW2025116444} of the multi-dimensional generalization of the two-sample K–S test described by \cite{1987MNRAS.225..155F} to assess the overall concordance between the light curve parameters and host property distributions for each pair of generated subsamples. %To avoid confusion between the $p$-values obtained from comparing individual property distributions of calibration and HF subsamples and those obtained using the multi-dimensional test, 
We refer to the latter as $p_{MD}$ throughout this work. Other metrics are explored %could also be applied which is why we provide an additional analysis 
in appendix~\ref{ap:other_metrics} showing always consistent results.
%, where the procedure is repeated using alternative metrics to evaluate overall distributional agreement. In all cases, the results are consistent with those obtained using the multi-dimensional generalization of the K-S test. 
We also computed the rolling median of the parameter distributions as it reduces the discrete nature of binning and provides a statistically more reliable trend than comparing medians obtained from equally-populated or equally-spaced bins \citep{binning}. Although not shown here, the rolling median clearly confirms that there is no noticeable trend between the estimated parameters and the size of generated subsamples.

The median $p_{MD}$ estimated for all the generated subsamples is approximately 0.097, indicating a good overall agreement between the property distributions of the calibration and HF subsamples. The $H_{0}$ values estimated using subsamples with $p_{MD}$ above the 60th (violet circles) and 80th percentiles (orange open circles) are also shown, observing a slight reduction in the median $H_{0}$ as more consistent subsamples are used, although the values remain fully consistent within their uncertainties.  No significant dependence of the estimated parameters on the subsample size is observed, with the exception of $\sigma_{int}$: as shown in Figure \ref{fig:size_pvalues_subsamples_sigma}, it tends to decrease as the subsample size becomes smaller.

We note that we also generated subsamples only from the HF sample while keeping the full calibration sample fixed. However, using this approach, most comparisons of the log($M/M_{\odot}$) and
log(sSFRyr) distributions still yield low $p$-values, with the majority below 0.05.

%This would make the analysis very challenging, as most subsamples would fail to fully represent the properties of the calibration sample, similar to what occurs with the full sample. Interestingly, there was no subsample giving $H_{0}$ < 71 km s$^{-1}$ Mpc$^{-1}$ when we considered the full calibration sample, which can indicate that a subsample from the Hubble Flow sample capable of independently resolving the Hubble tension may not exist, regardless of its agreement with the entire calibration sample or the properties we consider.

\subsection{Matching effect on the estimated parameters}

\label{sub:matching}
\begin{figure}[h]
    \centering

        \includegraphics[width=0.8\linewidth]{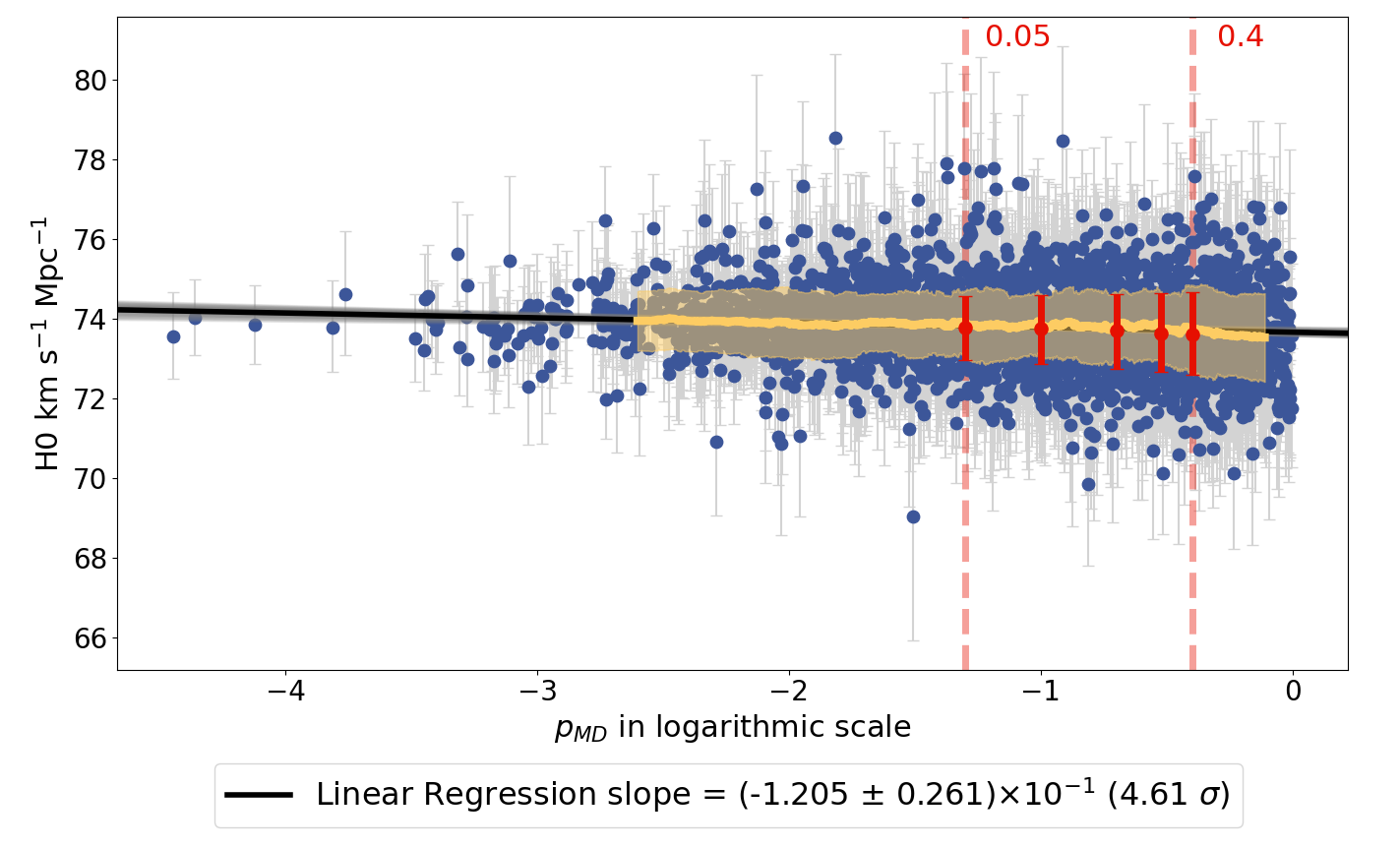}
    
    \caption{$H_{0}$ as a function of the $p_{MD}$ obtained for each generated calibration subsample when compared to the corresponding HF subsample. The yellow line represents the rolling median of each parameter, computed using a window size of 100 subsamples, while the shaded area represents its rolling standard deviation. The red dots correspond to the median values of the parameter distribution using subsamples with $p$-values higher than 0.05, 0.1, 0.2 and 0.4 . The red error bars indicate the difference between the 16th and 84th percentiles relative to the median values estimated for each bin. The dashed red line marks $p_{MD}$ = 0.05 and $p_{MD}$ = 0.4 on the logarithmic scale.}
    \label{fig:medianH0_cal}
\end{figure}

To investigate whether a better overall match of all $c$, $x_{1}$, log($M/M_{\odot}$) and
log(sSFRyr) distributions between the calibration and HF subsamples affects the estimated $H_{0}$, $M_{B}$, $\alpha$, $\beta$, $\Delta_{host}$ and $\sigma_{int}$, we performed a linear regression between the obtained parameters and the logarithm of their corresponding $p_{MD}$ using \textsc{linmix} \cite{Kelly_2007} to account for parameter uncertainties in the fit. In addition, we analyzed the median values of these parameters considering only subsamples with $p_{MD}$ above 0.05, 0.1, 0.2, 0.3, and 0.4, as illustrated in the Figure \ref{fig:medianH0_cal} for the $H_{0}$ case. While the median values of the considered parameters across subsamples with different $p$-value thresholds remain consistent within their uncertainties, the linear regression analysis reveals statistically significant slopes not only for $H_{0}$ which decreases with better subsample matching, but also for other parameters, as shown in Table \ref{tab:slopes}.

\begin{table}[h]
\centering

\begin{tabular}{ccc}

\hline
Parameter &
Linear Regression Slope & Sig.\\ 

\hline

\addlinespace
 
$H_{0}$ & (-1.205 $\pm$ 0.261) $\times$ 10$^{-1}$ & (4.61 $\sigma$) \\

\addlinespace

\hline

\addlinespace
 
$M_{B}$ & (-3.916 $\pm$ 0.632) $\times$ 10$^{-3}$ & (6.19 $\sigma$) \\

\addlinespace

\hline

\addlinespace
 
$\alpha$ & (9.109 $\pm$ 3.075) $\times$ 10$^{-4}$ & (2.96 $\sigma$) \\

\addlinespace

\hline

\addlinespace
 
$\beta$ & (1.613 $\pm$ 0.444) $\times$ 10$^{-2}$ & (3.63 $\sigma$) \\

\addlinespace

\hline

\addlinespace
 
$\Delta_{host}$ & (-4.740 $\pm$ 3.485) $\times$ 10$^{-4}$ & (1.36 $\sigma$) \\

\addlinespace

\hline

\addlinespace
 
$\sigma_{int}$ &  (-1.619 $\pm$ 0.293) $\times$ 10$^{-3}$ & (5.52 $\sigma$) \\

\addlinespace

\hline

\end{tabular}

\caption{Slopes and uncertainties resulting from the linear regression between $H_{0}$, $M_{B}$, $\alpha$, $\beta$, $\Delta_{host}$, and $\sigma_{\mathrm{int}}$ and the the logarithm of $p_{MD}$, along with the corresponding significance levels. The fitting procedure is performed using \textsc{linmix} \cite{Kelly_2007}.}
\label{tab:slopes}

\end{table}

\begin{figure}[h]
    \centering

        \includegraphics[width=0.8\linewidth]{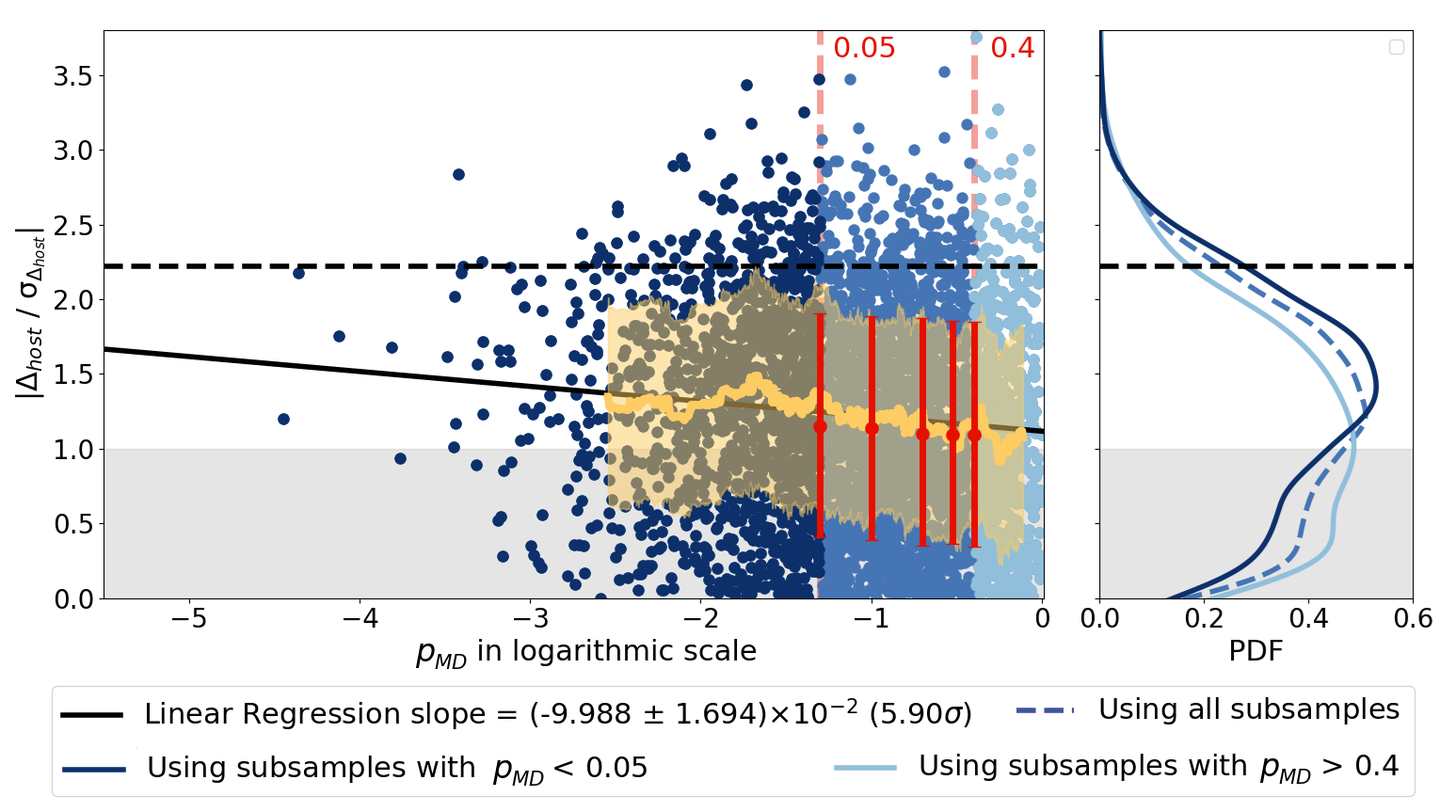}
    
    \caption{Similar to Fig. \ref{fig:medianH0_cal} for $\Delta_{\rm host}/\sigma_{\Delta_{\rm host}}$, where the dark blue dots represent subsamples with $p_{MD}$ below 0.05, medium blue those with $p_{MD}$ between 0.05 and 0.4, and in light blue those with $p_{MD}$ above 0.4. The black dashed line indicates the value of $\Delta_{\rm host}/\sigma_{\Delta_{\rm host}}$ estimated using the full SNe sample, which is 2.22$\sigma$. The grey shaded area highlights the region where the mass-step is consistent with 0 within $1\sigma$.}
    \label{fig:medianmass_step_cal}
\end{figure}

If distinct SN Ia subpopulations exist, characterized by their stretch \citep[e.g.,][]{10.1093/mnras/stad2590,Ginolin_2024}, and correlating strongly with their host galaxy stellar mass \citep[e.g.,][]{Sullivan2010,Ginolin_2024} and sSFR \citep[e.g.,][]{Sullivan2010,Nicolas_2021}, it is possible that one the two calibration/HF samples is predominantly composed of SNe from a specific subpopulation, as suggested by \cite{wojtak2025stretchstretchdustdust}. Improving thus the match in these property distributions between the calibration and HF samples could indeed influence the estimates of $H_{0}$ and $M_{B}$. The significant decrease in $\sigma_{\mathrm{int}}$ observed with improved matching of the subsamples properties further supports this interpretation. In fact, when performing the linear regression between these parameters and the one dimensional $p$-values obtained from comparing individually each of the $c$, $x_{1}$, $\log(M/M_{\odot})$, and $\log(\mathrm{sSFRyr})$ distributions of calibration vs HF samples, we find significant slopes ($>$ $3\sigma$) for $\log(M/M_{\odot})$ and $\log(\mathrm{sSFRyr})$. However, the most prominent relation arises for the $x_{1}$ distributions, with significances greater than $8\sigma$ in both cases.

We did not find a significant relation between the mass-step magnitude term, $\Delta_{host}$, and the increasing $p_{MD}$ . However, when accounting for the uncertainty in the estimated parameter, we observe that the fraction of subsamples with $\Delta_{host}$ consistent with 0 within $1\sigma$  increases as the match between the property distributions improves. In fact, if we apply a linear fitting between the consistency of the mass-step magnitude with 0 given by $|\Delta_{host}$/$\sigma_{\Delta_{host}}|$ and the $p_{MD}$, we obtain a fitted slope of (-9.998 $\pm$ 1.694) $\times$ 10$^{-2}$ with a significance of 5.90 $\sigma$, as we can see in Figure \ref{fig:medianmass_step_cal}. Furthermore, the probability density function (PDF) estimated for the subsamples with a multidimensional $p$-value greater than 0.4 shows a higher density of subsamples located in the region below $1\sigma$ consistency compared to the distribution estimated from all subsamples, and even more so compared to that of subsamples with $p$-values below 0.05. This trend is accompanied by a progressive decrease in the median value of $|\Delta_{\text{host}}/\sigma_{\Delta_{\text{host}}}|$ and their corresponding differences between the 16th and 84th percentiles. Specifically, the median decreases from $1.326^{+0.689}_{-0.790}$ when considering subsamples with $p_{MD}$ lower than 0.05, to $1.092^{+0.753}_{-0.748}$ when restricting to subsamples with $p$-values higher than 0.4.

Nevertheless, the apparent consistency with zero observed with increasing subsample matching may also be a consequence of the reduction in subsample size. As shown in Fig. \ref{fig:size_pvalues_subsamples}, better-matched subsamples tend to be associated with smaller sizes. Such a reduction is expected to enhance statistical fluctuations and increase the associated uncertainties in the mass-step magnitude, thereby making the mass-step appear more consistent with zero due to the larger uncertainty. A more detailed discussion of the impact of subsample size on the inferred values of $\Delta_{\mathrm{host}}$ is provided in Subsection \ref{sub:mass-step-size}.

\subsection{Accessing SN subpopulations with random subsamples}
\label{sub:subsamples_props}

So far, we have been testing the consistency of mixtures of populations in the calibration and in the HF subsamples, but not necessarily individual populations. Knowing that SN populations are also characterized by distinct intrinsic properties distributions \citep[e.g.,][]{10.1093/mnras/stad2590}, we
analyze here how the estimated parameters
of the most consistent subsamples vary as a function of the median values of their property distributions.

\begin{figure*}[]
    \centering

        \includegraphics[width=\linewidth]{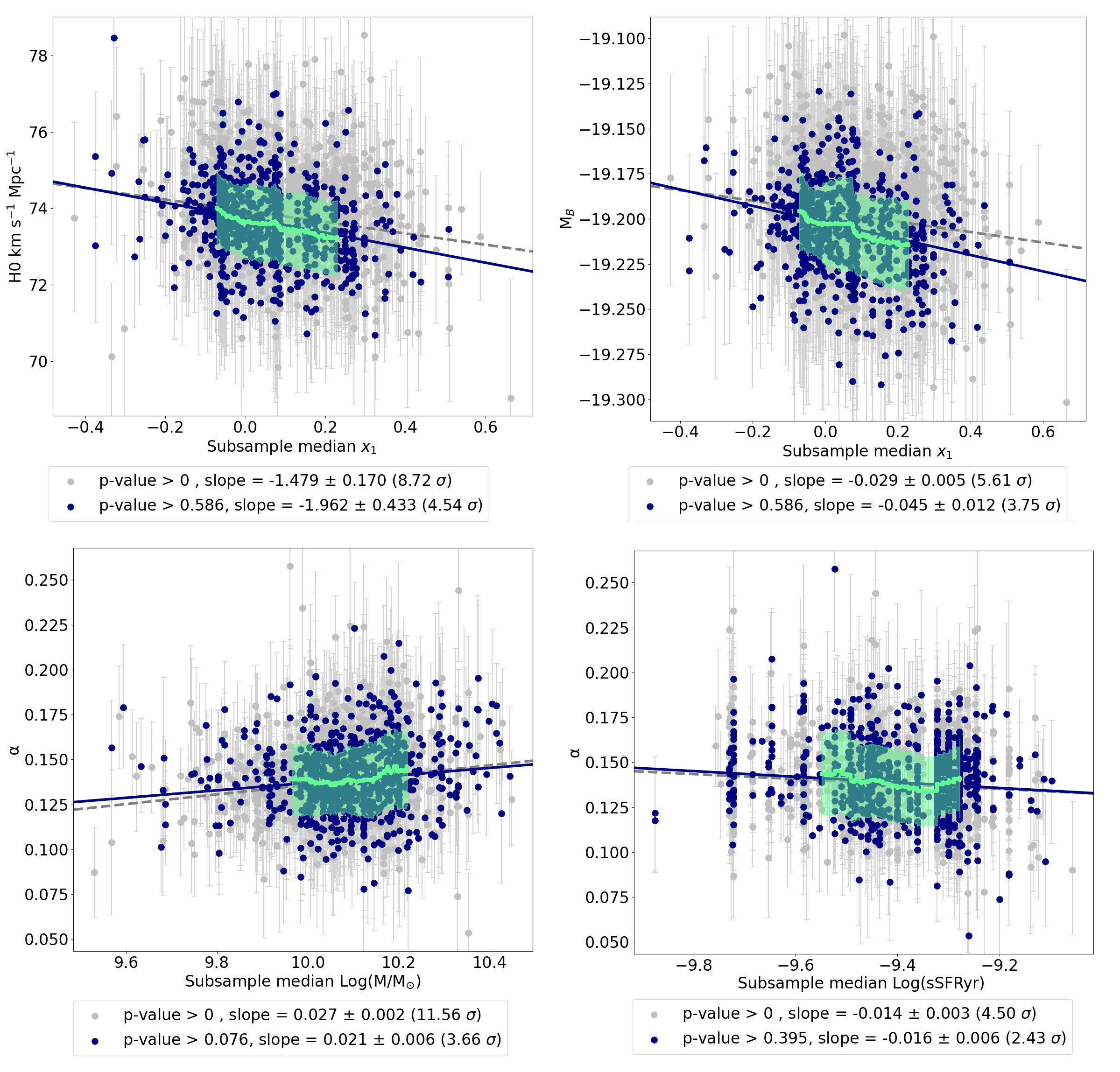}
    
    \caption{Upper panels: Show $H_0$ (left) and $M_B$ (right) as a function of the median value of the $x_1$ distribution for each generated subsample. Grey points represent all generated subsamples, each composed of one randomly drawn subsample from the calibration sample and one from the Hubble Flow sample. Blue points correspond to the best matching subsamples, defined as those whose calibration and Hubble Flow subsamples yield one dimensional $p$-values in the top 20th percentile when comparing their distributions in the considered property. The green line is the rolling median of the estimated parameters $H_0$ and $M_B$ from the best-matching subsamples, with the shaded region representing the corresponding standard deviation. Lower panels: Same analysis for $\alpha$ as a function of the median $\log(M/M_\odot)$ (left) and $\log(\text{sSFRyr})$ (right) . }
    \label{fig:stretch_prova}
\end{figure*}

We choose to show $H_0$ and $M_B$ as a function of the subsample median stretch in Figure \ref{fig:stretch_prova}, as this combination of estimated parameters and the SN light curve parameter show the strongest and most statistically significant relations. %that is clearly visible in the plots and supported by the rolling median, 
This interesting result may indicate a systematic trend where consistent subsamples, presumably composed mainly of SNe with higher $x_{1}$, tend to yield lower values of $H_{0}$ and $M_{B}$ compared to those with lower median $x_{1}$. Moreover, the fitted slope is steeper when considering subsamples whose $p$-values fall within the top 20th percentile of the $p$-value distribution. 

\begin{figure*}[]
    \centering

\includegraphics[width=\linewidth]{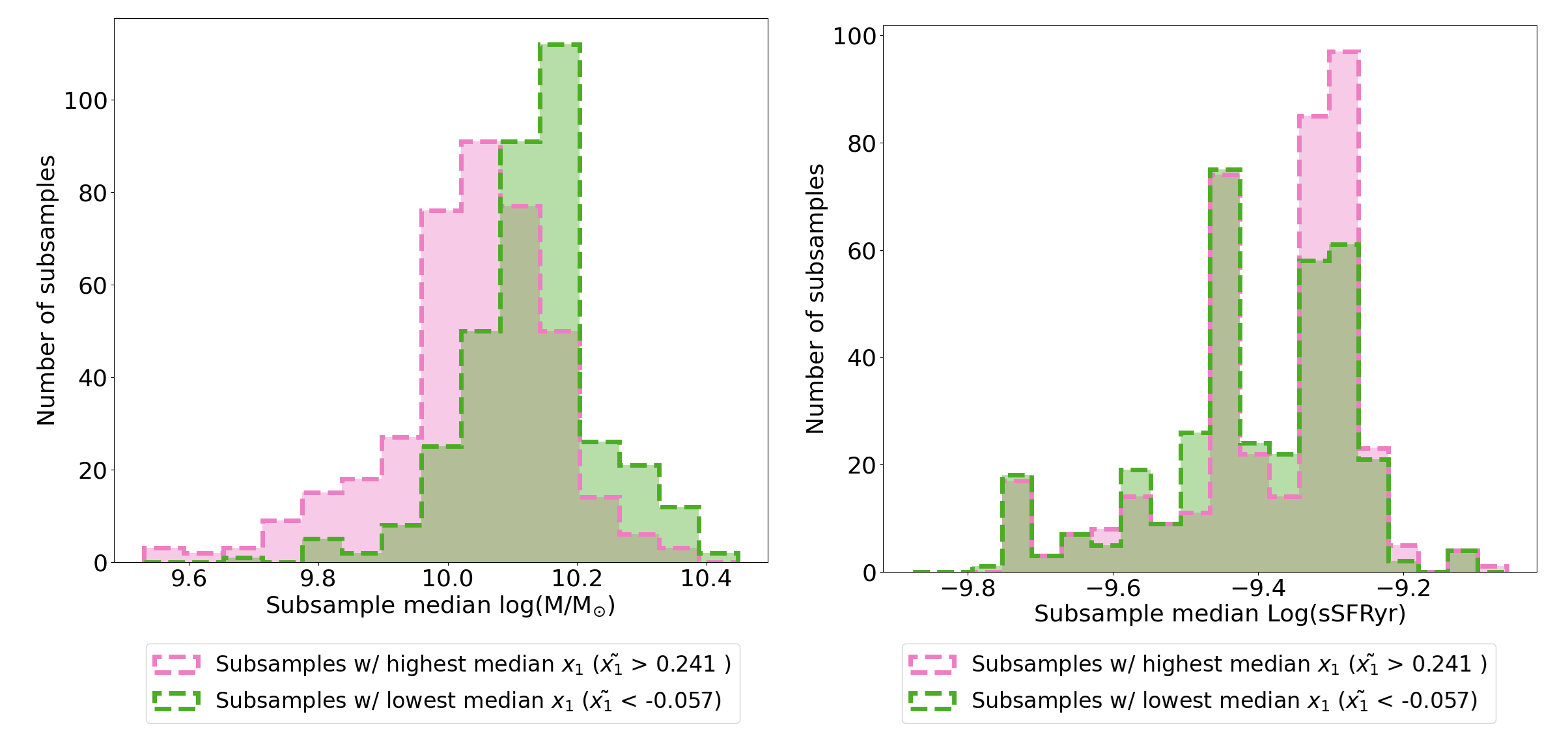}
    
    \caption{Distributions of the median $\log(M/M_\odot)$ (left) and log(sSFRyr) (right) for subsamples with the highest (pink) and lowest (green) median $x_{1}$ values, restricted to those whose calibration and HF components yield one dimensional $p$-values in the top 20th percentile when comparing their $x_{1}$ distributions. The highest and lowest $x_{1}$ bins are defined as those in the top and bottom 10th percentiles of the full subsample median $x_{1}$ distribution, respectively.}
    \label{fig:hist_prova}
    
\end{figure*}

There are also significant relations observed between $\alpha$ and log($M/M_\odot$), and between $\alpha$ and $\log(\text{sSFRyr})$, as shown in the lower panels of the Figure \ref{fig:stretch_prova}, with a slope of 0.027 $\pm$ 0.002 (11.56 $\sigma$) and $-$0.014 $\pm$ 0.003 (4.50 $\sigma$), respectively, when using all random subsamples. These trends may also reflect the influence of underlying SN Ia subpopulations on the luminosity standardization.  Given that SN Ia
stretch correlates with host galaxy properties such as log(M/M$_{\odot}$)
\citep[e.g.,][]{Sullivan2010,Uddin_2017}, this trend could
indeed result from an incomplete stretch–luminosity correction.
If this relation, normally described by a single $\alpha$ parameter, is
not accurately represented by a simple linear model, then residual dependencies may appear in the corrected SN Ia luminosities or in the correction parameters themselves as a function of
host galaxy characteristics rather than $x_{1}$ alone. Recent works by \cite{Larison_2024} and \cite{Ginolin_2024} support this
idea, proposing a non-linear shape–luminosity correction, in contrast to the linear form adopted in the standard SN Ia calibration
(equation \ref{trippformula}). Their findings suggest that low-stretch SNe Ia follow a much steeper magnitude–stretch relation (i.e., a higher $\alpha$) compared to high-stretch SNe Ia, which exhibit a weaker relation (i.e., a lower $\alpha$). Considering the slope fitted for $\alpha$ as a function of the median $\log(M/M_{\odot})$ and log(sSFRyr), our results may indicate a similar effect, although less pronounced and likely mediated by the host environment.

To test whether our subsamples may indeed recover the host stellar mass and sSFR dependence with stretch, we compared the median log($M/M_{\odot}$) and log(sSFRyr) distributions of subsamples with the highest (top 10th percentile) and lowest (bottom 10th percentile) median $x_{1}$ values, as a way of increasing the likelihood of selecting subsamples predominantly composed of SNe with higher or lower stretch, respectively. Starting by the stellar mass distributions, as shown in the left panel of Figure \ref{fig:hist_prova}, the subsamples with the lowest median $x_{1}$ tend to exhibit higher median stellar masses, with a median log(${\tilde{M}/M_{\odot}}$) = 10.142. This bin also shows a narrower mass distribution, as indicated by the difference between the 16th and 84th percentiles (0.162). In contrast, the subsamples with the highest median $x_{1}$ display a broader median mass distribution, spanning a wider range of stellar masses with a percentile difference of 0.227. These subsamples also include lower median mass values, resulting in a lower median of 10.059. This finding strengthens the proposed interpretation,
suggesting that the randomly generated subsamples do reflect the following trend: high-stretch SNe are found in both low- and high-mass galaxies and likely drive the lower $\alpha$ values measured for the low-mass hosts. In contrast, low-stretch SNe mainly occur in massive galaxies, leading to higher $\alpha$ values and thus a higher median of the parameter \citep[e.g.,][]{Ginolin_2024}. The simultaneous presence of both populations in massive galaxies is the most probable explanation for the broader dispersion of $\alpha$ observed at higher stellar masses. Similar findings are reported in \cite{Garnavich_2023}, showing that the estimated slopes become steeper with increasing host mass for fast-declining SNe Ia (defined in this work as
$x_{1}$ $>$ -1) while the $\alpha$ values estimated for slow-declining SNe ($x_{1}$ $<$ -1) remain consistently lower and show no significant dependence on host mass.

For the sSFR, we do not observe as clear a difference in Figure \ref{fig:hist_prova} as we do for the stellar mass. 
%This may be due to the fact that the log(sSFRyr) values could represent only lower limits on the true parameter values, which is a common interpretation in the literature, as previously discussed in Section \ref{sec:data}. 
Nevertheless, we still find that the median log(sSFRyr) estimated for the subsamples with the lowest median $x_{1}$ is lower (log(sSFRyr) = -9.400) than that obtained for the subsamples with the highest median $x_{1}$ (log(sSFRyr) = -9.322), as also found in the literature \citep[e.g.,][]{Sullivan2010,Uddin_2017}. The negative and significant slope found in the linear regression between $\alpha$ and sSFR can therefore be explained by a similar environmental dependence as that reported for stellar mass, since SNe with lower stretch tend to occur in less star-forming galaxies, while higher-stretch SNe are found in galaxies with higher sSFR \citep[e.g.,][]{Sullivan2010,Uddin_2017,Larison_2024}. This correlation can explain the decrease of the stretch–luminosity slope parameter with increasing
$\log(\text{sSFRyr})$. Despite the observed trends between the $\alpha$ parameter and the median environmental properties values typically associated with the different $x_{1}$ SN subpopulations, it remains surprising that no significant relation is observed with the median $x_{1}$ itself, for which a slope of -0.006 $\pm$ 0.003 (2.05 $\sigma$) is obtained. 

A significant slope of $1.161 \pm 0.239$ ($4.86\sigma$) is found when applying a linear regression between $H_{0}$ and $\log(M/M_{\odot})$ using all subsamples, becoming even steeper when only the more consistent subsamples are considered ($1.313 \pm 0.533$). Although this trend could also reflect the influence of underlying SN subpopulations characterized by the $x_{1}$ parameter and their dependence on stellar mass, it is also important to note that lowering the median $\log(M/M_{\odot})$ effectively reduces the mass-step threshold defined for that subsample, which in turn can lead to a reduction of the mass step itself \citep[see Figure 6 from][]{Duarte_2023}. Indeed, we estimate a significant negative slope of $-0.013 \pm 0.003$ ($4.98\sigma$) when comparing $\Delta_{\mathrm{host}}$ with the median $\log(M/M_{\odot})$, capturing this effect and likely contributing to the relation observed between $H_{0}$ and host stellar mass. However, when comparing our lower median value of $\log(M/M_{\odot}) = 9.53$ with the stellar mass interval explored in \cite{Duarte_2023}, the reduction in step magnitude relative to both the threshold yielding the most significant step ($\log(M/M_{\odot}) = 9.73$) and the threshold commonly adopted in the literature ($\log(M/M_{\odot}) \approx 10$) \citep[e.g.,][]{Betoule_2014,Johansson_2021} appears nearly negligible, as all stellar mass thresholds above our minimum median $\log(M/M_{\odot})$ seem to result in consistent step magnitudes. Therefore, this effect should not significantly impact our results and may reflect the influence of the mixed underlying SN subpopulations contributing to the origin of the mass step, as discussed in Section \ref{sub:mass-step-origin}. The relations involving other standardization parameters, as well as those involving different light-curve and host-galaxy properties, are much less clear, showing at most weak or marginal dependencies.

\subsection{Accessing SN subpopulations through property binning}
\label{sub:bins}
In order to minimize differences in the properties of SNe within and between the calibration and HF subsamples and thereby reduce the probability of including SNe from different subpopulations within the same random subsample, we divide the full calibration and HF samples into two distinct bins of SNe for each considered property, using the median value of that property distribution across the full sample as the dividing threshold.  We then compare the values of $H_{0}$, $M_{B}$, $\alpha$, $\beta$, $\Delta_{host}$ and $\sigma_{\text{int}}$ estimated using all the SNe from each bin. The median values obtained for each property were $\tilde{c}$ = - 0.022, $\tilde{x_{1}}$ = 0.087, $\tilde{\text{log}(M/M_{\odot}})$ = 10.019 and log($\tilde{\text{sSFR}}$yr) = -9.322. 

The most notable discrepancies are observed between the low- and high-stretch bins, with differences of 2.00 $\sigma$ in $H_{0}$ and 2.23 $\sigma$ in $M_{B}$ when using the entire sample, as shown in Table \ref{tab:sne_bins_median}. In contrast, bins defined by other properties such as $c$, log($M/M_{\odot}$) and log(sSFRyr), show much smaller variations in both parameters
, with differences remaining below 1.11 $\sigma$. No other notable variations are found in the estimated parameters across the different bins of any other property, with all results remaining highly consistent (see Tables \ref{tab:sne_bins1} and \ref{tab:sne_bins2} in Appendix \ref{ap:bin}).

\begin{table}[h]
\centering
\begin{tabular}{cccccc}

\hline
SNe bin  & Number of SNe &
$H_{0}$ &  $M_{B}$ \\ \hline

 \addlinespace

\addlinespace
 
$x_{1} < \tilde{x_{1}}$  & 126 & $75.266_{-1.166}^{+1.191}$ &$-19.155_{-0.033}^{+0.033}$\\

\addlinespace

$x_{1} \geq  \tilde{x_{1}}$  & 127 &
$71.247_{-1.546}^{+1.638}$ &  $-19.293_{-0.051}^{+0.052}$    \\ 

 \addlinespace

\hline

\end{tabular}

\caption{Estimated $H_{0}$ and $M_{B}$ using SNe from two different bins divided by the median $\tilde{x_{1}}$ = 0.087, and their respective difference to the 16th and 84th percentiles. The number of SNe in each bin is also provided.}
\label{tab:sne_bins_median}

\end{table}

The discrepancies found for $H_{0}$ and $M_{B}$ when dividing the SNe sample by its median $x_{1}$ value are unlikely to arise by chance ($P<0.5$\%) according to the random-partitioning test described in Appendix \ref{ap:by_chance}, contrary to those reported when dividing the SNe sample by other properties. Moreover, the value estimated for the high-stretch bin in this work is in great agreement with the $H_{0}$ estimate of  71.45 $\pm$ 1.03 km s$^{-1}$ Mpc$^{-1}$ reported by \cite{wojtak2025stretchstretchdustdust}, where the authors assume that the calibration sample consists exclusively of young, high-stretch supernovae and adopt the conservative assumption that all parameters of this population are the same in the Hubble flow and the calibration sample.

Interestingly, in both stretch bins, the mass step is consistent with zero within approximately 1$\sigma$ uncertainties, with $\Delta_{\mathrm{host}} = -0.016 \pm 0.012$ (1.29 $\sigma$) for the low-stretch bin and $\Delta_{\mathrm{host}} = -0.012 \pm 0.012$ (1.00 $\sigma$) for the high-stretch bin, a behavior that not observed for any other parameter. The choice of the median as a separation between both stretch samples is not necessarily ideal; however as shown in appendix~\ref{ap:mix}, this choice does not affect our main results.

When comparing each property distribution between the calibration and HF samples in both stretch bins using the K–S test, we still find that the $\log(M/M_{\odot})$ distribution of the calibration sample is not fully representative of that of the HF sample in both stretch bins, as shown in table \ref{tab:pval_highlow}.  Overall, the calibration and HF samples in the low-stretch bin exhibit a higher level of consistency than those in the high-stretch bin for the remaining properties. Furthermore, when applying the multi-dimensional adaptation of the K–S test, the resulting $p_{MD}$ fall below the significance threshold for the high-stretch mode, yielding a value of 0.025. 

\begin{table*}[h]
\centering
\renewcommand{\arraystretch}{1.4}
\resizebox{\columnwidth}{!}{%
\begin{tabular}{c|cccc|cccc|}
\cline{2-9}
                                                & \multicolumn{4}{c|}{Lower stretch bin}             & \multicolumn{4}{c|}{Higher stretch bin}            \\ \hline
\multicolumn{1}{|c|}{Division value}            & $c$ & $x_{1}$ & log($M/M_{\odot}$) & log(sSFRyr) & $c$ & $x_{1}$ & log($M/M_{\odot}$) & log(sSFRyr) \\
\hline

\multicolumn{1}{|c|}{$\tilde{x_{1}}$} & 0.986 
    & 0.763      &   0.007                  & 0.259
         & 0.417     &  0.198       &  0.048                   & 0.094           \\ \hline

\end{tabular}}
\caption{Estimated $p$-values by comparing the calibration and HF sample properties distributions for each stretch bin under the defined binning threshold of $\tilde{x_{1}}$ = 0.087.}
\label{tab:pval_highlow}
\end{table*}

To ensure that the calibration and HF samples are well matched within each stretch bin, we follow the same approach as in Subsection~\ref{sub:subsample_cal} and generate 600 random SNe subsamples\footnote{This number of subsamples is sufficient for robust statistics. The number of combinations that are significantly different is limited, given the small difference between the minimum number of 10 SNe per subsample and the 14 calibrators in the high-stretch bin, and we expect relatively little mixing of SN subpopulations within each bin. Hence, increasing the number of subsamples would likely not substantially affect the results.} from the calibration and Hubble Flow samples within each stretch bin to investigate how the improved matching of their property distributions affects the estimated parameters.

%\subsection{\textcolor{red}{Generating subsamples from each stretch bin}}
%\label{sub:strecth subsample}
%Given the dependence of SN subpopulations on their host properties (e.g., \cite{Sullivan2010,Ginolin_2024}), it is important to ensure that the calibration and HF samples are well matched within each stretch bin, thereby representing the same SN subpopulation. For this reason, and given that $H_{0}$, $M_{B}$, and some correction parameters appear to be influenced by the improved matching of the calibration and HF samples properties, as shown in Subsection~\ref{sub:subsample_cal}, we followed the same approach and generated 600 random SNe subsamples\footnote{This number of subsamples appears sufficient for robust statistics. The number of combinations that are significantly different is limited, given the small difference between the minimum number of 10 SNe per subsample and the 14 calibrators in each bin, and we expect relatively little mixing of SN subpopulations within each bin. Hence, increasing the number of subsamples would likely not substantially affect the results.} from the calibration and Hubble Flow samples within each stretch bin to investigate how the improved matching of their property distributions affects the estimated parameters.

\begin{figure}[]
    \centering

        \includegraphics[width=0.7\linewidth]{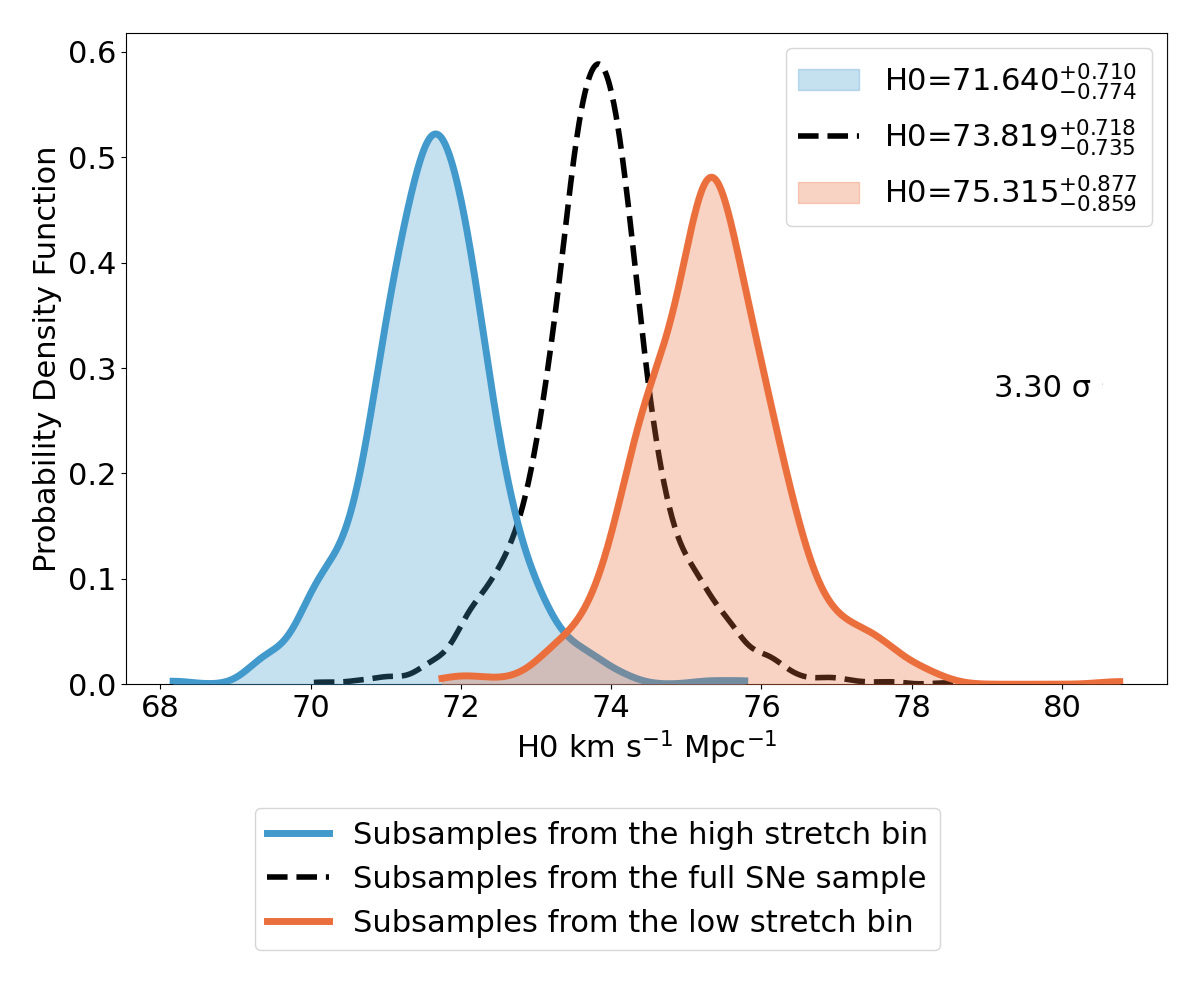}
    
    \caption{Probability density function (PDF) of the $H_0$ distribution estimated from SNe subsamples in the higher-stretch (blue) and lower-stretch (orange) bins defined using $\tilde{x_{1}}$ = 0.087. The black dashed line represent the PDF estimated from the subsamples generated in Section \ref{sub:subsample_cal} using the full SN sample. }
    \label{fig:H0_difpop}
\end{figure}

As shown in Figure~\ref{fig:H0_difpop}, we find a significant discrepancy of 3.30$\sigma$ between the median value of $H_{0}$ derived from the high-stretch subsamples and that obtained from the low-stretch subsamples. However, this difference do not seem to become more significant with the increasing of the $p_{MD}$. For the $M_{B}$ parameter, we also identify a notable difference of 2.90 $\sigma$, which do not seem to increase with the better overall matching of the calibration and HF subsamples, as well. The other estimated parameters remain consistent across subsamples, with differences below the $1\sigma$ level, without showing any relation with the increasing $p_{MD}$. 

\begin{figure}[]
    \centering

        \includegraphics[width=0.8\linewidth]{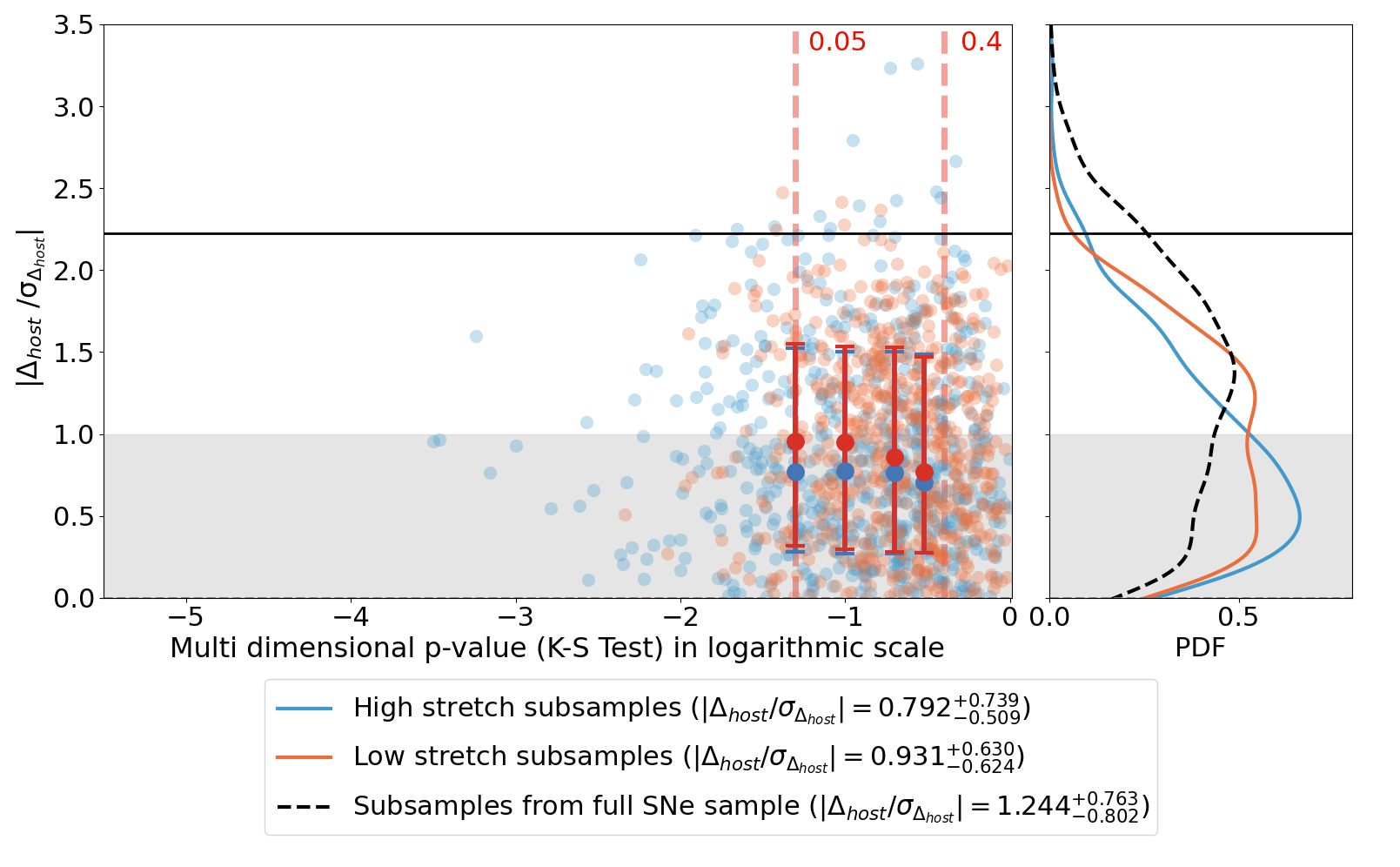}
    
    \caption{Same analysis as in the Figure \ref{fig:medianmass_step_cal}, with the orange dots representing subsamples drawn from the lower-stretch bin and blue dots representing those from the higher-stretch bin. Median values and corresponding uncertainties are not shown for subsamples with $p_{MD}$ $>$ 0.4 due to the low number of subsamples in that region for the low stretch bin. The black dashed line represent the $\Delta_{\rm host}/\sigma_{\Delta_{\rm host}}$ PDF estimated from the subsamples generated in Section \ref{sub:subsample_cal} using the full SN sample.}
    \label{fig:mass-step_difpop}
\end{figure}

When we look to the mass-step magnitudes estimated for both stretch bins and their corresponding uncertainties, we observe that the majority of the subsamples drawn from both stretch bins give $\Delta_{\rm host}$ values consistent with 0 under a 1$\sigma$ difference, independently of the better matching of the calibration and HF subsamples properties, as illustrated in Figure \ref{fig:mass-step_difpop}. In this case, both stretch modes present a mass step consistent with 0 within approximately 1$\sigma$ uncertainties, obtaining $\Delta_{host}$ = - 0.015 $\pm$ 0.013 (1.15 $\sigma$) for the low-stretch mode and $\Delta_{host}$ = - 0.015 $\pm$ 0.014 (1.07 $\sigma$) for the high stretch mode. All parameters estimated using this approach, along with the corresponding discrepancies, are shown in Tables \ref{tab:subsamples_1_median} and \ref{tab:subsamples_2_median}  (Appendix \ref{ap:bin_2}). 

From our analysis, the largest and most significant differences in the estimated $H_0$ and $M_B$ parameters between low- and high-stretch SNe primarily arise from the absence of additional bias correction terms \citep{Popovic_2023} and, most importantly, the neglect of a differentiation of the observed colour into
intrinsic and dust extinction \citep{Brout_2021}. By doing so, we are assuming that all SNe are intrinsically similar and that the dust extinction law (characterized by $R_V$) is consistent for all samples, with dust reddening affecting SN colours in the same way for all events. 
The significantly lower $H_0$ and $M_B$ values estimated using slow-declining (high-stretch) vs fast-declining (low-stretch) SNe suggest that these subpopulations may have %is most likely associated with a more consistent $R_V$ value. In contrast, the fast-declining (low-stretch) SNe subpopulation shows higher $H_0$ and $M_B$ values, likely due to the lack of corrections accounting for 
 intrinsic colour variations or dust extinction differences that are not properly corrected with the simple Tripp standardization law. A more detailed interpretation and discussion of these results, as well as of the discrepancies identified between the different SNe bins, is provided in \ref{sub:sne_pop}, where we also compare our findings with previous results in the literature.

Regarding the mass step, the fact that $\Delta_{\mathrm{host}}$ is repeatedly found to be consistent with zero within approximately $1\sigma$ uncertainty when the SN subpopulations are considered separately, contrary to the 2.22 $\sigma$ deviation from zero obtained using the full sample,  indicates that the mass step might arise from an over-correction of more than one subpopulation with different properties occurring in different environments which are not distinguished in the full SN sample. An additional noteworthy result that supports this conclusion is that all parameter estimates for the two stretch bins can be recovered without fitting the mass-step term. Nevertheless, it could be argued that the apparent consistency with zero could also be an effect of reducing the subsamples sizes, as noted in Subsection~\ref{sub:subsample_cal}. A more detailed discussion of the impact of subsample size on the inferred values of $\Delta_{\mathrm{host}}$, as well as on the possible origin of the mass step, are provided in Subsections \ref{sub:mass-step-size} and \ref{sub:mass-step-origin}, respectively.

%Although this may appear to be a simplistic approach to assessing SN subpopulations using their properties distributions 
%and more robust and precise methods have already been employed in the literature (e.g., \cite{Ginolin_2024}), we find that this method provides reliable and consistent results with those reported in previous studies, as we demonstrate throughout Subsection \ref{sub:sne_pop}. 
In appendix \ref{ap:dif_approaches}, we present alternative approaches for distinguishing the different SN subpopulations, along with the parameters derived for each stretch mode. These include adopting a different division threshold in the stretch distribution to improve the consistency of property distributions between the calibration and Hubble flow subsamples within each bin, as well as using a  Gaussian Mixture Models (GMM) based clustering algorithm to distinguish both SN subpopulations. 

\section{Discussion}   
\label{sec:discussion}

\subsection{Effect of the SN subpopulations}
\label{sub:sne_pop}
The $H_{0}$ and $M_{B}$ discrepancies observed for the log($M/M_{\odot}$) and log(sSFRyr) bins, seem to be simply reflecting the effect of the  underlying SNe subpopulations presented and partially identified throughout this study, which are essentially separated and identified by the stretch parameter $x_{1}$. 

SNe with higher stretch can be associated with both redder and bluer environments, and with both more and less massive hosts. They are also typically linked to star forming environments. In contrast, SNe with lower stretch are mainly associated with redder, high-mass galaxies with lower sSFR \citep[e.g.,][]{Ginolin_2024,Uddin_2017,Neill_2009}. As shown in Table \ref{tab:sne_bins1}, the bin composed of SNe hosted in less massive galaxies (typically with higher $x_{1}$) yields a lower $H_{0}$ compared to the bin that includes SNe from more massive galaxies (with both high and low $x_{1}$ SNe).
Similarly, we can see in the same Table that the Hubble constant estimated using SNe in more actively star-forming galaxies is lower than the one estimated using SNe from galaxies with lower sSFR.

We can also detect a 1.11 $\sigma$ difference when comparing the $H_{0}$ obtained using SNe in different colour bins, with the bluer SNe giving a lower value of $H_{0}$. The empirical shape-luminosity \citep{1993ApJ...413L.105P} and colour-luminosity \citep{Tripp1998} relations imply that intrinsically brighter SNe Ia are bluer and have slower declining lightcurves (higher stretch). According to \cite{Santiago_2021}, bluer SNe Ia are consistent with a lower colour–luminosity parameter $\beta$, contrary to redder SNe. This result is also recovered in this study, as we can see in Table \ref{tab:sne_bins2}, obtaining a difference of 1.10 $\sigma$ between the $\beta$ estimated using redder SNe and bluer ones. This can be explained with $\beta$ being a combination of a low intrinsic colour–luminosity relation dominant in bluer SNe and a higher extrinsic reddening relation dominant at redder colours. 

In the same study \citep{Santiago_2021}, the authors also suggest that bluer SNe in low-mass galaxies are the most homogeneous sample, with lower intrinsic scatter than the redder SNe.
Multiple studies also suggest less intrinsic scatter in local star-forming environments \citep[e.g.,][]{2013A&A...560A..66R}, as well as in global star-forming hosts \citep[e.g.,][]{2015Sci...347.1459K,Uddin_2017}. Given that intrinsically bluer SNe seems to occur in low-mass galaxies \citep[e.g.,][]{Pan_2020}, less luminosity scatter for blue SN colours in low-mass and higher local star-forming environments may partly be associated to the higher stretch-mode subpopulation of SNe, as this is the dominant subpopulation in low-mass environments and strongly associated to higher local sSFR \citep[e.g.,][]{Rigault_2020}.
The authors suggest that red SNe probably have larger luminosity scatter from a mixture of both a different intrinsic colour population and a larger extinction, not easily corrected with a simple linear colour–luminosity relation. \cite{Kelsey_2022} also observe the lowest scatter in the Hubble residual (HR) for blue SNe in low mass and blue environments, which coincides again with the subpopulation of SNe with the highest stretch. Despite previous reports in the literature, our study finds no significant difference in the intrinsic scatter $\sigma_{int}$ between the low- and high-stretch bins. This consistency holds even though we did not divide the SNe sample by stretch and colour, or by any environmental property and colour.

\begin{figure}[]
    \centering

        \includegraphics[width=0.7\linewidth]{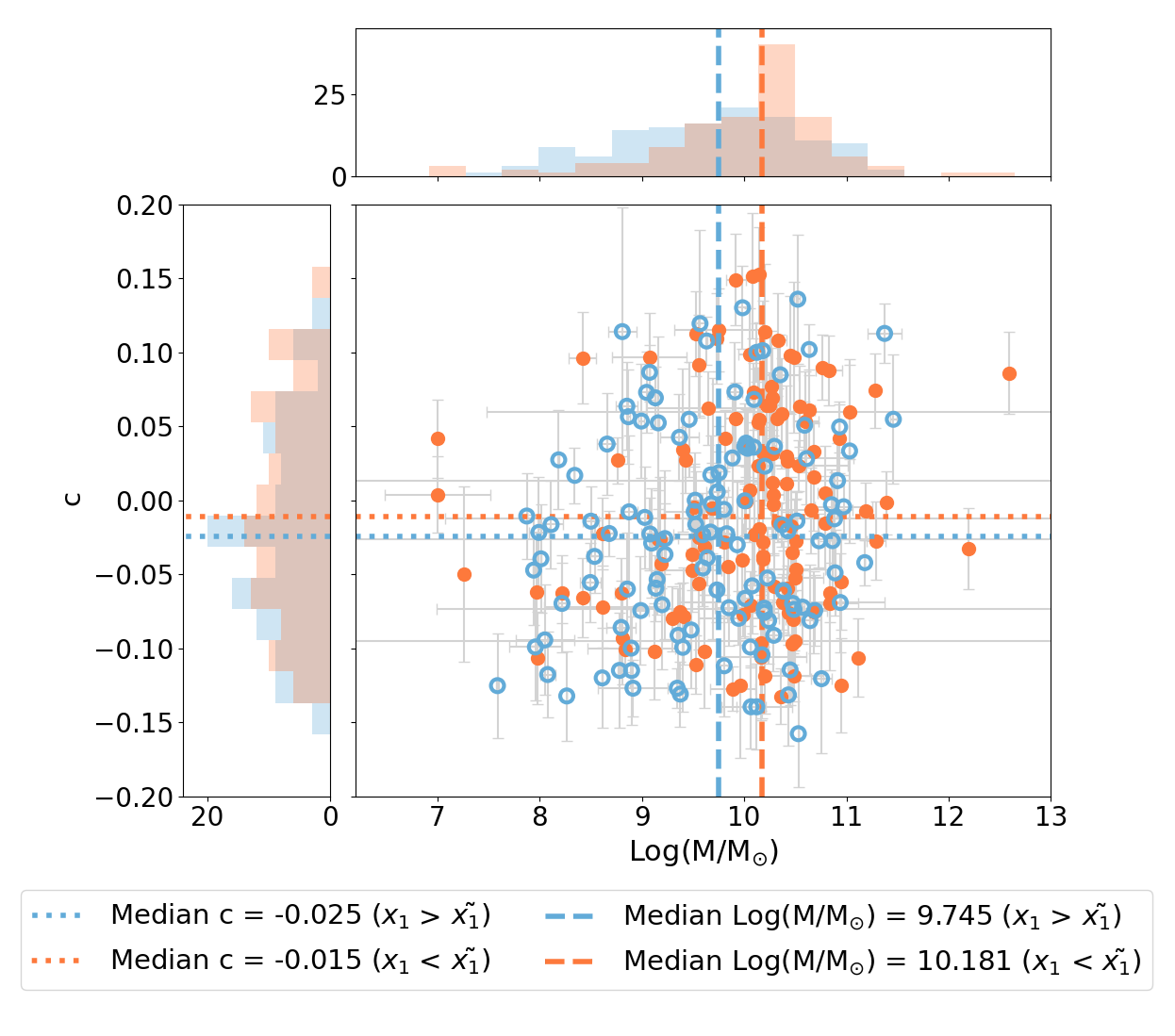}
    
    \caption{Colour ($c$) as a function of log($M/M_{\odot}$) for SNe from both the lower stretch bin (orange) and higher stretch one (blue) divided by the median stretch $\tilde{x_{1}}$. The dotted horizontal lines correspond to the median value of the $c$ distributions and the dashed vertical lines to the median value of the log($M/M_\odot$) distributions for each sample depending on the colour}
    \label{fig:x1_mass}
\end{figure}

According to \cite{10.1093/mnras/stad2590}, SNe from the population with higher average $x_{1}$ are found to be intrinsically bluer ($c$ $\approx$ -0.11) and twice as reddened by dust than those from the other population which is dominated  with $c$ $\approx$ -0.04. Although we do not have the intrinsic colour of the SNe, directly using the observable colour $c$, we can deduce from the cited studies that bluer SNe are the ones that are less affected by the dust reddening and thus more homogeneous between them, with the observable colour closer to the intrinsic one. As we can see in Figure \ref{fig:x1_mass}, a small difference between the median colour of the SNe from the high stretch bin can be seen when compared to the ones with lower stretch. However, we need to take into account that we can still have higher stretch SNe (presumably intrinsically bluer) in higher mass environments that will be observed with a redder colour due to the dust reddening \citep[e.g.,][]{Conley_2007,Duarte_2023}, limiting the ability to distinguish the two SN subpopulations based only on the observed colour.

For the stretch–luminosity relation slope $\alpha$, we do not observe any significant differences between the estimated values across the bins of any property, contrary to the differences found in other studies \citep[e.g.,][]{Garnavich_2023,Larison_2024,Ginolin_2024}.
The differing levels of significance reported in these studies compared to our work can primarily be attributed to variations in binning criteria and the methodological approaches adopted across the studies, as well as to the smaller SNe sample size used in our study. 

In the first two works by  \cite{Garnavich_2023} and \cite{Larison_2024}, the authors classify SNe Ia into fast-declining (lower $x_1$) and slow-declining (higher $x_1$) populations using a fixed threshold at $x_1 = -1$. Within each bin, they perform a weighted linear regression on the colour-corrected absolute magnitudes, adopting a fixed $\beta$ value, and model the dependence on $x_1$ independently for the two regimes. In contrast, in \cite{Ginolin_2024} the authors employ a broken-$\alpha$ standardization model aimed at reducing Hubble residuals. This approach introduces two distinct shape–luminosity slope parameters for the different stretch regimes and simultaneously fits the breaking point separating the two $x_1$ modes, as well as all other correction parameters included in the standardization formalism of \cite{Tripp1998}. However, the largest difference in the median values of $\alpha$ is found between the $x_1$ bins, whose associated uncertainties also span a much wider range compared to those obtained using SNe from bins of the other properties. Interestingly, when the lower and higher stretch bins are defined using SNe with $x_1$ values below the 30th percentile and above the 50th percentile of the full stretch distribution, thereby reducing the overlap of stretch distributions between subpopulations in each bin, the difference in the estimated $\alpha$ between the two bins becomes larger, but still consistent (with 0.77 $\sigma$).

Although we provide a theoretical basis for some of the observed discrepancies and their relation with SN subpopulations characterized by stretch, it is important to note that the discrepancies found when dividing the sample by the median values of properties other than $x_{1}$ may arise by chance (see Appendix \ref{ap:by_chance}) and are generally small. In contrast, the largest discrepancies observed when dividing the bins by the median stretch value appear unlikely to occur by chance.

At the end of Section \ref{sub:bins}, we concluded  that the most significant differences in the estimated $H_0$ and $M_B$ parameters between low- and high-stretch SNe primarily arise from the absence of additional bias correction terms \citep{Popovic_2023} and the neglect of any intrinsic or dust reddening effects in our luminosity corrections \citep{Brout_2021}. When we apply a simplified colour–luminosity correction defined by a single $\beta$ parameter, we assume that extinction and intrinsic colour effects are indistinguishable and identical for all SNe, which does not appear to be the case \citep[e.g.,][]{Santiago_2021}. By defining different SN bins according to their stretch, we assume that SNe within each bin are more likely to be intrinsically similar and that the dust extinction law (characterized by $R_V$) is more consistent within each bin.  However, if this assumption is not true, an incorrect dust correction from the $\beta$-colour relation that is not properly accounting for inconsistent dust properties between the calibration and HF samples within the same stretch bin, can cause the systematically higher $H_0$ and $M_{B}$ values obtained for the lower-stretch SNe bins defined in the previous Sections. 

\begin{figure}[h]
    \centering

        \includegraphics[width=0.7\linewidth]{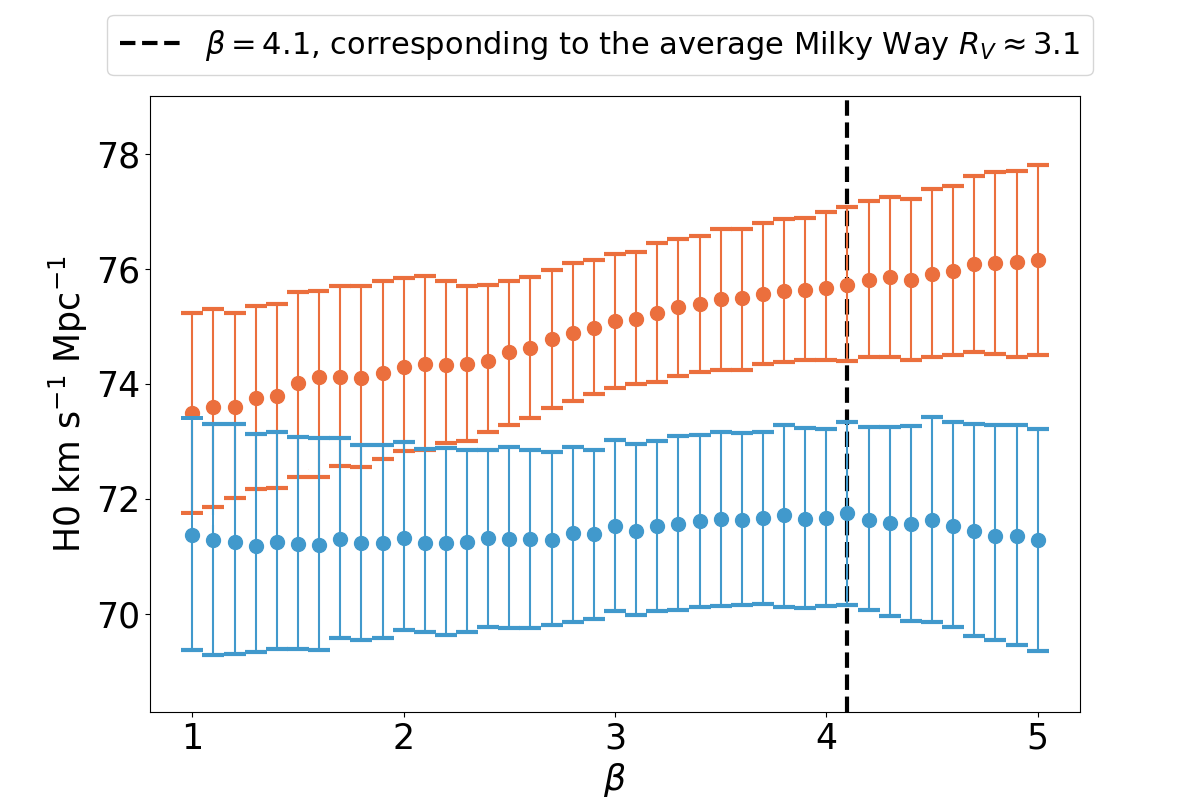}
    
    \caption[$H_{0}$ as a function of the fixed $\beta$ value estimated for different stretch bins]{$H_{0}$ as a function of the fixed $\beta$ value, estimated separately for the low-stretch (orange) and high-stretch (blue) SN bins defined with $\tilde{x_{1}}$ = 0.087. The black dashed line indicates the value of $\beta$ expected for the Milky Way dust law  if the colour–luminosity relation were driven solely by dust.}
    \label{fig:dif_betas}
\end{figure}

In Figure \ref{fig:dif_betas}, we estimate $H_{0}$ separately for the low- and high-stretch bins, while fixing $\beta$ to different values. All other parameters are left free, except for the mass-step magnitude ($\Delta_{host}$), which is fixed to 0. 

 If the colour–luminosity relation were driven solely by dust, then we would expect to have $\beta \approx R_{V} + 1$. In this scenario, a value of $\beta \approx 1$ would correspond to a very low $R_{V}$ and both stretch bins would be expected to respond similarly to variations in $\beta$, since changes in this parameter would directly correspond to changes in $R_{V}$. However, $\beta$ also includes a contribution from intrinsic colour variations, given by $\beta_{int}$. Therefore, if one stretch bin has relatively uniform dust properties (i.e., a well-defined $R_{V}$), changes in the fixed $\beta$ will have a smaller effect on its corrected luminosities and, consequently, on $H_{0}$. On the other hand, if one stretch bin contains SNe with more diverse or inconsistent dust properties (or alternatively a set of varying intrinsic colours), its inferred $H_{0}$ will be more sensitive to changes in $\beta$. This is exactly what we observe in Figure \ref{fig:dif_betas}. The value of $H_{0}$ inferred from the high-stretch bin does not show a strong dependence on the choice of the fixed $\beta$ parameter. In contrast, the low-stretch bin exhibits a much more pronounced variation, with $H_{0}$ decreasing as $\beta$ decreases. This behavior is consistent with the low-stretch population having more heterogeneous dust properties (or more heterogeneous intrinsic colour properties), while the high-stretch population appears to be more uniform in both its dust characteristics and intrinsic colour. This differing behavior between the two stretch populations, not only in the inferred values of $H_{0}$ and $M_{B}$ but also in their response to changes in the $\beta$ parameter, points to clear intrinsic and/or dust differences between the populations.

More recently, \cite{wojtak2025stretchstretchdustdust} probabilistically separated high-stretch and low-stretch SN Ia populations, and independently constrained the supernova and extinction properties within each population. They found that the calibration sample is consistent with being composed entirely of supernovae from the younger (high-stretch) population, which naturally results from the selection of late-type, star-forming host galaxies for Cepheid observations. In contrast, the Hubble flow sample exhibits a bimodal $x_{1}$ distribution, reflecting the presence of two underlying SN Ia populations that differ primarily in their mean absolute magnitudes $M_{B}$ and total-to-selective extinction coefficients $R_{B}$ . The authors estimate a mean $R_{B}$ for the young population that is consistent with the typical Milky Way value of $R_{B} \approx 4.3$ \citep[e.g.,][]{2007ApJ...663..320F}, suggesting that the dust properties in this population, and thus in the calibration sample, are similar to those observed in the Milky Way. In contrast, the older population exhibits a lower value of $R_{B} \approx 3$, indicating a mismatch in the sample's dust properties. This is consistent with our findings, as we observe a significant difference in the estimated absolute magnitude between the two subpopulations distinguished by stretch ($\Delta M_{B} \approx 0.138$) and since we do not apply corrections based on intrinsic colour variations or dust extinction models, our results suggest that the higher-stretch (younger) subpopulation exhibits more homogeneous dust properties compared to the lower-stretch (older) subpopulation.

\subsection{Effect of SN subpopulations on the $H_{0}$ uncertainty}
\label{sub:uncertainty}

\begin{figure*}[]
    \centering

\includegraphics[width=\linewidth]{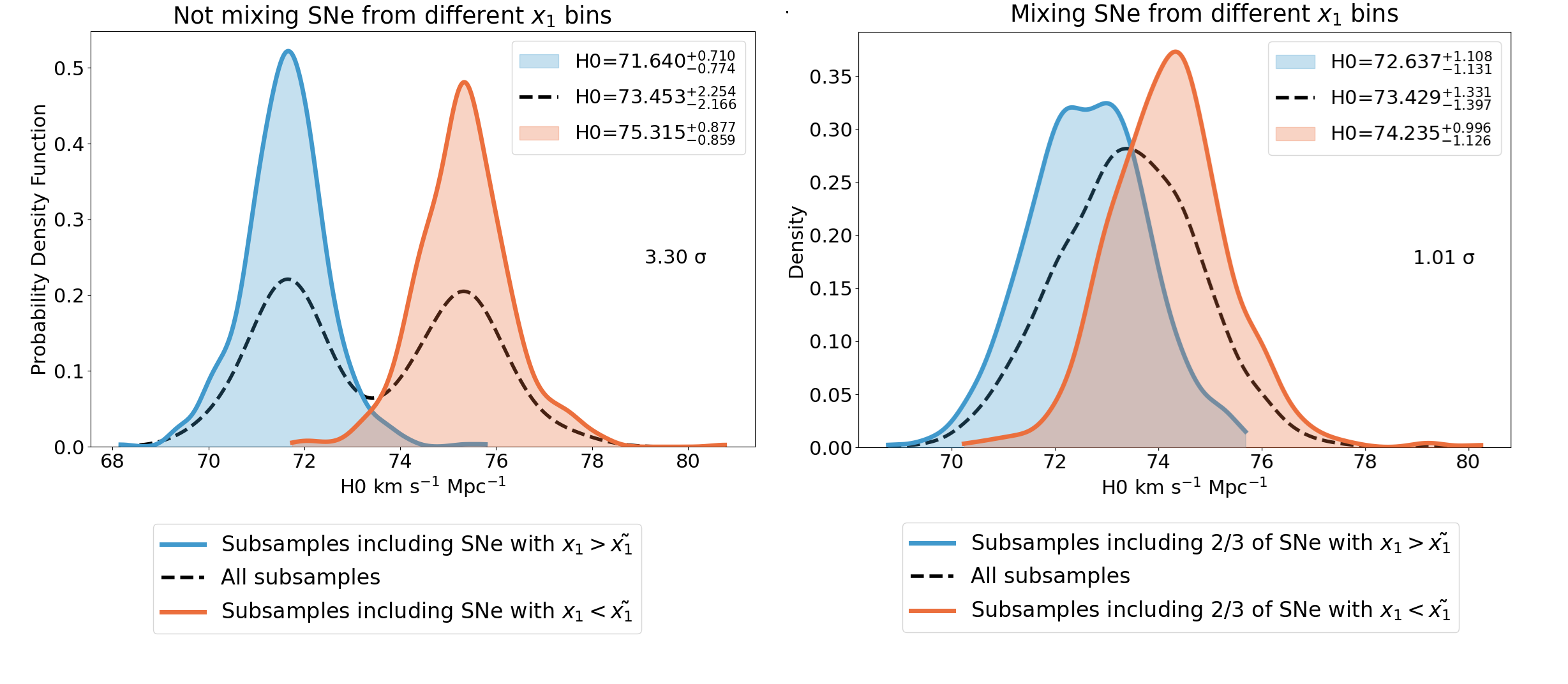}
    
\caption{The left panel shows the probability density function of the $H_{0}$ distribution from subsamples containing only SNe with $x_{1}$ above (blue) or below (orange) the median stretch value for the full sample distribution $\tilde{x_{1}}$, along with the total distribution from all subsamples (black dashed). The right panel shows the corresponding distributions when subsamples include 2/3 of SNe from one stretch bin and 1/3 from the opposite bin. In both cases, the discrepancy between the median $H_{0}$ values of the lower and higher stretch distributions is shown at the top of the plot.}

\label{fig:mix}
    
\end{figure*}

Generating subsamples from the higher and lower stretch SN bins defined in Subsection \ref{sub:bins}, regardless of the size of the calibration or HF subsamples, reveals a substantial difference between the $H_{0}$ distributions estimated for each bin, as we can see in Figure \ref{fig:mix}. This suggests that the broad $H_{0}$ distribution observed in Figure \ref{fig:size_pvalues_subsamples} when generating subsamples may primarily result from partially sampling different SN subpopulations present in both the calibration and Hubble Flow samples, which produce consistently distinct $H_{0}$ distributions, rather than from statistical fluctuations caused solely by the reduced subsample sizes. 

Since the subsamples are generated randomly, most contain SNe from both subpopulations, even though some can present a higher fraction of SNe from one specific SN subpopulation, leading to a higher or lower $\tilde{x_{1}}$. The mixing of SN subpopulations in the generated subsamples can cause the two $H_{0}$ distributions observed in the left panel of Figure \ref{fig:mix} to partially overlap, producing a single broader Gaussian shaped distribution. If we generate SN subsamples from the different $x_{1}$ bins divided by the median stretch of the full sample distribution ($\tilde{x_{1}}$ = 0.087), but replace one third of each subsample with SNe drawn from the opposite stretch bin, the discrepancy between the resulting distributions is reduced, with the $H_0$ distributions from both bins starting to become more disperse and to overlap, as we show in the right panel of the same figure, resembling the PDF obtained in the Figure \ref{fig:size_pvalues_subsamples}. For that reason, if both underlying SN subpopulations are present in the SNe sample from which we estimate a single value of $H_{0}$, the associated uncertainty should be increased relative to the values typically reported in the literature. 

When we treat the two stretch subpopulations separately and generate subsamples within each bin, the uncertainty in the inferred Hubble parameter increases significantly. In this case, we obtain $H_{0} = 73.453^{+2.453}_{-2.166}$ km s$^{-1}$ Mpc$^{-1}$ (as shown in the left panel of figure \ref{fig:mix}), with a 2.4 times larger uncertainty than the one obtained using all subsamples generated from the full SN sample, for which we obtain $H_{0} = 73.816^{+0.910}_{-0.898}$ km s$^{-1}$ Mpc$^{-1}$ (grey dots in Figure \ref{fig:size_pvalues_subsamples}). It is also about 2.3 times larger than the uncertainty estimated using the full SN sample, with $H_{0} = 73.781^{+0.970}_{-0.944}$ km s$^{-1}$ Mpc$^{-1}$ (see Table \ref{tab:parameters}). If we adopt the larger uncertainty obtained when treating the subpopulations independently as the true uncertainty of the full SN sample $H_{0}$ estimate, we obtain $H_{0} = 73.781 \pm 2.166$ km s$^{-1}$ Mpc$^{-1}$.  Under this assumption, the tension with the \cite{Planck_2018} measurement would be reduced from approximately $5.87\sigma$ to about $2.86\sigma$.

Although we do not make a direct comparison between the uncertainties estimated for $H_{0}$ in this study (where we consider the separate contributions of the two SN subpopulations characterized by $x_{1}$) and those reported in other works such as \cite{Riess_2022} due to methodological differences, and despite being a simplistic and approximate way of analyzing the error in $H_{0}$, this result is important to highlight that the contribution of both underlying SN subpopulations should be treated more carefully in future determinations of $H_{0}$ uncertainties.

\subsection{Effect of Subsample Size on the Mass Step} 

\label{sub:mass-step-size}

As concluded in Sections \ref{sub:subsample_cal} and \ref{sub:bins}, the mass-step magnitude tends to become more consistent with 0 as the overall matching between the calibration and HF subsamples improves, becoming effectively consistent with zero when SNe are analyzed separately by stretch mode. However, this apparent consistency may also partially reflect the reduction in subsample sizes, which can increase statistical fluctuations and the associated uncertainties in the mass-step magnitude.

Figure \ref{fig:size_pvalues_subsamples} shows that subsample matching generally improves as subsample sizes decrease. To assess whether correlation between these variables affects their individual relation with the  $\Delta_{\rm host}/\sigma_{\rm host}$, parameter, we computed their Variance Inflation Factor (VIF), which quantifies how much the variance of a regression coefficient is inflated due to correlation with other predictors \citep[see][]{marquardt1970generalized}.  Both the $p_{MD}$ and subsample size yielded a VIF of $\sim 1.2$, indicating that their contributions to the observed trend in $\Delta_{\rm host}/\sigma_{\Delta_{\rm host}}$ are largely independent. Consequently, the trend of the mass-step magnitude approaching zero with increasing $p_{MD}$ is not simply a proxy for the effect of subsample size.

We then modeled $\Delta_{\rm host}/\sigma_{\Delta_{\rm host}}$ as a linear combination of the $p_{MD}$ and subsample size using the Ordinary Least Squares (OLS) approach. The estimated coefficients indicate that the effect of subsample size, ($6.6 \pm 0.2) \times 10^{-3}$, is much more significant ($\sim$ 30$\sigma$) than that of the $p_{MD}$, $0.30 \pm 0.05$ ($\sim$ 6$  \sigma$). Overall, these results suggest that subsample size exerts a substantially stronger influence on $\Delta_{\rm host}/\sigma_{\rm host}$ than improved subsample matching, although both are significant. While the trend observed in Figure \ref{fig:medianmass_step_cal} is not purely a proxy effect of subsample size, we cannot completely rule out a minor contribution, since selecting subsamples with higher $p_{MD}$ tends to favor smaller subsample sizes. 

Interestingly, the R$^{2}$ obtained from this fitting shows that these two variables together explain only $\sim 19\%$ of the variance in the mass-step consistency. Considering the results from Subsections \ref{sub:bins} and \ref{sub:bins}, the specific properties of the SNe composing each subsample likely contribute significantly to the observed variation in $\Delta_{\rm host}/\sigma_{\Delta_{\rm host}}$. For instance, subsamples dominated by SNe with either higher or lower stretch are more likely drawn from the same SN subpopulation, yielding a mass step closer to zero. This could explain the absence of a significant linear correlation between $\Delta_{\rm host}$ and the subsample median stretch observed in Subsection \ref{sub:subsamples_props}.
On the other hand, the significant correlation between the median $\log(M/M_\odot)$ and $\Delta_{\rm host}$ in the same Subsection supports this interpretation. Subsamples with lower median stellar mass are less likely to contain SNe from multiple subpopulations and more likely to be dominated by higher-stretch SNe (see right panel from Figure \ref{fig:hist_prova}), resulting in $\Delta_{\rm host}$ values closer to zero.

\subsection{Mass-step possible origin}   

\label{sub:mass-step-origin}

Some studies have suggested that the observed mass step in SNe Ia luminosities may originate from the dust properties of their host galaxies. For instance, \cite{Brout_2021} find that the correlation between the corrected luminosity and the host galaxy mass can be attributed to differences in
the ratio of total-to-selective attenuation ($R_{V}$) values between low- and high-mass galaxies. More recently, \cite{10.1093/mnras/stad488} reported that the variation of $R_{V}$ with galaxy age can explain almost the entire observed mass step. Since the stellar mass of a galaxy seems to be correlated with the population age \citep[e.g.,][]{Kang_2016}, the host mass step may instead reflect a progenitor age effect on the standardized brightnesses of SNe Ia \citep[e.g.,][]{2020ApJ8898K,Lee_2020} rather than being mainly influenced by dust. Indeed, there is evidence to suggest that the mass step
cannot entirely be explained as a dust systematic \citep[e.g.,][]{Uddin_2020,Santiago_2021,Ponder_2021,10.1093/mnras/stac2714,Duarte_2023}.

In \cite{Chung_2023}  the authors suggest that the mass step arises from the bimodal nature of the host galaxies age distribution, based on the empirical galaxy colour–magnitude relation. Some studies have been reporting strong correlations between post-standardization SN Ia luminosities and their progenitor age \citep[e.g.,][]{10.1093/mnras/stac2840,chung2025strongprogenitoragebias}, with the implicit assumption that the stellar population from which the SN
progenitor originated shared these global properties, including the stellar population age. 
Some studies have also shown that the sSFR measured locally to the SN is a relatively pure tracer of the progenitor age \citep[e.g.,][]{10.1093/mnras/stad488} and that locally star forming environments are fainter after standardization than those from locally passive environments, suggesting that this luminosity step manifests itself most strongly when using the local SFR at the SN location, rather than the global SFR or the stellar mass of the host galaxy \citep[e.g.,][]{Rigault_2015,Kim_2018}. 

Interestingly, and as already introduced and discussed in  Subsection {\ref{sub:sne_pop}, the bimodal behavior of the $x_{1}$ distribution is found to be associated with the presence of underlying populations of SNe with dependence
on environment-based progenitor age \citep[e.g.,][]{Rigault_2015,Nicolas_2021}, with higher stretch SNe being associated to
higher local sSFR environments and younger progenitors, contrary to the lower stretch population \citep[e.g.,][]{Rigault_2020}. Not only with the star formation rate, but these SN populations also seem to be strongly correlated with their host galaxies masses, with less massive galaxies mostly hosting SNe with higher stretch, and the more massive galaxies hosting SNe from both stretch modes \citep[e.g.,][]{Sullivan2010,Ginolin_2024}. 

As shown across Section \ref{sec:analysis}, low-stretch SNe, typically associated with more massive hosts and older progenitors, are intrinsically fainter ($M_{B} \approx$ -19.155 mag) than high-stretch SNe, which are linked to younger progenitors and occur in both low- and high-mass galaxies ($M_{B} \approx$ -19.293 mag), yielding a difference of $\sim$ 0.138 mag and a mass step magnitude consistent with 0. Assuming all other correction parameters are consistent, we should expect that low-mass hosts dominated by the high-stretch subpopulation host SNe Ia would appear fainter, while high-mass hosts containing both subpopulations would host SNe that appear brighter after standardization, due to the inaccurate assumption of a universal SNe Ia intrinsic brightness. This suggests that the mass step may arise from an over-correction of more than one subpopulation with different properties occurring in different environments, in an attempt to account for their intrinsic absolute magnitude differences rather than treating the subpopulations separately. Some studies such as the ones by \cite{2013A&A...560A..66R} and \cite{2014MNRAS.445.1898C}, and more recently \cite{Rigault_2020}, also point that the stellar mass bias is, at least, partially caused by this local sSFR and, consequently, progenitor age bias.

On the other hand, as discussed in Section~\ref{sub:sne_pop} and reported in other studies such as \cite{10.1093/mnras/stad2590} and more recently \cite{wojtak2025stretchstretchdustdust}, SN subpopulations can also be associated with different mean total-to-selective extinction coefficients. The high-stretch SN subpopulation tends to yield higher mean $R_{B}$ values, closely matching the mean value estimated for the Milky Way ($R_{B} = R_{V} + 1 \approx 4$) \citep[e.g.,][]{2007ApJ...663..320F,Schlafly_2016}, while the low-stretch subpopulation appears to be associated with lower $R_{B}$ values. \cite{Brout_2021} predict that SNe in lower-mass galaxies have, on average, higher $R_{V}$ (and thus $R_{B}$) values than those in higher-mass galaxies. Similarly, \cite{Salim_2018} found that quiescent galaxies, typically more massive, have a mean $R_{V}$ of 2.61, while star-forming galaxies, which are on average less massive, have a mean $R_{V} = 3.15$. These trends are consistent with the host environmental dependence observed for each SN subpopulation characterized by their stretch parameter \citep[e.g.,][]{Sullivan2010,Ginolin_2024} and linked to progenitor age \citep[e.g.,][]{Rigault_2020} and, consequently, to the progenitor channel \citep[e.g.,][]{Brandt_2010}. 

Both SN subpopulations associated with different $R_{V}$ values \citep[e.g.,][]{wojtak2025stretchstretchdustdust} can be found in more massive galaxies, potentially increasing the average total-to-selective extinction coefficient. In less massive galaxies, we would expect a larger fraction of high-stretch SNe, which are associated with higher $R_{V}$ values. Moreover, high-stretch SNe are typically found in more star-forming galaxies, while low-stretch SNe are predominantly located in quiescent hosts, paralleling the $R_{V}$ trend observed by \cite{Salim_2018}. Together, these results might suggest that the two main hypotheses proposed to explain the origin of the mass step may, in fact, share a common physical origin: two different progenitor channels associated to different environments.

\section{Conclusions}   
\label{sec:conclusion}
In this work, we examine how improving the match between the SNe light curve parameters $c$ and $x_{1}$, and the corresponding host properties log($M/M_{\odot}$) and log(sSFRyr) distributions in subsamples drawn from the calibration and Hubble Flow SNe samples used in the \cite{Riess_2022} analysis affects the SN luminosity standarization, as well as the Hubble constant $H_{0}$ estimation. Using a full sample of 253 SNe, our main findings are as follows.\\

(i) The calibration sample is not fully representative of the Hubble Flow sample according to K-S test results, in terms of the stellar mass ($p$-value = 5.067 x 10$^{-4}$) and the specific star formation rate ($p$-value = 3.029 x 10$^{-2}$), as shown in Section \ref{sub:full_sample}. There is a clear difference between the log($M/M_{\odot}$) distributions of the two samples, with the calibration sample containing SNe hosted in more massive galaxies than those found in the HF sample. Regarding log(sSFRyr), the calibration sample includes SNe in galaxies with much higher sSFR values that are not present in the HF sample, while the latter shows lower values that are not seen in the SN calibration sample (see Figures \ref{fig:x1_vs_mass} and \ref{fig:c_vs_ssfr}).

(ii) A subsample from the HF capable of independently resolving the Hubble tension may not exist, regardless of its agreement with the entire calibration sample or the specific properties considered. In order to achieve a more accurate SN luminosity correction and potentially reduce the Hubble tension, intrinsic discrepancies in
the property distributions must be addressed in both the HF and calibration samples.

(iii)  We find statistical significant linear relations between the $H_{0}$ (4.61 $\sigma$), $M_{B}$ (6.19 $\sigma$), $\beta$ (3.63 $\sigma$) and $\sigma_{int}$ (5.52 $\sigma$) and the improved matching of the property distributions between the randomly generated calibration and HF subsamples, even though this relations do not strongly change any of the standardization parameters, nor the value of $H_{0}$, as shown in Table \ref{tab:slopes}.

(iv) Although the possible effects of subsample sizes (section \ref{sub:mass-step-size}) and the choice of mass threshold used to define the mass step may influence the relationship between the improved matching of subsample properties and the estimated $\Delta_{host}$ and corresponding uncertainties, our results indicate that more consistent subsamples tend to produce a mass step consistent with zero within 1$\sigma$ uncertainty, as shown in Figure \ref{fig:medianmass_step_cal}. 

(v) We find notable discrepancies between the estimated values of $H_{0}$ (2.00 – 3.30 $\sigma$) and $M_{B}$ (1.74 – 2.90 $\sigma$) when comparing different subpopulations of SN characterized by their $x_{1}$. The higher-stretch subpopulation yields consistently lower values of both $H_{0}$ and $M_{B}$ compared to the lower-stretch one, regardless of the method used to define the SN subpopulations (see Tables \ref{tab:sne_bins1}, \ref{tab:sne_lowhigh1},  \ref{table:sne_bins_1}, \ref{table:clusters1}).
These differences in the estimated $H_0$ and $M_B$ between low- and high-stretch SNe are likely due to the absence of additional bias correction terms and the simplified $\beta$-colour correction, which assumes that extinction and intrinsic colour effects are indistinguishable and identical for all SNe. However, these assumptions do not seem to be accurate, as inconsistencies in $R_V$ between the calibration and Hubble flow samples within the low-stretch bin, which are not fully accounted for by the $\beta$ colour term, appear to be the cause of the higher $H_0$ and $M_B$ values (see Figure \ref{fig:dif_betas}). This inconsistency may arise from the natural selection of late-type, star-forming host galaxies for Cepheid observations, which can result in a calibration sample predominantly composed of high-stretch SNe, as suggested by \cite{wojtak2025stretchstretchdustdust}. The differing behavior between the two stretch populations, not only in the inferred values of $H_{0}$ and $M_{B}$ but also in their response to changes in the $\beta$ parameter, points to clear intrinsic and/or dust differences between the populations.

(vi) Our findings suggest
that the $H_{0}$ and $M_{B}$  uncertainties commonly reported in the literature may themselves be underestimated, as
they do not account for the additional dispersion effect in both parameters distributions arising from different underlying SN subpopulations, as clearly illustrated in Figure \ref{fig:mix}. Considering the lower uncertainty obtained when treating the subpopulations independently as the true uncertainty of the full SN sample $H_{0}$ estimate (Table \ref{tab:parameters}), we obtain $H_{0} = 73.781 \pm 2.166$ km s$^{-1}$ Mpc$^{-1}$. Under this simplistic assumption, the tension with the \cite{Planck_2018} measurement would decrease from approximately $5.87\sigma$ to about $2.53\sigma$ (see Figure \ref{fig:tensions}), illustrating how sensitive the inferred tension is to the $H_{0}$ uncertainty, which itself seems to be underestimated.

(vii) The mass step seems to be consistent with 0 for both stretch subpopulations (see Tables \ref{tab:sne_bins2}, \ref{tab:sne_lowhigh2},  \ref{table:sne_bins_2} and Figure \ref{fig:mass-step_difpop}). This result, allied to the different $M_{B}$ and underlying $R_{V}$ found for both SN supopulations , suggests that this effect may arise from an over-correction of more than one subpopulation with different properties occurring in different environments, in an attempt to account for their intrinsic absolute magnitude and total-to-selective coefficients rather than treating the subpopulations separately, as suggested in Section \ref{sub:mass-step-origin}.

To conclude, we place into perspective the tension between the late-time measurements of $H_{0}$ obtained in this work and the early-time estimate of $H_{0}$ = 67.4 $\pm$ 0.5 km s$^{-1}$ Mpc$^{-1}$ from \cite{Planck_2018}, as shown in Figure \ref{fig:tensions}, alongside the tension between the latter and the late-time measurement of $H_{0}$ = 73.04 $\pm$ 1.04 km s$^{-1}$ Mpc$^{-1}$ reported by  \cite{Riess_2022}. In this work, we find $H_{0}$ = 71.25 $\pm$ 1.59 km s$^{-1}$ Mpc$^{-1}$ using SNe from the high-stretch bin (presumably representing the younger SN subpopulation), and $H_{0}$ = 75.27 $\pm$ 1.18 km s$^{-1}$ Mpc$^{-1}$ using SNe from the low-stretch bin (presumably representing the older SN subpopulation). We can see a reduction of the tension between the late-time and early-time measurements in Figure \ref{fig:tensions}, when only using SNe from the bin with higher stretch (2.52 $\sigma$) in relation to the value estimated using all the SNe from the full sample \citep[4.81 $\sigma$;][]{Riess_2022}. From this Figure and also comparing all the results from our work, it even seems that the full sample $H_{0}$ and $M_{B}$ estimations might result from the simultaneous contribution of the underlying old and younger SNe subpopulations characterized by a lower and higher values of $x_{1}$ respectively. Our estimation of $H_{0}$ =  73.78 $\pm$ 2.17 km s$^{-1}$ Mpc$^{-1}$ obtained using the full sample of SNe and accounting for uncertainties by treating the two subpopulations of stretch independently is also shown in Figure \ref{fig:tensions}, reducing the tension to a discrepancy of 2.86 $\sigma$.

Our study can be improved in several aspects, mainly in the modeling and distinction of the SNe mixed subpopulations. For future work, we aim to distinguish the two SN Ia subpopulations in a more rigorous manner by applying other clustering algorithms, such as the HDBSCAN \citep{mcinnes2017hdbscan}. We would also like to incorporate other host galaxy properties such as their morphology \citep[e.g.,][]{Baluta_2024,Suzuki_2012} and use local properties, as type Ia supernova luminosities appear to correlate more strongly with local rather than global host properties \citep[e.g.,][]{Kim_2018,Rose_2019,Rigault_2020}. Moreover, increasing the amount of data would be very useful to identify stronger and more significant relationships between SN luminosities and their environmental properties. It would also allow for a clearer separation of different SN subpopulations, particularly within the calibration sample, where the current number of SNe is still limited. Future analyses could also incorporate the volume-limited Zwicky Transient Facility (ZTF) SN Ia DR2 dataset \citep{rigault2025ztf}, as well as SN Ia light-curve data compiled by the Carnegie Supernova Project (CSP) I and II \citep{2017AJ....154..211K,2019PASP..131a4001P} and the upcoming data from the Vera Rubin Observatory’s Legacy Survey of Space and Time \citep[LSST;][]{Ivezić_2019}.

\begin{figure}[h]
    \centering

        \includegraphics[width=0.7\linewidth]{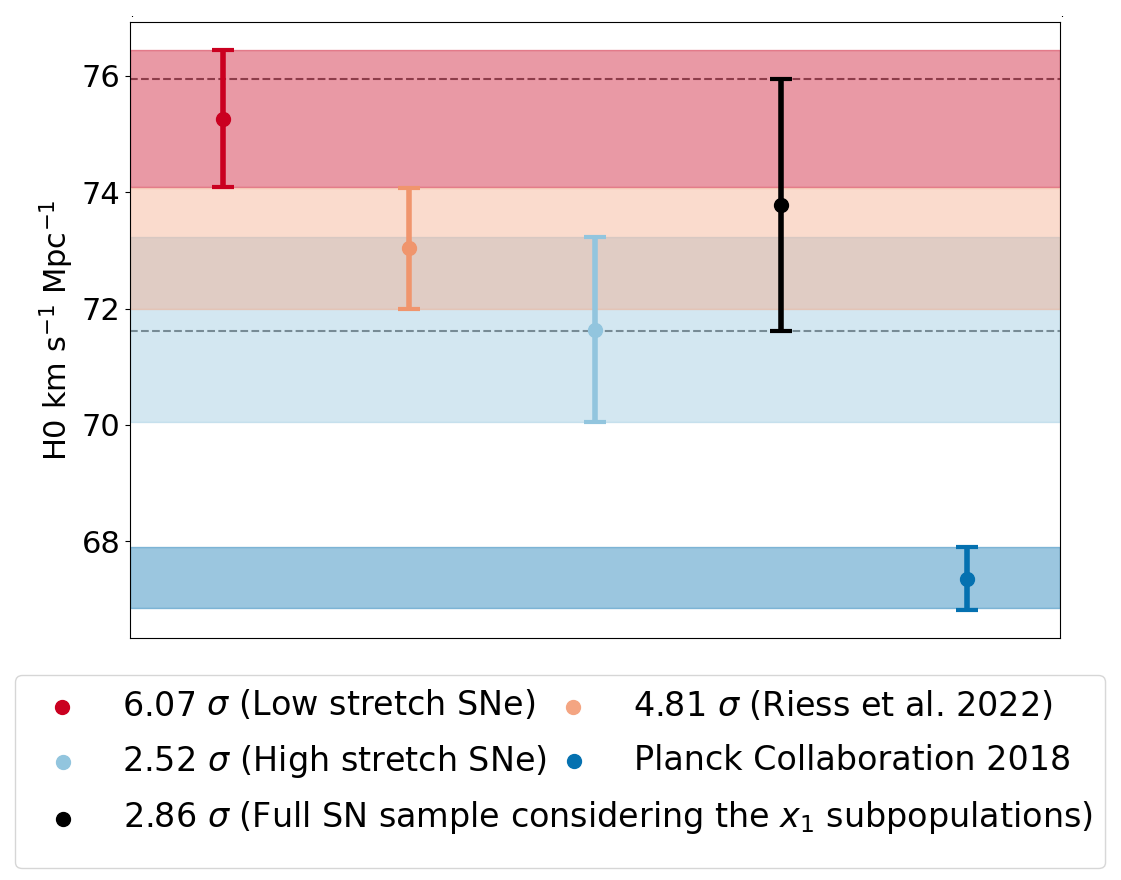}
    
    \caption{ $H_{0}$ estimates and their associated uncertainties obtained in this work using the high-stretch (light blue) and low-stretch (red) SNe bins, alongside the measurement from \cite{Riess_2022} (orange) and the early-time estimate from \cite{Planck_2018} (dark blue). The $H_{0}$ obtained in this work using the full sample
    of SNe and accounting for uncertainties by treating the two subpopulations of stretch independently (black) is also shown. All significance values reported in the legend correspond to the level of discrepancy between each late-time measurement of $H_{0}$ and the early-time value of the Hubble parameter (in blue). }
    \label{fig:tensions}

\end{figure}

\newpage

\appendix 

\section{Results from the property binning analysis}

\subsection{Dividing the bins by the median property value}
\label{ap:bin}
\begin{table}[h]
\centering
\resizebox{\textwidth}{!}{%
\begin{tabular}{cccccccc}

\hline
SNe bin  & Number of SNe &
$H_{0}$ & Sig. & $M_{B}$ & Sig. & $\alpha$ & Sig.\\ \hline

 \addlinespace

$c < \tilde{c}$  & 126 & $72.937_{-1.224}^{+1.217}$ & \multirow{2.5}{*}{1.11 $\sigma$} &$-19.204_{-0.040}^{+0.038}$ & \multirow{2.5}{*}{0.12 $\sigma$} &$0.127_{-0.013}^{+0.013}$ & \multirow{2.5}{*}{0.52 $\sigma$} \\

\addlinespace

$c \geq \tilde{c}$  & 127 & $75.045_{-1.456}^{+1.463}$ & & $-19.198_{-0.041}^{+0.040}$ & & $0.137_{-0.016}^{+0.017}$ &  \\ 

\addlinespace

\hline

\addlinespace
 
$x_{1} < \tilde{x_{1}}$  & 126 & $75.266_{-1.166}^{+1.191}$ & \multirow{2.5}{*}{2.00 $\sigma$} &$-19.155_{-0.033}^{+0.033}$ & \multirow{2.5}{*}{2.23 $\sigma$} &$0.128_{-0.023}^{+0.022}$ & \multirow{2.5}{*}{0.42 $\sigma$} \\

\addlinespace

$x_{1} \geq  \tilde{x_{1}}$  & 127 &
$71.247_{-1.546}^{+1.638}$ & & $-19.293_{-0.051}^{+0.052}$ & & $0.112_{-0.030}^{+0.029}$ &  \\ 

 \addlinespace
 
 \hline

 \addlinespace
 
log($M/M_{\odot}$) $<$ log($\tilde{M}/M_{\odot}$)  & 126 &  $72.627_{-1.638}^{+1.634}$ & \multirow{2.5}{*}{1.05 $\sigma$} &$-19.210_{-0.049}^{+0.047}$ & \multirow{2.5}{*}{0.21 $\sigma$} &$0.134_{-0.016}^{+0.016}$ & \multirow{2.5}{*}{0.31 $\sigma$}\\

 \addlinespace

log($M/M_{\odot}$) $\geq$ log($\tilde{M}/M_{\odot}$)  & 127 &  $74.711_{-1.142}^{+1.160}$ & & $-19.198_{-0.031}^{+0.031}$ & & $0.141_{-0.014}^{+0.014}$ & \\ 

\addlinespace
 
\hline

\addlinespace

log(sSFRyr) $<$ log($\tilde{\text{sSFR}}\text{yr}$)  & 123 &  $74.702_{-1.193}^{+1.332}$ & \multirow{2.5}{*}{1.13 $\sigma$} &$-19.182_{-0.034}^{+0.036}$ & \multirow{2.5}{*}{0.84 $\sigma$} &$0.136_{-0.013}^{+0.013}$ & \multirow{2.5}{*}{0.35 $\sigma$}\\

\addlinespace

log(sSFRyr) $\geq$ log($\tilde{\text{sSFR}}\text{yr}$)  & 130 &  
$72.526_{-1.434}^{+1.513}$ & & $-19.228_{-0.042}^{+0.042}$ & & $0.129_{-0.017}^{+0.017}$ &  \\
\addlinespace

\hline

\end{tabular}
}
\caption[Estimated $H_{0}$, $M_{B}$ and $\alpha$ using SNe from two different bins divided by the median value for each property c, $x_{1}$, log($M/M_{\odot}$), and log(sSFRyr) distribution of the full sample]{Estimated $H_{0}$, $M_{B}$ and $\alpha$ using SNe from two different bins divided by the median value for each property $c$, $x_{1}$, log($M/M_{\odot}$), and log(sSFRyr) distribution of the full sample and their respective difference to the 16th and 84th percentiles. The significances of the differences between the estimated parameters for each bin of the considered property (Sig.) are also shown.}
\label{tab:sne_bins1}

\end{table}

\begin{table}[h]
\centering
\resizebox{\textwidth}{!}{%
\begin{tabular}{cccccccc}

\hline
SNe bin & Number of SNe &
$\beta$ & Sig. & $\Delta_{host}$ & Sig. & $\sigma_{int}$ & Sig. \\ \hline

 \addlinespace

$c < \tilde{c}$ & 126 & $2.837_{-0.235}^{+0.241}$ & \multirow{2.5}{*}{1.10 $\sigma$} &$-0.020_{-0.011}^{+0.010}$ & \multirow{2.5}{*}{ 0.39 $\sigma$} &$0.073_{-0.017}^{+0.015}$ & \multirow{2.5}{*}{0.72 $\sigma$} \\

\addlinespace

$c \geq \tilde{c}$  & 127 &  $3.220_{-0.252}^{+0.258}$ & & $-0.013_{-0.014}^{+0.014}$ & & $0.090_{-0.016}^{+0.016}$ &  \\ 

\addlinespace

\hline

\addlinespace
 
$x_{1} < \tilde{x_{1}}$ & 126 & $2.861_{-0.147}^{+0.144}$ & 
\multirow{2.5}{*}{0.94 $\sigma$} &$-0.016_{-0.012}^{+0.012}$ & \multirow{2.5}{*}{ 0.23 $\sigma$} &$0.083_{-0.013}^{+0.013}$ & \multirow{2.5}{*}{ 0.92 $\sigma$} \\

\addlinespace

$x_{1} \geq \tilde{x_{1}}$ & 127 & $2.633_{-0.193}^{+0.192}$ & & $-0.012_{-0.012}^{+0.012}$ & & $0.100_{-0.013}^{+0.013}$ &  \\ 

 \addlinespace
 
 \hline

 \addlinespace
 
log($M/M_{\odot}$) $<$ log($\tilde{M}/M_{\odot}$) & 126 &  $2.910_{-0.170}^{+0.178}$ & 
\multirow{2.5}{*}{0.75 $\sigma$}
&$-0.017_{-0.012}^{+0.012}$ & \multirow{2.5}{*}{0.18 $\sigma$} &$0.079_{-0.016}^{+0.015}$ & \multirow{2.5}{*}{ 0.68 $\sigma$} \\

 \addlinespace

log($M/M_{\odot}$) $\geq$ log($\tilde{M}/M_{\odot}$) & 127 & $2.736_{-0.155}^{+0.159}$ & & $-0.020_{-0.012}^{+0.012}$ & & $0.092_{-0.012}^{+0.013}$ &  \\ 

\addlinespace
 
\hline

\addlinespace

log(sSFRyr) $<$ log($\tilde{\text{sSFR}}\text{yr}$) & 123 &   $2.751_{-0.163}^{+0.158}$ & \multirow{2.5}{*}{0.16 $\sigma$} &$-0.024_{-0.012}^{+0.012}$ & \multirow{2.5}{*}{0.90 $\sigma$} &$0.089_{-0.012}^{+0.013}$ & \multirow{2.5}{*}{0.22 $\sigma$} \\

\addlinespace

log(sSFRyr) $\geq$ log($\tilde{\text{sSFR}}\text{yr}$) & 130 &  $2.787_{-0.163}^{+0.180}$  & &
$-0.008_{-0.013}^{+0.013}$ & & $0.093_{-0.013}^{+0.014}$ & \\
\addlinespace

\hline

\end{tabular}
}
\caption[Estimated $\beta$, $\Delta_{host}$ and $\sigma_{int}$ using SNe from two different bins divided by the median value for each property c, x1, log($M/M_{\odot}$), and log(sSFRyr) distribution of the full sample]{Same as Table \ref{tab:sne_bins1} for the remaining parameters $\beta$, $\Delta_{host}$ and $\sigma_{int}$.}
\label{tab:sne_bins2}

\end{table}

\newpage

\subsection{Using subsamples from the high and low stretch bins defined with $\tilde{x_{1}}$}

\label{ap:bin_2}
\begin{table}[h]
\centering

\begin{tabular}{ccccccc}

\hline
SNe bin &
$H_{0}$ & Sig. & $M_{B}$ & Sig. & $\alpha$ & Sig.\\ 

\hline

\addlinespace
 
$x_{1} < \tilde{x_{1}}$ & $75.315_{-0.859}^{+0.877}$ & \multirow{2.5}{*}{ 3.30$\sigma$} &$-19.156_{-0.025}^{+0.022}$ & \multirow{2.5}{*}{ 2.90 $\sigma$} &$0.129_{-0.025}^{+0.021}$ & \multirow{2.5}{*}{ 0.21 $\sigma$} \\

\addlinespace

$x_{1} \geq \tilde{x_{1}}$ &
$71.640_{-0.774}^{+0.710}$ & & $-19.276_{-0.030}^{+0.033}$ & & $0.118_{-0.037}^{+0.046}$ &  \\ 

 \addlinespace

\hline

\end{tabular}

\caption{Estimated median values of $H_{0}$, $M_{B}$, and $\alpha$, with differences relative to the 16th and 84th percentiles, for SNe subsamples drawn from the two stretch bins defined by $\tilde{x_{1}}$. The significances of the differences between the estimated parameters for each bin (Sig.) are also shown.}
\label{tab:subsamples_1_median}

\end{table}

\begin{table}[h]
\centering

\begin{tabular}{ccccccc}

\hline
SNe bin &
$\beta$ & Sig. & $\Delta_{host}$ & Sig. & $\sigma_{int}$ & Sig. \\ \hline

 \addlinespace

$x_{1} < \tilde{x_{1}}$ & $2.901_{-0.116}^{+0.174}$ & 
\multirow{2.5}{*}{ 0.91 $\sigma$} &$-0.015_{-0.013}^{+0.013}$ & \multirow{2.5}{*}{0 $\sigma$} &$0.078_{-0.019}^{+0.012}$ & \multirow{2.5}{*}{ 0.36 $\sigma$} \\

\addlinespace

$x_{1} \geq \tilde{x_{1}}$ & $2.633_{-0.293}^{+0.269}$ & & $-0.015_{-0.014}^{+0.014}$ & & $0.089_{-0.029}^{+0.025}$ &  \\ 

 \addlinespace
 
 \hline

\end{tabular}

\caption{Same as Table \ref{tab:subsamples_1_median} for $\beta$, $\Delta_{host}$ and $\sigma_{int}$.}
\label{tab:subsamples_2_median}

\end{table}

\section{Minimizing the subpopulations mixing}
\label{ap:mix}
For the stretch parameter, \cite{10.1093/mnras/stad2590} identify distinct, though overlapping, SN populations that can be primarily distinguished by their mean stretch values, obtained by fitting a bimodal Gaussian distribution. Using the Pantheon+ light-curve parameters \citep{Brout_2022}, they find mean values of $\bar{x_1} \approx -1.22$ and $\bar{x_1} \approx 0.42$ for the two subpopulations.
In our sample, the mean values for each bin defined by stretch are higher, with the low-stretch bin yielding a value of $\bar{x_1} \approx -0.45$ and the high-stretch bin a value of $\bar{x_1} \approx 0.74$. 
In \cite{Ginolin_2024}, the authors  employ a broken-$\alpha$ standardization model aimed at reducing Hubble residuals. In this work, they fit the break point separating the two $x_1$ modes and obtain a value of $x_{1} = -0.48$, significantly lower than the one we use to separate our bins ($x_{1} = 0.087$).  Additionally, they also fit a bimodal Gaussian distribution to the stretch parameter, finding mean values of $\bar{x_1} \approx 0.42$ for the high-stretch mode and $\bar{x_1} \approx -1.24$ for the low-stretch mode.

\begin{figure}[h]
    \centering

        \includegraphics[width=0.6\linewidth]{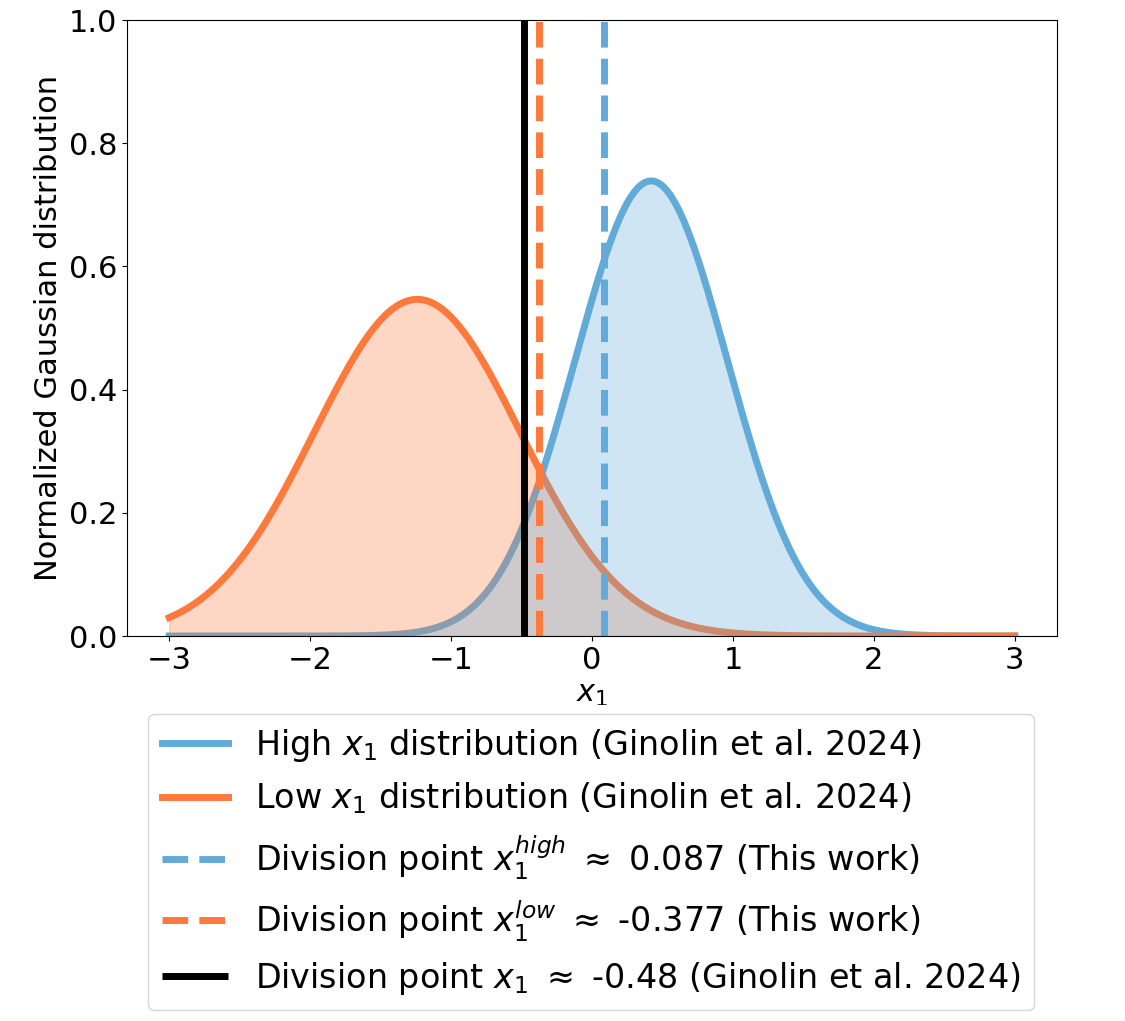}
    
    \caption{Normalized Gaussian distributions representing the high-stretch (blue) and low-stretch (orange) SN populations, using the fitted parameters from \cite{Ginolin_2024}. The vertical solid line indicates the $x_1$ break point between the two stretch modes obtained in the cited study, while the vertical dashed lines mark the splitting points adopted in this study to reduce the overlap of the two SN subpopulations. The shaded region highlights the overlap between the two distributions.}
    \label{figure: mixing_sub}
\end{figure}

As we can see in Figure \ref{figure: mixing_sub}, by choosing a higher $x_1$ splitting value ($\tilde{x_1} = 0.087$ in our case), as opposed to the lower thresholds used in the literature (e.g., $x_1 = -1$ in \cite{Garnavich_2023} and \cite{Larison_2024}, or $x_1 \approx -0.5$ in \cite{Ginolin_2024}), we are guaranteeing that our higher stretch bin is mainly assessing the higher-stretch SN population, typically associated with younger progenitors \citep[e.g.,][]{Nicolas_2021}. This choice reduces the risk of contamination from the low-stretch mode, typically associated to older progenitors, in our high-stretch bin. In contrast, the defined lower stretch bin in our study may include an overlap between the two subpopulations, especially in the $x_1$ range between approximately -1 and 0. While this overlap may have only a modest impact, it might not be negligible.  To mitigate this contamination, one possible alternative would be to adopt a three-bin strategy, as in \cite{Kelly2010}, where the authors divide the stretch parameter into $x_1 < -1$, $-1 < x_1 < 0$, and $x_1 > 0$. This method allows a cleaner separation between the two populations, particularly in the overlapping region. However, this can reduce the statistical precision of the estimated parameters for each bin due to the limited number of calibrators. Instead, we redefine the bins by selecting SNe below the 30th percentile ($x_{1}^{\mathrm{low}} = -0.377$) and above the median ($x_{1}^{\mathrm{high}} = 0.087$) of the stretch distribution, as also presented in Figure \ref{figure: mixing_sub}.  Although this approach does not completely exclude the overlapping region, it reduces the number of SNe that may be misclassified between the high- and low-stretch bins while maintaining sufficient calibrators in both cases.

Overall, the results remain consistent with those obtained using the median stretch $x_{1}$ value ($\bar{x_{1}}$ = 0.087) as a division point of the stretch distribution. The most notable differences are again found for $H_{0}$ and $M_{B}$, as shown in Table \ref{tab:sne_lowhigh1}, although the discrepancy in the latter has decreased compared to the median-based division (see Table \ref{tab:sne_bins_median}), with differences of
2.00 $\sigma$ and 1.75 $\sigma$, respectively.

\begin{table}[h]
\centering
\resizebox{\columnwidth}{!}{%
\begin{tabular}{cccccccc}

\hline
SNe bin & Number of SNe &
$H_{0}$ & Sig. & $M_{B}$ & Sig. & $\alpha$ & Sig.\\ 

\hline

\addlinespace
 
$x_{1} < {x_{1}^{\text{low}}}$& 76 & $75.863_{-1.875}^{+1.687}$ & \multirow{3}{*}{ 2.00 $\sigma$} &$-19.169_{-0.050}^{+0.051}$ & \multirow{3}{*}{1.74 $\sigma$} &$0.149_{-0.037}^{+0.037}$ & \multirow{3}{*}{ 0.77 $\sigma$} \\

\addlinespace

$x_{1} \geq x_{1}^{\text{high}}$ & 127 &
$71.249_{-1.567}^{+1.576}$ & & $-19.291_{-0.052}^{+0.049}$ & & $0.113_{-0.028}^{+0.029}$ &  \\ 

 \addlinespace

\hline

\end{tabular}}

\caption{Estimated $H_{0}$, $M_{B}$ and $\alpha$ using SNe from two different stretch bins. The low-stretch bin includes SNe with $x_{1}$ below $x_{1}^{\text{low}}$ = -0.377, while the high-stretch bin includes SNe with $x_{1}$ above the $x_{1}^{\text{high}}$ = 0.087 threshold. The significances of the differences between the estimated parameters for each bin (Sig.) are also shown.}
\label{tab:sne_lowhigh1}

\end{table}

This reduction may partly result from the smaller number of SNe included in the low-stretch bin, while the number of SNe in the high-stretch bin remains unchanged for the two approaches. Again, no other notable variations are found for the remaining parameters estimated from the two different stretch bins and the mass step is nearly consistent with 0 within 1$\sigma$ uncertainties for both bins, obtaining $\Delta_{\mathrm{host}} = -0.012 \pm 0.017$ (0.71 $\sigma$) for the low-stretch bin and $\Delta_{\mathrm{host}} = -0.013 \pm 0.012$ (1.08 $\sigma$) for the high-stretch bin. A complete summary of all parameters estimated using this approach, along with the corresponding discrepancies, is provided in Tables \ref{tab:sne_lowhigh1} and \ref{tab:sne_lowhigh2}.

\begin{table}[h]
\centering
\resizebox{\columnwidth}{!}{%
\begin{tabular}{cccccccc}

\hline
SNe bin & Number of SNe &
$\beta$ & Sig. & $\Delta_{host}$ & Sig. & $\sigma_{int}$ & Sig. \\ \hline

 \addlinespace

$x_{1} < x_{1}^{\text{low}}$ & 76 & $2.829_{-0.178}^{+0.185}$ & 
\multirow{3}{*}{ 0.77 $\sigma$} &$-0.012_{-0.017}^{+0.017}$ & \multirow{3}{*}{ 0.05 $\sigma$} &$0.096_{-0.016}^{+0.017}$ & \multirow{3}{*}{ 0.23 $\sigma$} \\

\addlinespace

$x_{1} \geq x_{1}^{\text{high}}$ & 127 & $2.626_{-0.193}^{+0.194}$ & & $-0.013_{-0.012}^{+0.012}$ & & $0.101_{-0.013}^{+0.013}$ &  \\ 

 \addlinespace
 
 \hline

\end{tabular}}

\caption[Estimated $\beta$, $\Delta_{host}$ and $\sigma_{int}$ using SNe from two different stretch bins defined by $x_{1}^{ low}$ = -0.377 and $x_{1}^{high}$ = 0.087 ]{Same as Table \ref{tab:sne_lowhigh1} for the remaining parameters $\beta$, $\Delta_{host}$ and $\sigma_{int}$.}
\label{tab:sne_lowhigh2}

\end{table}

\section{Accessing SN subpopulations using other approaches}

\label{ap:dif_approaches}

\subsection{Maximizing the  property distributions matching}

\label{ap:other_binning}
As a way of trying to obtain a more accurate threshold for dividing the stretch distribution and distinguishing the two SN subpopulations, we estimated the value of $x_{1}$ that, when used as a division point in the full stretch distribution, maximizes the agreement between the calibration and Hubble flow distributions across all properties for the two stretch populations, hereafter denoted as $x_{1}^{*}$. To do that, and taking into account the division points between stretch modes reported in previous studies such as $x_{1}$= - 1 for \cite{Larison_2024} and $x_{1}$ = - 0.48 for \cite{Ginolin_2024}, we defined an interval for $x_{1}^{*}$ ranging from -1.5 to 0.5 in steps of 0.001. For each candidate threshold, we performed K–S tests comparing the distributions of $c$, $\log(M/M_{\odot})$, and $\log(\mathrm{sSFRyr})$ between the calibration and Hubble flow samples within each bin divided by $x_{1}^{*}$. The final value of $x_{1}^{*}$ was chosen as the one that maximized the resulting product of all the estimated $p$-values.

We obtained a value of $x_{1}^{*} = - 0.334$, which produced a notable improvement in the agreement between the calibration and Hubble flow samples for the $\log(M/M_{\odot})$ distribution in the lower-stretch bin and for the log(sSFRyr) distribution in the higher-stretch bin, relative to the values obtained using the median value of $\tilde{x_{1}} = 0.087$ as a division threshold, as shown in Table \ref{tab:pvals}. On the other hand, adopting this improved threshold also led to a significant decrease in the agreement of the stretch distributions within the lower-stretch bin. 

\begin{table}[h]
\centering
\renewcommand{\arraystretch}{1.4}
\begin{adjustbox}{max width=\textwidth}
\begin{tabular}{c|cccc|cccc|}
\cline{2-9}
                                                & \multicolumn{4}{c|}{Lower stretch bin}             & \multicolumn{4}{c|}{Higher stretch bin}            \\ \hline
\multicolumn{1}{|c|}{Division value}            & $c$ & $x_{1}$ & log($M/M_{\odot}$) & log(sSFRyr) & $c$ & $x_{1}$ & log($M/M_{\odot}$) & log(sSFRyr) \\
\hline
 
\multicolumn{1}{|c|}{$\tilde{x_{1}}$}           &  0.986 &  0.763      &    0.007 &   0.259        & 0.417    &  0.198         &   0.048                  &   0.094        \\ \hline
\multicolumn{1}{|c|}{$x_{1}^{\text{low/high}}$} & 0.733 
    & 0.305      &   0.056                  & 0.252
         & 0.417     &  0.198       &  0.048                   & 0.094           \\ \hline
\multicolumn{1}{|c|}{$x_{1}^{*}$} &  0.982
    &  0.187      &  0.062                  &  0.298          &  0.534    &  0.212         &  0.037                 & 0.336            \\ \hline
\end{tabular}
\end{adjustbox}
\caption{Estimated $p$-values comparing the calibration and Hubble Flow sample property distributions across different stretch binning thresholds: the median of the full sample ($\tilde{x}_{1} = 0.087$), the 30th ($x_{1}^{\mathrm{low}} = -0.377$) and 50th percentiles ($x_{1}^{\mathrm{high}} = 0.087$) of the $x_{1}$ distribution, and the value that maximizes the overall agreement between the calibration and Hubble Flow samples ($x_{1}^{*} = -0.334$).}
\label{tab:pvals}

\end{table}

As expected, the $p$-values for the lower-stretch bin upper limited by $x_{1}^{\text{low}} = -0.377$ are very similar to those obtained using $x_{1}^{*}$, given the close proximity of the division values, which results in nearly identical SNe composing both bins and thus similar property distributions. Conversely, the $p$-values for the higher-stretch bin lower limited by $x_{1}^{\text{high}} = 0.087$ are equal to those derived using $\tilde{x_{1}}$.  This reproduces the previously observed increase in the $p$-values for the $\log(\mathrm{sSFRyr})$ distributions in the high-stretch bin when the division point is changed from $\tilde{x_{1}}$ to $x_{1}^{*}$, as well as the observed decrease in the $p$-value for the $x_{1}$ parameter in the lower-stretch bin. All other properties exhibit consistent $p$-values across the bins defined by the different division points with only the $\log(M/M_{\odot})$ distribution in the higher-stretch bin, yielding $p$-values below the significance threshold of 0.05 for both cases.

Comparing the values of $H_{0}$, $M_{B}$, $\alpha$, and $\beta$ estimated using all SNe from the two bins defined by $x_{1}^{*}$, presented in Tables \ref{table:sne_bins_1} and \ref{table:sne_bins_2}, we find results broadly consistent with those obtained when using either the median of the full stretch distribution or the 30th and 50th percentiles of the full SN sample stretch distribution as division thresholds, with 
the largest discrepancies being found for the $H_{0}$ and $M_{B}$ estimated from the different bins. However, except for the $H_{0}$ case, all the discrepancies seem to be less significant when using the $x_{1}^{*}$ as the division value to separate the high and low stretch modes bins. We also note that both $H_{0}$ and $\alpha$ take slightly higher values under this separation. These differences  may result from the lower division value of $x_{1}^{*}$ compared to the division values defined across Section \ref{sub:bins}, which, while reducing the potential misclassification of SNe from the higher-stretch subpopulation into the lower-stretch bin, can increase the overlap between the two subpopulations within the defined high-stretch bin, as shown in Figure \ref{figure: mixing_sub}. This behavior is similar to what we observed for the SN bins obtained by dividing the stretch distribution using $\tilde{x_{1}}$, although in that case the subpopulation mixing primarily affects the lower-stretch bin. In contrast, the effect on the estimated $H_{0}$ and $\alpha$ is the opposite, as the parameters are lower in both stretch bins.

\begin{table*}[h]
\centering
\begin{adjustbox}{max width=\textwidth}
\begin{tabular}{cccccccc}

\hline
SNe bin & Number of SNe &
$H_{0}$ & Sig. & $M_{B}$ & Sig. & $\alpha$ & Sig. \\ 
\hline

 \addlinespace
 
$x_{1} < x_{1}^{*}$ & 80 & $76.386^{+1.602}_{-1.548}$ & \multirow{2.5}{*}{2.30 $\sigma$} & $-19.162^{+0.045}_{-0.046}$ & \multirow{2.5}{*}{1.42 $\sigma$}	& $0.153^{+0.034}_{-0.034}$ & \multirow{2.5}{*}{0.34 $\sigma$} \\

 \addlinespace
 
$x_{1} > x_{1}^{*}$ & 173 & $71.794^{+1.250}_{-1.240}$ & &	$-19.246^{+0.037}_{-0.037}$ & & $0.140^{+0.018}_{-0.018}$ & \\ 

 \addlinespace
 
 \hline

\end{tabular}
\end{adjustbox}
\caption[Estimated $H_{0}$, $M_{B}$ and $\alpha$ using SNe from two different bins divided by 
$x_{1}^{*}$ = -0.334]{Estimated $H_{0}$, $M_{B}$, $\alpha$ and $\beta$ using SNe from two different bins divided by 
$x_{1}^{*}$ = - 0.334 and their respective difference to the 16th and 84th percentiles.  The significance of the differences between the estimated parameters for each bin of the considered property (Sig.) are also shown.}
\label{table:sne_bins_1}

\end{table*}

\begin{table*}[h]
\centering
\begin{adjustbox}{max width=\textwidth}
\begin{tabular}{cccccccc}

\hline
SNe bin & Number of SNe &
$\beta$ & Sig. & $\Delta_{host}$ & Sig. & $\sigma_{int}$ & Sig.\\ \hline

 \addlinespace
 
$x_{1} < x_{1}^{*}$ & 80 &  $2.849^{+0.169}_{-0.168}$ & \multirow{2.5}{*}{0.21 $\sigma$} & $-0.014^{+0.016}_{-0.015}$ & \multirow{2.5}{*}{ 0.32 $\sigma$}	& $0.092^{+0.017}_{-0.016}$ & \multirow{2.5}{*}{0.15 $\sigma$}  \\ 

 \addlinespace
 
$x_{1} > x_{1}^{*}$ & 173 & $2.801^{+0.150}_{-0.152}$ & &	$-0.008^{+0.010}_{-0.010}$ & & $0.089 ^{+0.012}_{-0.011}$ &   \\ 

 \addlinespace
 
 \hline

\end{tabular}
\end{adjustbox}
\caption[Estimated $\beta$, $\Delta_{host}$ and $\sigma_{int}$ using SNe from two different bins divided by 
$x_{1}^{*}$ = -0.334]{Same as Table \ref{table:sne_bins_1} for the remaining parameters $\beta$, $\Delta_{host}$ and $\sigma_{int}$}
\label{table:sne_bins_2}

\end{table*}

\subsection{Applying Gaussian Mixture Models}
\label{ap:dif_divs}
The Gaussian Mixture Model (GMM) is a probabilistic method that assumes that the data is generated from a mixture of several Gaussian distributions with unknown parameters, each corresponding to a distinct cluster \citep{mclachlan2000finite}. Typically, the algorithm begins by initializing the means, covariances, and weights of the Gaussian distributions, often using k-means clustering for preliminary labeling \citep{steinhaus1956division}, and then iteratively refines these parameters by fitting the GMM to the data with the Expectation-Maximization (EM) algorithm \citep{477e7e2b-4ded-3369-981e-9b40850a2701}. One of the main limitations of Gaussian Mixture clustering is that the number of clusters must be specified in advance. In our case, however, this is not a critical drawback, since we aim to identify and distinguish the two already documented SN subpopulations characterized by their stretch distribution and strong host environment dependence \citep[e.g.,][]{Nicolas_2021,Ginolin_2024}, which naturally motivates the choice of two clusters. 

To evaluate and compare the quality of the clustering results, we computed the Silhouette score \citep{ROUSSEEUW198753} for each group of clusters, which quantifies how well each data point fits within its assigned cluster and how clearly it is separated from other clusters. Both the GMM algorithm and the Silhouette score were implemented using the \textit{scikit-learn} python package \citep{scikit-learn}. We applied the GMM clustering algorithm to different combinations of $c$, $x_{1}$, log($M/M_{\odot}$), and log(sSFRyr) distributions in an attempt to distinguish the different SN subpopulations. Besides all possible combinations of lighcurve and host properties, meaningful and scientifically motivated clustering was only obtained when applying GMM clustering to the $x_{1}$ and log($M/M_{\odot}$) distributions together, obtaining the clusters whose density contours are presented in Figure \ref{fig:clusters}.

\begin{figure*}[h]
    \centering

\includegraphics[width=\linewidth]{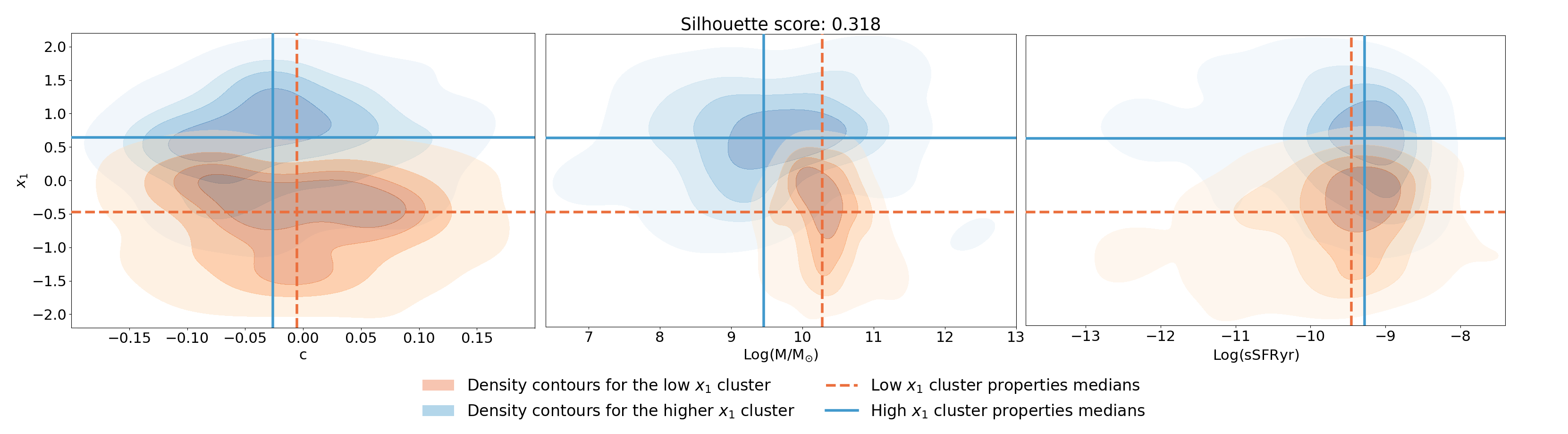}
    
    \caption[Density contours of $x_{1}$ as a function of $c$, $\log(M/M_\odot)$, and $\log(\mathrm{sSFRyr})$ for the two identifiedclusters, representing the high-stretch and low-stretch  SN subpopulations]{Density contours of $x_{1}$ as a function of $c$ (left), $\log(M/M_\odot)$ (middle), and $\log(\mathrm{sSFRyr})$ (right) for the two identified clusters, representing the high-stretch (blue) and low-stretch (orange) SN subpopulations. Dashed lines indicate the median values of each property distribution for the low-$x_{1}$ cluster, while solid lines indicate the corresponding medians for the high-$x_{1}$ cluster. The median $x_{1}$ value for the low-stretch cluster is $\tilde{x_{1}} = -0.475$ while for the high-stretch cluster it is $\tilde{x_{1}} = 0.640$.} 
    
    \label{fig:clusters}
    
\end{figure*}

We then proceed to estimate the cosmological parameters for each cluster, along with the standardization parameters listed in Tables \ref{table:clusters1} and \ref{table:clusters2}. The largest discrepancies arise in the estimated $H_{0}$ and $\beta$ between the lower and higher stretch SN clusters, with significances of 1.23 $\sigma$,  and 1.22 $\sigma$ respectively.  All parameters appear consistent with the results and conclusions of the previous analysis presented in \ref{ap:other_binning} and discussed in Subsection \ref{sub:bins}, despite showing much less significant discrepancies. Several factors may account for this behavior.

\begin{table}[h]
\centering
\resizebox{\columnwidth}{!}{%
\begin{tabular}{cccccccc}

\hline
Cluster & Number of SNe &
$H_{0}$ & Sig. & $M_{B}$ & Sig. & $\alpha$ & Sig. \\ 
\hline

 \addlinespace
 
Low-stretch (orange) & 108 & $74.769^{+1.325}_{-1.318}$ & \multirow{2.5}{*}{ 1.19 $\sigma$} & $-19.185^{+0.036}_{-0.035}$ & \multirow{2.5}{*}{0.87  $\sigma$}	& $0.143^{+0.022}_{-0.021}$ & \multirow{2.5}{*}{0.67 $\sigma$} \\

 \addlinespace
 
High-stretch (blue) & 145 & $72.473^{+1.399}_{-1.390}$ & &	$-19.231^{+0.040}_{-0.042}$ & & $0.125^{+0.017}_{-0.018}$ & \\ 

 \addlinespace
 
 \hline

\end{tabular}}

\caption[Estimated $H_{0}$, $M_{B}$ and $\alpha$ using SNe from two different clusters obtained using GMM clustering]{Estimated $H_{0}$, $M_{B}$ and $\alpha$ using SNe each cluster obtained using the GMM clustering algorithm and their respective difference to the 16th and 84th percentiles. The right-hand columns report the significance of the differences between the estimated parameters for each bin of the considered property.}
\label{table:clusters1}

\end{table}

\begin{table}[h]
\centering
\resizebox{\columnwidth}{!}{%
\begin{tabular}{cccccccc}

\hline
Cluster & Number of SNe &
$\beta$ & Sig. & $\Delta_{host}$ & Sig. & $\sigma_{int}$ & Sig.\\ \hline

 \addlinespace
 
Low-stretch (orange) & 108 & $2.692^{+0.156}_{-0.153}$ & \multirow{2.5}{*}{1.13 $\sigma$} & $-0.024^{+0.013}_{-0.013}$ & \multirow{2.5}{*}{  0.12 $\sigma$}	& $0.102^{+0.014}_{-0.014}$ & \multirow{2.5}{*}{1.41 $\sigma$}  \\ 

 \addlinespace
 
High-stretch (blue) &  145 & $2.964^{+0.176}_{-0.182}$ & &	$-0.026^{+0.011}_{-0.011}$ & & $0.074^{+0.014}_{-0.015}$ &   \\ 

 \addlinespace
 
 \hline

\end{tabular}}

\caption[Estimated $\beta$, $\Delta_{host}$ and $\sigma_{int}$ using SNe from two different cluster obtained using GMM clustering]{Same as Table \ref{table:clusters1} for the remaining parameters $\beta$, $\Delta_{host}$ and $\sigma_{int}$}
\label{table:clusters2}

\end{table}

First, when comparing our stretch distributions for each SN subpopulation with those reported in previous studies based on larger samples \citep[e.g.,][]{Wojtak_2022,Ginolin_2024,wojtak2025stretchstretchdustdust} who found nearly coincident Gaussian distributions describing the stretch populations, we find that while the higher-stretch distribution in our analysis closely agrees with the literature, the lower-stretch distribution exhibits substantially higher values than those previously reported, as we can see in Figure \ref{fig:clusters2}. Assuming that the $x_{1}$ distributions proposed in the literature are more closely representative of the expected behavior for each SN subpopulation, not only because of the consistency across multiple studies but also due to the larger sample sizes and methodology employed, our lower-stretch cluster is likely contaminated by SNe belonging to the higher-stretch population.

\begin{figure}[h]
    \centering

\includegraphics[width=0.9\linewidth]{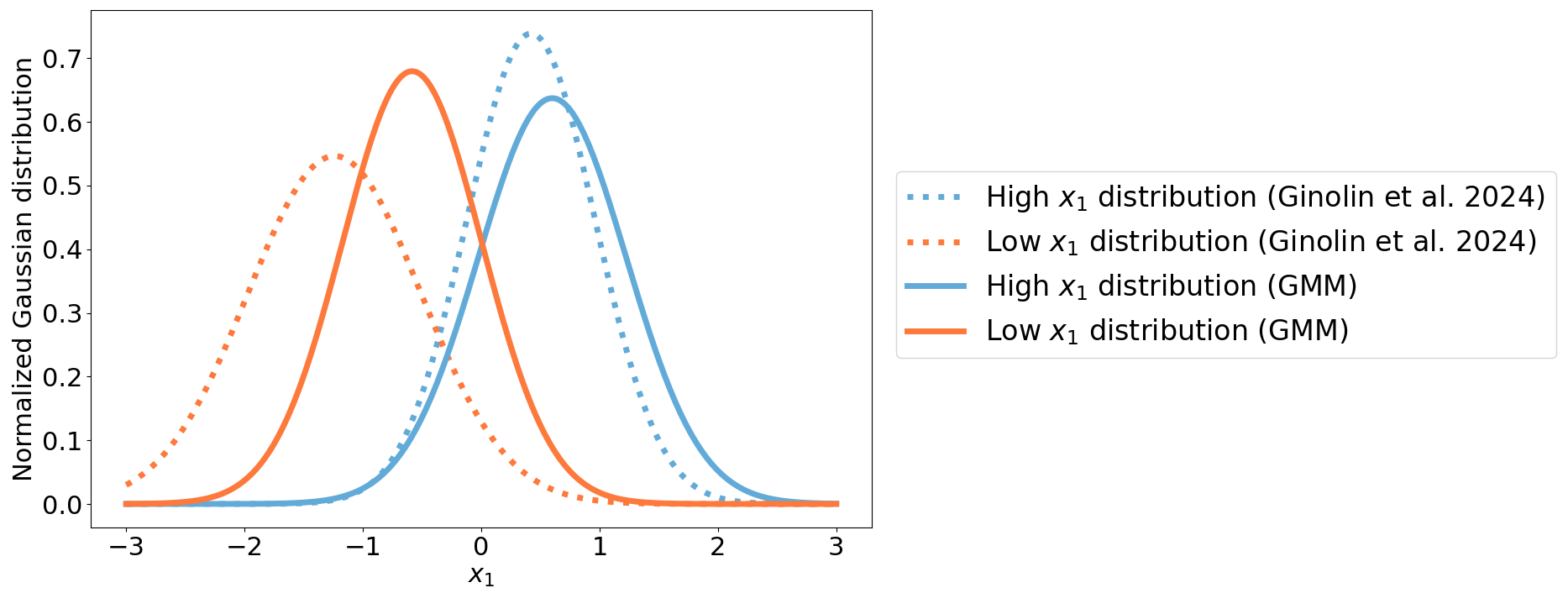}
    
    \caption{Normalized Gaussian distributions representing the high-stretch (blue) and low-stretch (orange) SN populations, using the fitted parameters from  \cite{Ginolin_2024} (dotted lines) and the parameters estimated in this work using the GMM clustering algorithm (solid lines).} 
    
    \label{fig:clusters2}
    
\end{figure}

In addition, our Gaussian Mixture fitting relies solely on the $x_{1}$ and $\log(M/M_{\odot})$ distributions. However, several other host-environmental properties, such as the local host colour, sSFR \citep[e.g.,][]{Sullivan2010,Ginolin_2024} and environment-based progenitor age \citep[e.g.,][]{Nicolas_2021}, are also strongly correlated with stretch and could provide additional constraints on the underlying distributions of both SN subpopulations. The GMM clustering also have its own limitations. The algorithm is somewhat sensitive to outliers, as Gaussian mixture models assume that the features are normally distributed and assign all data points to the identified clusters. Moreover, contamination of the data by noise or outliers can further lead to misclassification of the underlying subpopulations \citep{kasa2023avoiding}.

\label{ap:GMM}
\section{Using other matching metrics}
\label{ap:other_metrics}

In this chapter, we explore alternative metrics to the multi-dimensional K-S test used throughout this study, discussing their potential advantages and disadvantages. We also present the slopes obtained from linear regressions between the estimated parameters ($H_{0}$, $M_{B}$, $\alpha$, $\beta$, $\Delta_{host}$ and $\sigma_{int}$) and the different metrics used to evaluate the consistency between the calibration and Hubble Flow subsamples.

\subsection{Sum and product of $p$-values}

\label{ap:sum_vs_prod}

To evaluate the influence of the overall distribution concordance, we decided to plot the estimated parameters as a function of the logarithm of the sum and product of the $p$-values obtained when comparing each property distribution of the calibration and Hubble Flow subsamples using the one dimensional K-S Test, as shown in the left and right panels of Figure \ref{fig:sum_and_prod}, respectively.

\begin{figure*}[h]
    \centering
    \includegraphics[width=\linewidth]{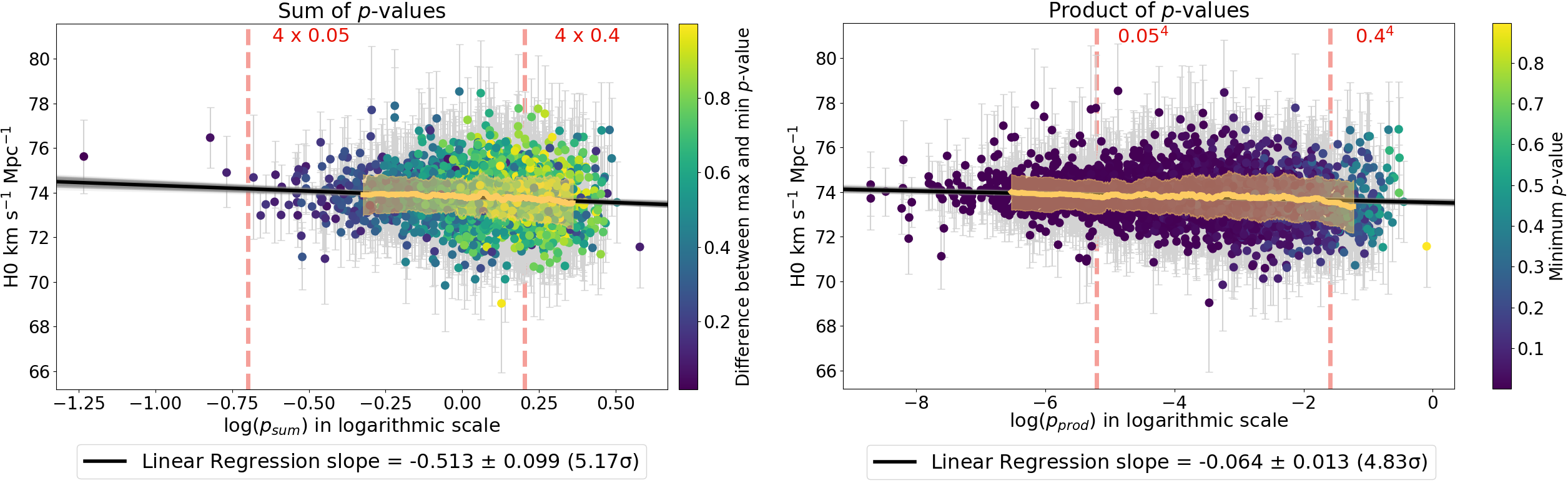}
    
    \caption{Similar to analysis shown in Figure \ref{fig:medianH0_cal} for $H_{0}$ when using the sum (left) or the product (right) of the individual $p$-values as the overall matching metric. The colormap in the left panel represents the difference between the highest and lowest $p$-values among the $c$, $x_{1}$, log($M/M_{\odot}$) and log(sSFRyr) distributions estimated for each subsample, while the colormap in the right panel represents the lowest $p$-value of each subsample. The red dashed lines indicate the obtained metric values corresponding to a subsample where all $p$-values are 0.05 or 0.4.}
    \label{fig:sum_and_prod}
\end{figure*}

When generating random subsamples from the calibration and Hubble Flow, it is expected that some property distributions may closely match those of the calibration subsample, resulting in very high one-dimensional $p$-value, while others may not, resulting in very low $p$-value. This can lead to large discrepancies between the individual $p$-values estimated for different properties distributions within the same subsample. By using only the sum of all  $p$-values, the higher 
$p$-values will dominate the result, overshadowing the contribution of the smaller values. This effect is evident in the left panel of Figure \ref{fig:sum_and_prod}, which shows that the total sum of $p$-values increases with the difference between the highest and lowest $p$-values obtained for each subsample. This can give the misleading impression that most subsamples are consistent across all properties when 
$p_{sum}>$ 0.05, while this apparent consistency is primarily driven by the highest $p$-value among the properties within each subsample

On the other hand, using the direct product of the one-dimensional
$p$-values will cause extremely high or low values to dominate the result.  As shown in the right panel of Figure~\ref{fig:sum_and_prod}, most subsamples have a low minimum $p$-value (mostly below 0.05). However, their $p$-value product is still higher than that of a subsample in which the distributions of $c$, $x_1$, $\log(M/M_\odot)$, and $\log({\rm sSFRyr})$ in the calibration and Hubble flow subsamples all match, with all one-dimensional $p$-values above the 0.05 significance level. Despite having a higher $p$-value product, these subsamples are not in full concordance with each other, since at least one of their individual $p$-values falls below the significance threshold. While the product metric includes information from all parameters, its sensitivity to extremely higher or lower values limits its ability to assess whether both subsamples are in full agreement.

The slopes and corresponding uncertainties obtained from the linear regressions between the estimated parameters and the two metrics are presented in Table \ref{tab:relations_sum_prod}. These results are consistent with those shown in Figure~\ref{fig:medianH0_cal} and discussed in Section \ref{sub:matching}.

\begin{table*}[h]
\centering
\renewcommand{\arraystretch}{1.4}
\begin{tabular}{c|c|c|c|c|}
\cline{2-5}
                                                & \multicolumn{2}{c|}{Sum of $p$-values ($p_{sum}$)}             & \multicolumn{2}{c|}{Product of $p$-values ($p_{prod}$)}            \\ \hline
\multicolumn{1}{|c|}{Parameter}  &          
 Linear Regression Slope & Sig. & Linear Regression Slope & Sig \\
\hline
 
\multicolumn{1}{|c|}{$H_{0}$}           & (-5.126 $\pm$ 0.991) $\times$ 10$^{-1}$ & 5.17 $\sigma$ & (-6.432 $\pm$ 1.333) $\times$ 10$^{-2}$  &  4.83 $\sigma$         \\ \hline
\multicolumn{1}{|c|}{$M_{B}$} &   (-1.541 $\pm$ 0.325) $     \times$ 10$^{-2}$ & 4.74 $\sigma$ &  (-2.253 $\pm$ 0.367) $   \times$ 10$^{-3}$ & 6.15 $\sigma$        \\ \hline
\multicolumn{1}{|c|}{$\alpha$} &  (4.419 $\pm$ 1.438) $\times$ 10$^{-3}$ & 3.07 $\sigma$ & (7.416 $\pm$ 1.484) $\times$ 10$^{-4}$ & 5.00 $\sigma$    \\ \hline

\multicolumn{1}{|c|}{$\beta$} &  (2.543 $\pm$ 1.797) $\times$ 10$^{-2}$ & 1.42 $\sigma$ & (5.997 $\pm$ 1.854) $\times$ 10$^{-3}$ & 3.23 $\sigma$    \\ \hline

\multicolumn{1}{|c|}{$\Delta_{host}$} &  (-7.200 $\pm$ 14.161) $\times$ 10$^{-4}$ & 0.51 $\sigma$ &  (-1.966 $\pm$ 1.637) $ \times$ 10$^{-4}$ & 1.20 $\sigma$     \\ \hline

\multicolumn{1}{|c|}{$\sigma_{int}$} &  (-4.210 $\pm$ 1.324) $\times$ 10$^{-3}$ & 3.18 $\sigma$ & (-8.561 $\pm$ 1.691) $\times$ 10$^{-4}$ &  5.06 $\sigma$     \\ \hline

\end{tabular}
\caption{Slopes and uncertainties resulting from the linear regression between $H_{0}$, $M_{B}$, $\alpha$, $\beta$, $\Delta_{host}$, and $\sigma_{\mathrm{int}}$ and either the sum (left column) or the product (right column) of the $p$-values obtained from comparing $c$, $x_{1}$, $\log(M/M_{\odot})$, and $\log(\mathrm{sSFRyr})$ distributions between the calibration and HF samples using the one-dimensional K-S test, along with the corresponding significance levels. The fitting procedure is performed using \textsc{linmix} \cite{Kelly_2007}.}
\label{tab:relations_sum_prod}
\end{table*}

\subsection{Minimum $p$-value}

\label{ap:sum_vs_prod}

We also defined the minimum $p$-value as the lowest among the four values obtained from the one-dimensional K-S tests
comparing the distributions of the light curve parameters and host properties between each calibration and HF subsample, providing a general metric to assess the overall concordance between their distributions.  Indeed, we are expecting that increasing the minimum $p$-value should result in subsamples whose properties distributions match better and better as a higher minimum $p$-value necessarily implies that all individual $p$-values exceed that threshold, ensuring a level of concordance across all
parameters. This allows us to use a single metric to evaluate the overall concordance across all properties distributions . However, a low minimum $p$-value may not  necessarily indicate poor matching across all the considered distributions of $c$, $x_{1}$, log($M/M_{\odot}$), and log(sSFRyr), as a subsample with a small minimum $p$-value could still exhibit high $p$-values for the remaining three properties, which gives some limitations on the use of the minimum $p$-value as a metric for overall distributional concordance. 

As shown in  Figure \ref{fig:minpvalue}, most subsamples have a minimum $p$-value below 0.05, which would suggest poor matching based on this metric. However, their median $p$-values are generally above 0.05, indicating that the low minimum $p$-value often reflects disagreement in just one property rather than across all properties.
Nevertheless, the results obtained using this metric are consistent with the previous findings reported in Section \ref{sub:matching}, as shown in Table \ref{tab:minpvalue}.

\begin{figure}[h]
\centering
\includegraphics[width=0.7\linewidth]{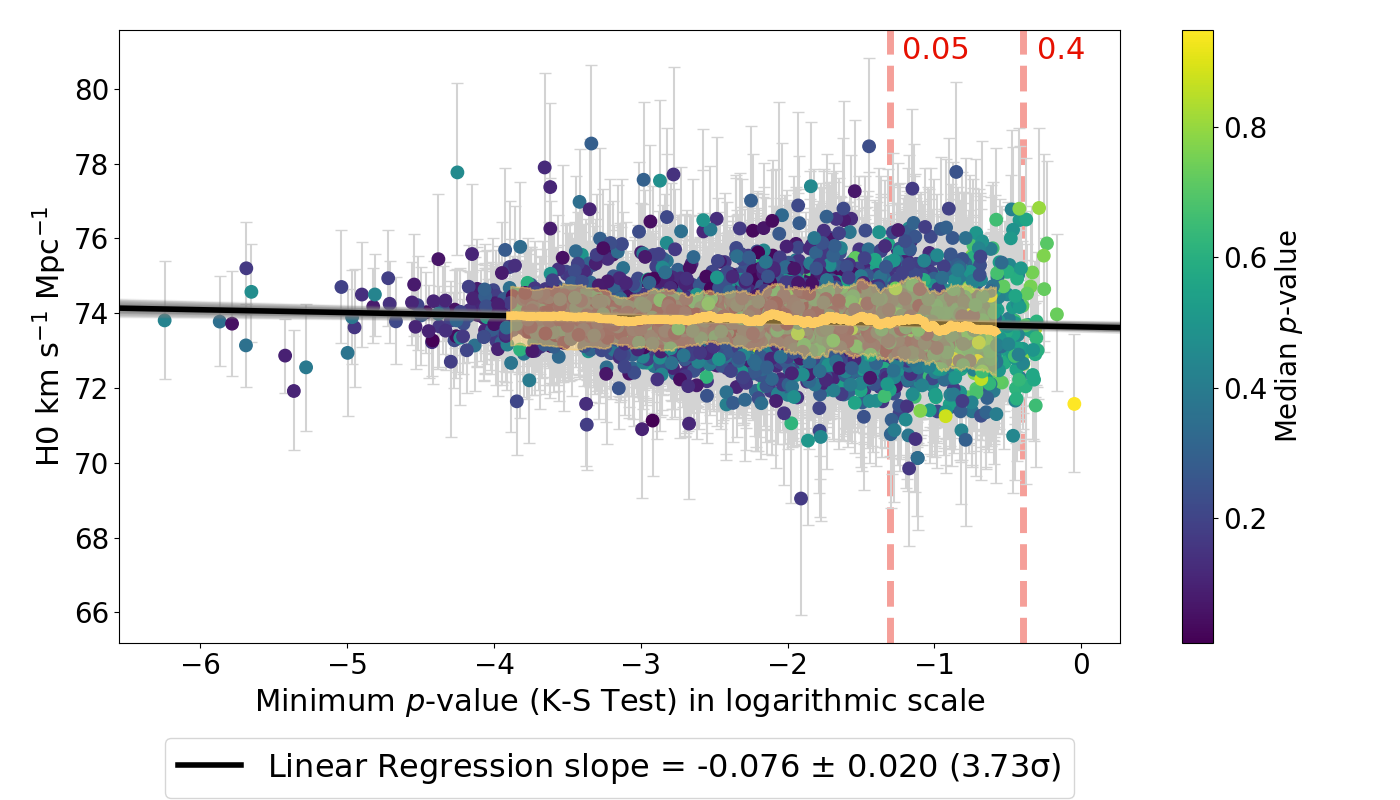}
\caption{Similar to the analysis shown in Figure \ref{fig:medianH0_cal} for $H_{0}$ when using the minimum $p$-values as the overall matching metric. The colormap represents the median $p$-value estimated from the $c$, $x_{1}$, log($M/M_{\odot}$) and log(sSFRyr) distributions estimated for each subsample.}
\label{fig:minpvalue}
\end{figure}

\begin{table}
    \centering
    \renewcommand{\arraystretch}{1.4}
    \begin{tabular}{c|c|c|}
    \cline{2-3}
                                                    & \multicolumn{2}{c|}{Minimum $p$-value}            \\ \hline
    \multicolumn{1}{|c|}{Parameter}  & Linear Regression Slope & Sig. \\ \hline
    \multicolumn{1}{|c|}{$H_{0}$} & $(-7.626 \pm 2.044)\times 10^{-2}$ & $3.73$ $\sigma$ \\ \hline
    \multicolumn{1}{|c|}{$M_{B}$} & $(-2.846 \pm 0.579)\times 10^{-3}$ & $4.91$ $\sigma$ \\ \hline
    \multicolumn{1}{|c|}{$\alpha$} & $(1.186 \pm 0.235) \times 10^{-3}$ & $5.05$ $\sigma$ \\ \hline
    \multicolumn{1}{|c|}{$\beta$} & $(8.404 \pm 2.503)\times 10^{-3}$ & $3.36$ $\sigma$ \\ \hline
    \multicolumn{1}{|c|}{$\Delta_{host}$} & $(-1.195 \pm 2.488)\times 10^{-4}$ & $0.48$ $\sigma$ \\ \hline
    \multicolumn{1}{|c|}{$\sigma_{int}$} & $(-1.020 \pm 0.257)\times 10^{-3}$ & $3.97$ $\sigma$ \\ \hline
    \end{tabular}
    \label{tab:relations_min}

\caption{Slopes and uncertainties from the linear regression between $H_{0}$, $M_{B}$, $\alpha$, $\beta$, $\Delta_{host}$, and $\sigma_{\mathrm{int}}$ and the minimum $p$-values obtained from comparing $c$, $x_{1}$, $\log(M/M_{\odot})$, and $\log(\mathrm{sSFRyr})$ distributions between the calibration and HF samples. The fitting procedure is performed using \textsc{linmix} \cite{Kelly_2007}.}
\label{tab:minpvalue}
\end{table}

\subsection{Energy distance based on the Wasserstein metric}

The Wasserstein distance measures how different two probability distributions are by quantifying the minimum “work” required to transform one into the other, where work is defined as the amount of probability mass moved times the distance it is transported. This allows it to capture differences in both shape and location of the distributions \citep[see][]{panaretos2019statistical}. The concept is rooted in optimal transport theory, introduced by \cite{monge1784memoire}, which seeks the least-cost mapping between two distributions. In the discrete case, the Wasserstein distance corresponds to the minimal total cost of an optimal transport plan that reallocates mass from one distribution to match the other. In this approach, we used the python implementation of this algorithm\footnote{\url{https://docs.scipy.org/doc/scipy/reference/generated/scipy.stats.wasserstein_distance_nd.html}} provided by the SciPy library \citep{2020SciPy-NMeth} that allow us to compute the 1st Wasserstein distance between two N-dimensional discrete distributions.

The Energy Distance is a statistical measure used to quantify the dissimilarity between two probability distributions \citep[see][]{rizzo2016energy}. It is based on the idea of comparing the expected pairwise distances between samples, capturing both the separation between the distributions, as well as their internal variability. In this method, we use the Wasserstein metric to estimate the expected pairwise distances between samples, contrary to the typically Euclidian distance. To estimate the Energy distance, we used the python implementation\footnote{\url{https://docs.scipy.org/doc/scipy/reference/generated/scipy.stats.energy_distance.html}} of the energy distance algorithm provided by the SciPy library \citep{2020SciPy-NMeth}.

To assess whether the differences measured by these distances are statistically significant, we apply a permutation test \citep{fisher1935, pitman1937}. This non-parametric test evaluates how likely it is that the observed difference between two samples could arise by chance. By repeatedly reshuffling the data labels and recalculating the distance-based statistic, we build a reference distribution that represents the null hypothesis that the two samples are drawn from the same population. The fraction of reshuffled datasets that produce a difference as large as the observed one yields the $p$-value, allowing us to determine whether the samples differ significantly without assuming any specific underlying distribution.

This method offers several advantages over previous approaches, including the multi-dimensional K-S test. In particular, the Wasserstein distance allows us to quantify the distance between all points in two different distributions. The Wasserstein distance evaluates the differences across the entire distribution, rather than only the maximum difference between cumulative distribution functions as the K-S test does. This provides a more complete measure of the difference between the samples and allows for better distinction between them. Furthermore, \cite{ramirez2024novel} shows that optimal transport interpolation generates spectral time series that closely match the originals, with errors increasing more slowly than with linear interpolation. These findings suggest that the Wasserstein distance not only captures differences in the spatial displacement of distributions, but also preserves differences in their shapes with high accuracy.

However, despite its potential advantages, this methodology is still under development, as computing the energy distance using the Wasserstein distance for multi-dimensional samples remains highly computationally demanding \citep[see][]{panaretos2019statistical}. Since the multi-dimensional K-S test already provides a robust approach for comparing distributions and the metric discussed in this Section is not yet fully established, we chose not to adopt it as the primary comparison metric throughout this work. Nevertheless, we include in Table \ref{tab:energy_distance_slopes} the slopes obtained from linear regressions between the estimated parameters and the 
$p$-values computed when comparing the property distributions of the calibration and Hubble Flow samples using the energy distance with the Wasserstein metric. The permutation test used to calculate these $p$-values was based on 1000 permutations for each subsample.

\begin{table}[h]
\centering
\renewcommand{\arraystretch}{1.4}
\begin{tabular}{c|c|c|}
\cline{2-3}
                                                & \multicolumn{2}{c|}{Energy distance $p_{E}$}            \\ \hline
\multicolumn{1}{|c|}{Parameter}  &          
 Linear Regression Slope & Sig. \\
\hline
 
\multicolumn{1}{|c|}{$H_{0}$}           & $(-4.613 \pm 2.143) \times 10^{-2}$ & $2.15$ $\sigma$  \\ \hline
\multicolumn{1}{|c|}{$M_{B}$} &  $(-2.429 \pm 0.544) \times 10^{-3}$ & $4.46$ $\sigma$  \\ \hline
\multicolumn{1}{|c|}{$\alpha$} & $(1.147 \pm 0.263) \times 10^{-3}$ & $4.36$ $\sigma$  \\ \hline

\multicolumn{1}{|c|}{$\beta$} &  $(9.463 \pm 2.030) \times 10^{-3}$ & $4.66$ $\sigma$ \\ \hline

\multicolumn{1}{|c|}{$\Delta_{host}$} & $(-0.952 \pm 2.456) \times 10^{-6}$ & $0.39$ $\sigma$  \\ \hline

\multicolumn{1}{|c|}{$\sigma_{int}$} & $(-1.663 \pm 0.262) \times 10^{-3}$ & 6.34 $\sigma$ \\ \hline

\end{tabular}
\caption[Slopes and uncertainties resulting from the linear regression between $H_{0}$, $M_{B}$ , $\alpha$, $\beta$,
$\Delta_{host}$ and $\sigma_{int}$, and $p_{E}$] {Slopes and uncertainties resulting from the linear regression between $H_{0}$, $M_{B}$, $\alpha$, $\beta$, $\Delta_{host}$, and $\sigma_{\mathrm{int}}$ and the $p_{E}$ obtained from comparing the calibration and HF samples using the Energy distance with the Wasserstein metric, along with the corresponding significance levels. The fitting procedure is performed using \textsc{linmix} \cite{Kelly_2007}.}
\label{tab:energy_distance_slopes}
\end{table}

\newpage

\section{Effects of including the lower-sSFR SNe}
\label{ap:ssfr_effect}

As specified in section \ref{sec:data}, some SNe presented values of log(sSFRyr) $<$ - 15, which seemed to be outliers or lower limits on the true values of this parameter. We removed these SNe given their outlier sSFR values, and their small number (7 SNe) relative to the total sample size (260 SNe) should not have a statistically significant impact, nor affect the final conclusions. If we compare the log(sSFRyr) distributions of the calibration and HF samples without removing any SNe, we obtain a one-dimensional $p$-value of 0.018, which is very similar to the $p$-value of 0.030 presented in Fig.~\ref{fig:c_vs_ssfr} for the sample without the 7 SNe. This leads us to conclude that the observed mismatch between the sSFR distributions of the calibration and HF samples is not caused by these extreme values. Nevertheless, due to the extreme values of this parameter for some host galaxies ($<-22$) reported in the Pantheon+SH0ES compilation, including these values could affect the comparison of sSFR distributions in subsamples with a small number of SNe.

When comparing the parameter estimates for the full sample with and without the exclusion of SNe with log(sSFRyr) $<$ -15 (Table~\ref{table:2}), we observe a slight difference in the estimated value of $H_0$, that seems to be primarily driven by the removal of SNe from the calibration sample. Comparing the full sample with no SNe removed (I) to the sample restricted to SNe with log(sSFRyr) $>$ -15 (II), we find no significant discrepancy ($\sim$ $0.18\sigma$ in $H_0$).
We can also see in Table~\ref{table:2} that removing these SNe from the Hubble Flow sample while keeping all calibration SNe (IV) has a negligible impact on the parameter estimates. In contrast, excluding only the two identified SNe from the calibration sample (III) results in parameter values that differ from the full sample and align more closely with those obtained using the restricted sample adopted in this study, although the estimates of $H_0$ and $M_B$ remain highly consistent. Likewise, the inclusion of the previously removed low-sSFR SNe does not change the results obtained using the SN bins defined by the sample median $x_{1}$, as shown in Tables~\ref{tab:median1_lowerssfr} and \ref{tab:median2_lowerssfr}. The parameter values remain highly consistent with those estimated after excluding these 7 SNe (Tables \ref{tab:sne_bins1} and \ref{tab:sne_bins2}).

\begin{table}[]
\centering

\begin{tabular}{ccccccc}

\hline
Sample &
$H_{0}$ & M$_{B}$ & $\alpha$ & $\beta$  & $\Delta_{host}$  & $\sigma_{int}$  \\ \hline

\\ I &
74.029$^{+0.947}_{-0.939}$& -19.192$^{+0.026}_{-0.026}$	& 
0.135$^{+0.010}_{-0.010}$& 
2.783$^{+0.119}_{-0.112}$& 
-0.020$^{+0.009}_{-0.008}$& 
0.092$^{+0.009}_{-0.009}$ \\ 

\\ \hline

\\ II &
73.781$^{+0.970}_{-0.944}$& -19.199$^{+0.027}_{-0.027}$	& 
0.136$^{+0.010}_{-0.010}$& 
2.777$^{+0.116}_{-0.111}$& 
-0.020$^{+0.009}_{-0.008}$& 
0.092$^{+0.009}_{-0.009}$ \\

\\ \hline

\\ III &
73.795$^{+0.936}_{-0.957}$& -19.199$^{+0.026}_{-0.027}$	& 
0.136$^{+0.010}_{-0.010}$& 
2.777$^{+0.120}_{-0.110}$& 
-0.020$^{+0.009}_{-0.008}$& 
0.091$^{+0.009}_{-0.009}$\\
\\ \hline

\\ IV &
74.132$^{+0.900}_{-0.930}$& -19.190$^{+0.025}_{-0.025}$	& 
0.135$^{+0.010}_{-0.010}$& 
2.780$^{+0.116}_{-0.112}$& 
-0.020$^{+0.009}_{-0.009}$& 
0.091$^{+0.009}_{-0.009}$\\ 

\\ \hline

\end{tabular}

\caption{Estimated parameter values and their differences relative to the 16th and 84th percentiles, obtained for: (I) the full Calibration and Hubble Flow sample, (II) the full sample with SNe having log(sSFRyr) $<$ -15 removed from both samples (same as the one defined in section \ref{sec:data} and used through this study), (III) SNe removed only from the calibration sample, and (IV) SNe removed only from the Hubble Flow sample.}
\label{table:2}
\end{table} 

\begin{table}[h]
\centering
\centering
\resizebox{\columnwidth}{!}{%
\begin{tabular}{cccccccc}

\hline
SNe bin & Number of SNe &
$H_{0}$ & Sig. & $M_{B}$ & Sig. & $\alpha$ & Sig.\\ 

\hline

\addlinespace
 
$x_{1} < \tilde{x_{1}}$ & 130 & $75.489_{-1.181}^{+1.202}$ & \multirow{2.5}{*}{ 2.00$\sigma$} &$-19.151_{-0.032}^{+0.032}$ & \multirow{2.5}{*}{ 2.14 $\sigma$} &$0.129_{-0.022}^{+0.021}$ & \multirow{2.5}{*}{ 0.33 $\sigma$} \\

\addlinespace

$x_{1} \geq \tilde{x_{1}}$ & 130 &
$71.626_{-1.495}^{+1.526}$ & & $-19.276_{-0.049}^{+0.049}$ & & $0.117_{-0.029}^{+0.029}$ &  \\ 

 \addlinespace

\hline

\end{tabular}}

\caption{Estimated median values of $H_{0}$, $M_{B}$, and $\alpha$, with differences relative to the 16th and 84th percentiles, for SNe subsamples drawn from the two stretch bins defined by $\tilde{x_{1}}$ including the 7 lower-sSFR SNe. The significances of the differences between the estimated parameters for each bin (Sig.) are also shown.}
\label{tab:median1_lowerssfr}

\end{table}

\begin{table}[h]
\centering
\resizebox{\columnwidth}{!}{%
\begin{tabular}{cccccccc}

\hline
SNe bin & Number of SNe &
$\beta$ & Sig. & $\Delta_{host}$ & Sig. & $\sigma_{int}$ & Sig. \\ \hline

 \addlinespace

$x_{1} < \tilde{x_{1}}$ & 130 & $2.867_{-0.139}^{+0.154}$ & 
\multirow{2.5}{*}{ 0.87 $\sigma$} &$-0.014_{-0.012}^{+0.012}$ & \multirow{2.5}{*}{0.17 $\sigma$} &$0.082_{-0.013}^{+0.013}$ & \multirow{2.5}{*}{ 1.08 $\sigma$} \\

\addlinespace

$x_{1} \geq \tilde{x_{1}}$ & 130 & $2.658_{-0.200}^{+0.195}$ & & $-0.011_{-0.012}^{+0.013}$ & & $0.102_{-0.013}^{+0.013}$ &  \\ 

 \addlinespace
 
 \hline

\end{tabular}}

\caption{Same as Table \ref{tab:median1_lowerssfr} for $\beta$, $\Delta_{host}$ and $\sigma_{int}$.}
\label{tab:median2_lowerssfr}

\end{table}

\section{Effect of Bias Corrections and Intrinsic Scatter modeling}
\label{ap:bias}

As introduced in section \ref{sec:methodology}, cosmological studies should include an additional bias correction term $\delta_{bias}$ in equation \ref{eq:mu} to account for
observational selection biases determined from simulations, as well as second order corrections arising
from full forward modelling of SN colour parameters and observed magnitudes based on physically
motivated models of dust and supernova intrinsic color (e.g., \cite{Scolnic_2016,Popovic_2021}). In this work, we exclude the correction term $\delta_{bias}$, as we do not attempt to correct for residual
effects associated with population-dependent selection biases, and instead aim to minimize these effects
by using more consistent SNe samples.

One common form of selection bias is the Malmquist bias \cite{malmquist1922some}, which describes the
tendency of astronomical surveys to preferentially detect intrinsically brighter objects. This bias implies
that bluer and longer-lasting SNe Ia are more likely to be detected, since they are intrinsically brighter.
Consequently, the observed sample becomes biased toward SNe above a certain apparent magnitude threshold, leading to the exclusion of fainter events and an overestimation of the mean absolute magnitude of the selected sample. Regarding this bias,  which is also included in this bias correction term \cite{Scolnic_2016} but not provided in isolation, we expect its effects to be minimal since the sample is restricted to low redshifts (in this case z $\leq$ 0.15) (see \cite{10.1111/j.1365-2966.2005.09306.x,2011ApJS..192....1C}

The statistical covariance matrix provided in the Pantheon+SH0ES compilation (\cite{Riess_2022,Brout_2022}) also includes a term describing intrinsic brightness fluctuations and a color-independent
floor, which represents the intrinsic scatter in SN Hubble residuals. This uncertainty denoted as $\sigma_{\text{floor}}$ is 
computed for the best-fit model of dust and supernova intrinsic colors adopted in the Pantheon+ SNe compilation, with its dependence on stellar mass implemented via the step function defining the stellar
mass bins (for more details see
\cite{Brout_2022,Popovic_2021}). Therefore, we removed this term and reintroduced it as a free parameter in Equation \ref{eq:sigma}, denoted by $\sigma_{\mathrm{int}}$.

To test the impact of including these terms on the estimation of $H_0$, $M_B$, $\alpha$, and $\beta$ from the full SNe sample, we repeat the analysis described at the beginning of Section \ref{sub:full_sample}, either adopting the $\delta_{bias}$ and $\sigma_{\text{floor}}$ values from the Pantheon+SH0ES compilation or not including one or both of them in the fitting. As shown in Table \ref{tab:bias_fullsample}, the estimated $H_0$, $M_B$ and $\alpha$ values are consistent with those obtained without including the $\delta_{bias}$ term or the $\sigma_{\text{floor}}$ term in the fitting, and with the $H_0$ and $M_B$ values reported in \cite{Riess_2022}. However, the color–luminosity parameter $\beta$ increases when the $\delta_{bias}$ term is adopted, becoming more consistent with the $\beta \sim 3.1$ value reported in \cite{Brout_2022}, independently of whether $\sigma_{\text{floor}}$ is included. Comparing the value of this parameter obtained with $\delta_{bias}$ to that obtained without it, we observe a difference in the range of $\sim 1.9 - 2.2\sigma$. This shift likely arises because the bias correction reduces colour-dependent trends in the luminosity residuals (see \cite{popovic_bbc}) which would otherwise be partly absorbed by the $\beta$ parameter in the fit.

\begin{table}[h]
\centering
\resizebox{\columnwidth}{!}{%
\begin{tabular}{ccccccc}

\hline
Included values &
$H_{0}$ & M$_{B}$ & $\alpha$ & $\beta$  & $\Delta_{host}$  & $\sigma_{int}$ \\ \hline

\\ None &
73.781$^{+0.970}_{-0.944}$& -19.199$^{+0.027}_{-0.027}$	& 
0.136$^{+0.010}_{-0.010}$& 
2.777$^{+0.116}_{-0.111}$& 
-0.020$^{+0.009}_{-0.008}$& 
0.092$^{+0.009}_{-0.009}$ \\ 

\\ \hline

\\ Only $\sigma_{   \text{floor}}$  &
73.892$^{+0.910}_{-0.873}$& -19.192$^{+0.026}_{-0.025}$	& 
0.138$^{+0.010}_{-0.010}$& 
2.817$^{+0.107}_{-0.097}$& 
-0.022$^{+0.008}_{-0.008}$ & 
- \\ \\ \hline

\\ Only $\delta_{bias}$  &
73.647$^{+0.906}_{-0.905}$& -19.191$^{+0.026}_{-0.026}$	& 
0.137$^{+0.010}_{-0.010}$& 
3.129$^{+0.120}_{-0.114}$& 
- & 
0.079$^{+0.010}_{-0.010}$ \\ \\ \hline

\\ $\delta_{bias}$ and $\sigma_{\text{floor}}$ &
73.580$^{+0.922}_{-0.866}$& -19.189$^{+0.025}_{-0.025}$	& 
0.138$^{+0.010}_{-0.010}$& 
3.098$^{+0.115}_{-0.108}$& 
- & 
- 
\\ \\ \hline

\end{tabular}
}
\caption{Median parameter values obtained using different fitting approaches, either including or excluding the $\delta_{\rm bias}$ and $\sigma_{\rm floor}$ values from the Pantheon+SH0ES compilation, for the full calibration and Hubble flow sample used in this work.
Row 1: Fit excluding both $\delta_{\rm bias}$ and $\sigma_{\rm floor}$ (same results as Table \ref{tab:parameters}).
Row 2: Fit including $\sigma_{\rm floor}$ but excluding $\delta_{\rm bias}$.
Row 3: Fit including $\delta_{\rm bias}$ but excluding $\sigma_{\rm floor}$.
Row 4: Fit including both $\delta_{\rm bias}$ and $\sigma_{\rm floor}$.}
\label{tab:bias_fullsample}
\end{table}
}

For the different stretch bins divided by the median $\tilde{x_{1}}$ = 0.087, we still observe a discrepancy between the estimated $H_{0}$ and $M_{B}$ when including the $\delta_{bias}$ and $\sigma_{\text{floor}}$ values in the fitting. This discrepancy is slightly less pronounced for the former when compared to the one reported in Section~\ref{sub:bins} and Table~\ref{tab:sne_bins_median}, as shown in Table~\ref{tab:bins_with_bias}. Each $H_{0}$ estimate shifts by $\sim$ 0.3 km s$^{-1}$ Mpc $^{-1}$ closer to each other when including these values in the fitting. The main change is again seen in the color–luminosity parameter $\beta$, which remains consistently higher for both $x_{1}$ bins (Table \ref{tab:bins_with_bias}) when using the $\delta_{bias}$ and $\sigma_{\text{floor}}$ from the Pantheon+SH0ES compilation, compared to the estimates obtained without these terms (Table~\ref{tab:sne_bins2}).

\begin{table}[h]
\centering
\resizebox{\columnwidth}{!}{%
\begin{tabular}{cccccccc}

\hline
SNe bin & Number of SNe &
$H_{0}$ & Sig. & $M_{B}$ & Sig. & $\beta$ & Sig. \\ \hline

 \addlinespace

$x_{1} < \tilde{x_{1}}$ & 126 & $74.937_{-1.213}^{+1.225}$ & 
\multirow{2.5}{*}{1.81 $\sigma$} &$-19.146_{-0.035}^{+0.036}$ & \multirow{2.5}{*}{2.20 $\sigma$} &$3.205_{-0.154}^{+0.169}$ & \multirow{2.5}{*}{1.11 $\sigma$} \\

\addlinespace

$x_{1} \geq \tilde{x_{1}}$ & 127 & $71.561_{-1.403}^{+1.418}$ & & $-19.268_{-0.043}^{+0.043}$ & & $2.953_{-0.147}^{+0.165}$ &  \\ 

 \addlinespace
 
 \hline

\end{tabular}}

\caption{Estimated $H_{0}$, $M_{B}$ and $\beta$ using SNe from two different bins divided by the median $\tilde{x_{1}}$ = 0.087 and  including the $\delta_{\rm bias}$ and $\sigma_{\rm floor}$ values from the Pantheon+SH0ES compilation in the fitting. The number of SNe in each bin is also provided.}
\label{tab:bins_with_bias}

\end{table}

\section{Significance of Discrepancies under Random Partitioning}
\label{ap:by_chance}

To test whether the discrepancies observed between the high- and low-stretch bins defined in Section \ref{sub:bins} could arise by chance, we performed a random-partitioning test analogous to a permutation test. In this procedure, the SN sample was randomly divided 1000 times into two equally sized bins. For each partition, we estimated $H_{0}$ and $M_{B}$ for both bins and computed their differences, producing a  distribution of discrepancies. The $p$-value (denoted here as $P$) was defined as the fraction of random partitions in which the discrepancy in $H_{0}$ and $M_{B}$ was greater than or equal to the observed discrepancy between the high- and low-stretch bins (Table \ref{tab:sne_bins_median}).

\begin{figure}[h]
    \centering

        \includegraphics[width=\linewidth]{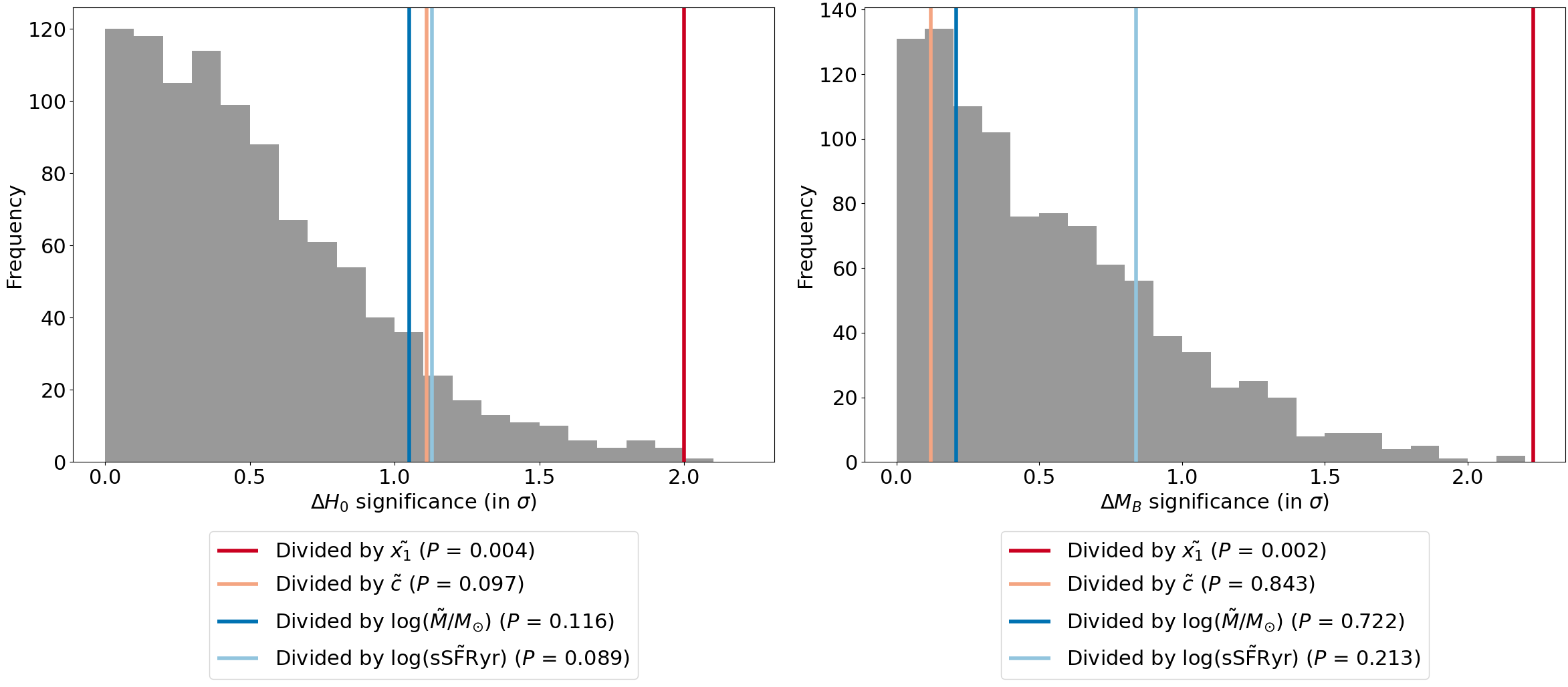}
    
    \caption{Distributions of the 1000 random-partition discrepancies (in units of 
$\sigma$) for $H_{0}$ (Left)
and 
$M_{B}$ (Right). The vertical lines represent the observed discrepancies obtained by splitting the SN sample at the median value of $c$ (yellow), $x_{1}$ (red), log($M/M_{\odot}$) (dark blue) and log(sSFRyr) (light blue) with legends reporting the corresponding 
$P$ values.}
 \label{fig:by_chance}

\end{figure}

As we can see in figure \ref{fig:by_chance}, the $P$ estimated for the $H_{0}$ and $M_{B}$ discrepancies obtained dividing the SNe sample by its median $x_{1}$ are 0.004 and 0.002, respectively. Adopting a significance threshold of $P$ = 0.05, these values indicate that the observed discrepancies in Section \ref{sub:bins} are unlikely to arise from random sampling noise and instead suggest the presence of a population-dependent effect predominantly driven by the stretch parameter. On the other hand, the discrepancies obtained when dividing the SN sample by the median values of the other properties considered in Table \ref{tab:sne_bins1} are consistent with arising by chance, with $P$ $>$ 0.05.

\acknowledgments

G. M., J. D., and A. M. thank the Fundação para a Ciência e Tecnologia (FCT) for the financial support to the Center for Astrophysics and Gravitation (CENTRA/IST/ULisboa) through grant No. UID/PRR/00099/2025 (\url{https://doi.org/10.54499/UID/PRR/00099/2025}) and grant
No. UID/00099/2025 (\url{https://doi.org/10.54499/UID/00099/2025}). J. D. also acknowledges support by FCT under the PhD grant 2023.01333.BD, with DOI \url{https://doi.org/10.54499/2023.01333.BD}. S.G.G. acknowledges support from the ESO Scientific Visitor Programme. All authors thank R. Santos for their insightful comments and suggestions on the work.

%-------------------------------------------------------------------

\bibliographystyle{JHEP}
\bibliography{bibliography.bib}

\end{document}